\documentclass[reprint,aps,prd,twocolumn, amsmath,amssymb,nofootinbib]{revtex4-2}

\usepackage{graphicx}% Include figure files
\usepackage{dcolumn}% Align table columns on decimal point
\usepackage{bm}% bold math

\usepackage[colorlinks=true,
            linkcolor=blue,
            citecolor=blue,
            urlcolor=blue,
            filecolor=blue]{hyperref}
\newcommand*{\Grad}{\vec{\nabla}}
\newcommand*{\Div}{\vec{\nabla}\cdot}

\renewcommand*{\vec}[1]{\ensuremath{\boldsymbol{#1}}}

\usepackage{soul}
\begin{document}
\allowdisplaybreaks

\preprint{APS/123-QED}

\title{Magnetothermal evolution of neutron star cores in the `weak-coupling' regime: implications of ambipolar diffusion for the quiescent X-ray luminosity of magnetars}% Force line breaks with \\
\author{
N.~A. Moraga$^{1\,*}$,
F. Castillo$^{2}$,
D.~D. Ofengeim$^{3,4}$,
A. Reisenegger$^{2}$,
J.~A. Valdivia$^{1}$,
M.~E. Gusakov$^{4}$,
E.~M. Kantor$^{4}$,
A.~Y. Potekhin$^{4}$
}

\affiliation{$^1$Departamento de F\'{\i}sica, Facultad de Ciencias, Universidad de Chile, Las Palmeras 3425, \~Nu\~noa, Santiago, Chile}
\email{nicolas.ventmga@gmail.com}

\affiliation{$^2$Departamento de F\'{\i}sica, Facultad de Ciencias B\'asicas, Universidad Metropolitana de Ciencias de la Educaci\'on, Av. Jos\'e Pedro Alessandri 774, \~Nu\~noa, Santiago, Chile}
\affiliation{$^3$Racah Institute of Physics, The Hebrew University, Jerusalem 91904, Israel}
\affiliation{$^4$Ioffe Institute, Politekhnicheskaya 26, St. Petersburg, 194021, Russia}

\date{\today}

\begin{abstract}
The high quiescent X-ray luminosity observed in some magnetars is widely attributed to the decay and evolution of their ultra-strong magnetic fields. Several dissipation mechanisms have been proposed, each operating with different efficiencies depending on the region of the star. In this context, ambipolar diffusion, i.~e., the relative motion of charged particles with respect to neutrons in the neutron star core, has been proposed as a promising candidate due to its strong dependence on magnetic field strength and its capacity to convert magnetic energy into heat. We perform axisymmetric magnetohydrodynamic simulations to study the long-term magnetic evolution of a NS core composed of normal (non-Cooper paired) matter under the influence of ambipolar diffusion. The core is modeled as a two-fluid system consisting of neutrons and a charged-particle fluid (protons and electrons), coupled to the magnetic field. Simulations are performed both at constant and variable temperatures. In the latter case, a strategy that decouples the magnetic and thermal evolution is employed, enabling efficient thermal modeling across a range of initial magnetic field strengths. At constant temperature, we obtained the expected result where neutrons reach diffusive equilibrium, the Lorentz force is balanced by chemical potential gradients of charged particles, and the magnetic field satisfies a non-linear Grad–Shafranov equation. When thermal evolution is included, fields $B \gtrsim 5\times10^{15}\,\mathrm{G}$ can balance ambipolar heating and neutrino cooling, delaying the evolution over $\sim 10^{3}\,[B/(5\times10^{15}\,\mathrm{G})]^{-6/5}$ yr. Although the surface luminosity is enhanced compared to passive cooling, the heating from ambipolar diffusion alone is insufficient to fully explain the persistent X-ray emission observed in magnetars.

\end{abstract}

\maketitle

\section{Introduction\label{sec:Int} }

Magnetars are a subclass of neutron stars (NSs) that possess the strongest magnetic fields observed in the universe. Initially observed as soft gamma repeaters and anomalous X-ray pulsars, the surface dipole magnetic field strength inferred from their spin-down can be as large as a few times $10^{15}\text{~G}$ \cite{mereghetti2015}, with the possibility of even stronger internal fields. The phenomenology exhibited by magnetars -- strong quiescent emission, short bursts, large outbursts, giant flares, and quasi-periodic oscillations (QPOs), often coupled with intriguing timing behaviors \citep{kaspi2017magnetars} -- makes them a major focus of study. All of these phenomena are thought to be a direct consequence of the evolution and decay of an ultrastrong magnetic field, which is believed to provide the main energy budget \citep{DuncanThompson1995,DuncanThompson1996}. 

In recent decades, significant advances in X-ray astronomy have enabled more precise measurements of quiescent X-ray luminosities, reducing uncertainties. Additionally, kinematic ages have been determined for some sources, making the problem of explaining the magnetar's strong luminosity an excellent opportunity to test and compare magnetothermal simulations with observational data.

In this context, several mechanisms have been proposed (see, e.~g., Ref.~\cite{beloborodov2016}), with the magnetothermal evolution focusing solely on the crust being particularly successful. In the crust, ions have very restricted mobility, so the evolution depends only on the motion of the electrons. Thus, the long-term mechanisms that control the magnetothermal evolution in this region are Ohmic diffusion, i.~e., current dissipation by electric resistivity; and  Hall drift, which corresponds to advection of the magnetic field lines by the electron fluid motion. Although the Hall drift by itself does not dissipate magnetic energy, it changes the magnetic field configuration, creating small-scale structures that dissipate more quickly \citep{goldreich1992magnetic}, possibly explaining the luminosities of some magnetars  \citep{vigano2013}. Additionally, the anisotropy of heat conduction caused by strong magnetic fields can account for the morphology of surface hot spots \citep{dehman20233d, ascenzi2024}, and when the presence of strong toroidal fields is considered, not only can the surface luminosity be explained, but phase-resolved light curves can also reproduce the observational data \citep{igoshev2021strong}. Despite these encouraging results, determining the hot-spot morphology and light curves remains a degenerate problem, as different magnetic field configurations can produce similar surface luminosities and phase-resolved light curves \citep{ascenzi2024}. Furthermore, there is a need to clarify the role that the core plays, as most of these authors make simplified assumptions about the core's magnetic field, either assuming it has been completely expelled (see, e.~g., Ref.~\cite{lander2024meissner}) or that it remains frozen throughout the evolution. In the former scenario, the characteristic length scale of the magnetic field decreases, leading to a stronger crustal field compared to the external dipolar component. This intensifies Joule heating and accelerates the overall crustal evolution \citep{Pons2009,vigano2013,deGrandis2020, deGrandis2021, vigano2021,igoshev2021strong,gourgouliatos+2022magnetic, 
igoshev2023Offsetdipole,dehman20233d, Igoshev2025}. Conversely, in the frozen-field approximation, the magnetic field in the core has a stabilizing effect, significantly slowing down the magnetic evolution in the crust and resulting in minimal heating \citep{vigano2013,vigano2021}.

The magnetothermal evolution in the core is much more complicated than in the crust, as the physics behind it is poorly understood. 
The NS core is a mixture of neutrons (mostly), protons, and electrons, joined by muons and possibly other more exotic species at increasing densities. Also, neutrons and protons are likely to become superfluid and superconducting, respectively, relatively early in the star's life at core temperatures $T\sim 10^{8}-10^{10}\,\text{K}$ (as first suggested by Ref.~\cite{migdal1959superfluidity}; see, e.g., Ref.~\cite{page2013stellar}, for review). 
The presence of superfluidity and superconductivity leads to very complex dynamics as a result of the macroscopic manifestation of the interactions between quantized neutron vortices and magnetic flux tubes \citep{glampedakis2011magnetohydrodynamics, gd16, Gusakovetal2020,bransgrove2024giant}. However, in the present work, we consider the case of a ``normal'' core, i.~e., non-superconducting and non-superfluid, which is likely to be realistic for young magnetars with high temperatures and strong magnetic fields.

The magnetic field in a NS reaches an equilibrium configuration shortly after the proto-neutron star phase (see, e.~g., Ref.~\cite{braithwaite2004fossil,Becerra22a}). Because of frequent collisions between different species of particles, the matter in the NS core initially behaves as a single fluid that is stably stratified due to a radial gradient of the relative abundances, which causes buoyancy forces opposing convective motions \citep{pethick1992,Reisenegger92,goldreich1992magnetic}. On the other hand, due to the high electrical conductivity, the magnetic field lines are effectively frozen into the fluid \citep{blandford2008applications}. 

In order for the magnetic field to evolve, the charged particles must move, overcoming the stable stratification. As explained in Ref.~\cite{Moraga2024}, this can happen by two different mechanisms, each of which dominates in a different temperature range:
\begin{enumerate}
    \item \textbf{Strong coupling regime}: At high core temperatures ($T\gtrsim 5\times10^{8}\,\mathrm{K}$), frequent collisions strongly couple all particle species, causing charged particles and neutrons to move as a single fluid, whose buoyancy forces are overcome by ``Urca reactions'' \citep{shapiro1983physics} that convert neutrons into protons and electrons and vice versa, adjusting the chemical composition \citep{Reisenegger2005,reisenegger2009stable,ofegeim2018}. Contrary to simplified models extensively used in the literature \citep{goldreich1992magnetic,hoyos2008magnetic,beloborodov2016,Passamonti2017ambipolar,igoshev2023,Skiathas2024}, this bulk motion does not depend on a relative motion of neutrons and charged particles and is therefore not affected by inter-particle collisions. 
    
    Numerical simulations in axial symmetry \citep{Moraga2024} show that, in principle, the NS core will evolve to a state in which the matter is in chemical equilibrium and the magnetic field satisfies the Grad-Shafranov equation. However, when the thermal evolution is considered together with the magnetic field evolution, it is found that the rapid decline of the Urca reaction rates does not allow this state to be reached. Even for magnetar-strength fields, the magnetic feedback on the thermal evolution in this regime is negligible, and the magnetic field remains essentially in its initial equilibrium state.

    \item \textbf{Weak coupling regime}: At lower core temperatures ($T\lesssim 5\times 10^{8}\,\mathrm{K}$), Urca reaction rates drop significantly, but the reduced collisional coupling allows ambipolar diffusion (relative motion between charged particles and neutrons) \citep{goldreich1992magnetic,hoyos2008magnetic,reisenegger2009stable,gusakov2017evolution,castillo2020twofluid,Moraga2024}. This relative motion is critical to allow further evolution of the magnetic field, although the latter remains dominated by the bulk motion \citep{ofegeim2018,castillo2020twofluid}.
    This process is likely critical to understanding magnetar activity, due to its dependence on magnetic field intensity \citep{DuncanThompson1995,DuncanThompson1996}.
    The weak coupling regime, particularly its magnetothermal evolution, is the primary focus of this study.

\end{enumerate}

In the development of magnetar models, \citet{DuncanThompson1995, DuncanThompson1996} first highlighted the potentiality of ambipolar diffusion to explain a magnetar's quiescent luminosity through its strong dependence on magnetic field strength and through its ability to convert magnetic energy into internal energy, thereby heating the NS core. Later, \citet{hoyos2008magnetic,Hoyos2010} conducted the first one-dimensional simulations that evolved both the magnetic field and the small density perturbations it induces in charged particles and neutrons, driven by ambipolar diffusion and non-equilibrium Urca reactions within the NS core at constant temperature. Subsequently, \citet{Castillo2017} reported the first axially symmetric simulations of the magnetic field evolution due to ambipolar diffusion, using a one-fluid model in which the neutrons are fixed and only the charged particles can move. This work showed that the magnetic field is transported by a large-scale, mostly solenoidal flow of the charged particles controlled by their collisions against the neutrons. Although it considered a more realistic spherical geometry, it ignored Urca reactions and assumed the friction coefficients to be constant in time (implying a constant temperature), and did not accurately capture the multifluid nature of the core by neglecting the motion of the neutrons. Later, \citet{castillo2020twofluid} addressed these limitations by incorporating the motion of the neutrons and distinct density profiles for neutrons and charged particles (using a toy-model equation of state), accounting for stable stratification. Again, the magnetic field is transported by a solenoidal flow of the charged particles, but now accompanied by a very similar flow of the neutrons, causing a faster evolution, as was predicted by Ref.~\cite{ofegeim2018}. However, neither the effect of Urca reactions nor the thermal evolution were considered in this work.

Recently, the first three-dimensional simulations of ambipolar diffusion within the one-fluid model were reported in Ref.~\cite{igoshev2023}, partially accounting for crustal effects by incorporating a thin shell dominated by Ohmic dissipation. Their study focused on a purely poloidal magnetic field, which was found to be unstable. Similarly, the coupled magnetic evolution of the core and crust at a constant temperature, accounting for the interplay between ambipolar diffusion and the Hall effect, was investigated in Ref.~\cite{Skiathas2024}. However, this study was also conducted within the one-fluid model framework.

The inclusion of the thermal evolution was addressed by Ref.~\cite{beloborodov2016} and \cite{Tsuruta2023} within a simplified one-dimensional, one fluid model with fixed neutrons. This model cannot capture the multifluid nature of the core and the large-scale, mostly solenoidal flows of stellar matter that naturally emerge in the two-fluid model in two (and potentially three) dimensions. Consequently, the conclusions drawn from the one-fluid model require reconsideration, which is the purpose of the present work. 

The present article can be considered as a continuation of \citet{Moraga2024}, where we reported the first simulations of the strong-coupling regime using a new numerical scheme within the anelastic approximation, where time derivatives in the continuity equations are neglected. We note that the inertial terms in the Euler equations had already been neglected in previous work, replacing them with a fictitious friction force \citep{hoyos2008magnetic,Hoyos2010,Castillo2017,castillo2020twofluid,Moraga2024,castillo2025AA}\footnote{Unknown to these authors, this method had been used as ``magneto-frictional method'' in different contexts by Ref.~\cite{Chodura1981,Yang1986,Roumeliotis1994}; and \cite{Vigano2011}.}. The validity of this scheme was thoroughly studied in Ref.~\cite{castillo2025AA}, who compared its results with rigorous semi-analytical calculations and found good convergence in the limit where the fictitious friction force is negligible.
These approximations eliminate sound and Alfv\'en waves, which are damped on timescales much shorter than the times of interest for NS field evolution. 
Here, we apply this approach to the regime of weak coupling, which is astrophysically more interesting than the strong coupling regime. As in Ref.~\cite{Moraga2024}, this leaves only two relevant timescales in the problem, in this case, the physically relevant ambipolar diffusion time, $t_{ad}\propto B^{-2}$, and a fictitious friction time, which can be chosen to be a fixed small fraction of the former. This choice allows us to scale a given simulation to different magnetic field strengths by just rescaling time. It furthermore leaves the collisional coupling strength as the only temperature-dependent physical parameter in the magnetic field evolution, allowing us to decouple the calculation of the magnetic and thermal evolution and to obtain the magnetothermal behavior based on the results obtained from constant-temperature simulations.

This paper is organized as follows: In Sec.~\ref{sec_main_equations}, we present the main equations and assumptions governing the physical model and discuss the time scales associated with the problem. In Sec.~\ref{sec_numerical_method}, we describe the application of the numerical scheme to simulate the weak-coupling regime. In Sec.~\ref{sec_constantT}, we present the outcomes of our simulations at constant temperature, characterizing and analyzing the final equilibrium state. Then, in Sec.~\ref{sec_Tev}, we
describe the same strategy already implemented in Ref.~\cite{Moraga2024} to include the thermal evolution and present the outcomes of the magnetothermal evolution. Finally, in Sec.~\ref{sec_conclusions},
we summarize our results and outline the main conclusions.

%%%%%%%%%%%%%%%%%%%%%%%%%%%%%%%%%%%%%%%%%%%%%%%%%%%%%%%%%%%%%%%%%%%%%%%%%%%%%%%%%%%%%%%%%%%%%%%%%%%%%%%%

\section{Main equations}\label{sec_main_equations}

\subsection{Background stellar model}\label{sec:background}

After the supernova explosion and the proto-neutron star phase, the magnetic field settles into a hydromagnetic equilibrium configuration (see, e.~g., the simulations reported in Ref.~\cite{braithwaite2004fossil} and \cite{Becerra22a}), within a few Alfv\'en crossing timescales
\begin{align}
t_{\text{Alf}}&\sim \dfrac{(4\pi \rho)^{1/2}R}{B}\\
&\sim 7\times 10^{-2}\,\left(\dfrac{R}{12\,\text{km}}\right)\left(\dfrac{\rho}{\rho_{\text{nuc}}}\right)^{\frac{1}{2}}\left(\dfrac{10^{15}\,\text{G}}{B}\right)\,\text{s}\,,
\end{align}
where $\rho$ represents the stellar mass density, $\rho_{\text{nuc}}=2.8\times 10^{14}\text{g\,cm}^{-3}$ is the nuclear saturation mass density, $R$ denotes the stellar radius, and $B$ is the magnetic field strength. 
The NS structure in this hydromagnetic equilibrium state differs only slightly from the spherically symmetric, unmagnetized equilibrium state since the ratio between the magnetic pressure and the typical fluid pressure $P$ in a NS is $B^{2}/8\pi P \lesssim 10^{-6}\,(B/10^{15}\,\mathrm{G})^{2}$ \citep{reisenegger2009stable}. 
Hence, we can consider a background stellar model where we adopt the standard static and spherically symmetric metric, defined through the space-time interval
\begin{equation}
    ds^{2}= -c^{2}\mathrm{e}^{2\Phi(r)/c^{2}}dt^{2}+\mathrm{e}^{2\Lambda(r)}dr^{2}+r^{2}d\theta^{2}+r^{2}\sin^{2}\theta \,d\phi^{2},
\label{ds2}
\end{equation}
where $\mathrm{e}^{\Lambda(r)}=[1-2Gm(r)/(c^{2}r)]^{-1/2}$, $m(r)$ is the gravitational mass enclosed within a radius $r$, i.~e.,
\begin{equation}
    m(r) = 4\pi \int_{0}^{r}\rho(\tilde{r})\tilde{r}^{2}d\tilde{r},
\label{m_r}
\end{equation}
$\rho$ is the mass density, $c$ is the speed of light, and $G$ is the gravitational constant. The lapse function $\mathrm{e}^{\Phi(r)/c^{2}}$ 
accounts for redshift factors, and is determined by
\begin{equation}\label{eqDphiDr}
    \dfrac{d \Phi}{dr} = \dfrac{G m}{r^{2}}\left(1+\dfrac{4\pi r^{3}P}{c^{2}m}\right)\left(1-\dfrac{2Gm}{c^{2}r}\right)^{-1}
\end{equation}
with the boundary condition $\mathrm{e}^{2\Phi(R)/c^{2}} = 1-2GM/(c^{2}R)$, where $M\equiv m(R)$ is the total gravitational mass of the star. We note that $\Phi(r)$ reduces to the gravitational potential in the Newtonian limit ($|\Phi| \ll c^{2}$, i.~e., $\mathrm{e}^{\Phi/c^{2}}\approx \mathrm{e}^{\Lambda}\approx 1$). 
We take the composition of this background model to consist only of electrons ($e$), protons ($p$), and neutrons ($n$) in chemical and hydrostatic equilibrium, i.~e.,
\begin{gather}
\mu_p(r)+\mu_e(r)=\mu_n(r)\equiv\mu(r),\\
 \dfrac{d\mu^{\infty}(r)}{dr}= 0,\label{eqEqBaackground}
\end{gather}
where $\mu_i$ ($i=n,p,e$) represent the chemical potentials of the three particle species, and the superscript $\infty$ denotes quantities redshifted to infinity; here $\mu^{\infty} \equiv \mathrm{e}^{\Phi(r)/c^{2}} \mu(r)$. Eq.~(\ref{eqEqBaackground}) is equivalent to the Tolman--Oppenheimer--Volkoff (TOV) equation \citep{tolman1939static, TOV},
\begin{equation}\label{eq_TOV}
    \dfrac{dP}{dr} = -\dfrac{\varepsilon+P}{c^{2}}\dfrac{d\Phi}{dr},
\end{equation}
to which it reduces with the aid of the thermodynamic relations
\begin{align}
    \varepsilon+P &= \sum_{i =n,p,e} n_{i}\mu_i,\\
    dP &=\sum_{i =n,p,e} n_{i}d\mu_i,
\end{align}
valid for strongly degenerate matter, in addition to the condition of charge neutrality, $n_p(r)=n_e(r)\equiv n_c(r)$. Here, $n_i$ and $\varepsilon = \rho c^2$ correspond to the number density of each particle species and the energy density, respectively.

Thus, the chemical potentials and particle number densities of the magnetized star can be split into a time-independent, spherically symmetric background and a much smaller, time-dependent perturbation induced by the magnetic field:
\begin{align}
	 n_i(\vec{r},t) & = n_i(r) + \delta n_i(\vec{r},t),\quad  (i=n,c),\label{eqDeltan} \\ 
	 \mu^{\infty}_i(\vec{r},t) & =\mu^{\infty}(r) + \delta \mu^{\infty}_i(\vec{r},t),\quad  (i=n,c),\label{eqDeltamu}
\end{align}
where we again used the condition of local charge neutrality, 
\begin{equation}
    n_p(\vec r, t)=n_e(\vec r, t)\equiv n_c(\vec r, t),
\end{equation}
and defined $\delta\mu_c\equiv\delta\mu_p+\delta\mu_e$. In Eqs.~(\ref{eqDeltan}) and (\ref{eqDeltamu}), the number density perturbations are related to the chemical potential perturbations by 
\begin{equation}
       \begin{pmatrix}
    \delta n_{c}  \\
    \delta n_{n} 
\end{pmatrix} = \begin{pmatrix}
    K_{cc}       & K_{cn}  \\
    K_{cn}      & K_{nn} 
\end{pmatrix}^{-1}\begin{pmatrix}
    \delta \mu_{c}  \\
    \delta \mu_{n} 
\end{pmatrix},
\end{equation}
where $K_{ij}\equiv\partial \mu_{i}/\partial n_{j}=K_{ji}\, (i,j=n,c)$. 

In this work, we will use the same microphysical input as in \cite{Moraga2024}, i.~e., we will use the HHJ equation of state (EoS) \citep{Heiselberg_1999} for a NS with a total mass $M=1.4\,M_\odot$; radius $R=12.2\,\text{km}$; core radius $R_{\mathrm{core}}=11.2\,\text{km}$; central mass density $\rho_0= 9.3\times 10^{14}\,\text{g}\,\text{cm}^{-3}$; central pressure $P_0 = 1.2\times 10^{35}\,\text{erg}\,\text{cm}^{-3}$; central number densities $n_{n0}=4.7\times 10^{38}\,\text{cm}^{-3}$ and $n_{c0}=4.2\times 10^{37}\,\text{cm}^{-3}$; and central matrix elements $K_{cc,0}=4.5\times 10^{-42}\,\text{erg}\,\text{cm}^{3}$, $K_{nn,0}=1.3\times 10^{-42}\,\text{erg}\,\text{cm}^{3}$, and $K_{nc,0}= K_{cn,0} = 7.3\times 10^{-43}\,\text{erg}\,\text{cm}^{3}$.

The perturbations in Eqs.~\eqref{eqDeltan} are  also subject to the global constraint of total baryon number conservation,
\begin{equation}
    \delta N_b = \int_{\mathcal{V_{\mathrm{core}}}}\delta n_b\,d\mathcal{V}=0,\label{eq_baryon}
\end{equation}
where $\delta n_b = \delta n_c +\delta n_n$, and $d\mathcal{V}\equiv 2\pi \mathrm{e}^{\Lambda(r)}r^{2}\sin\theta\,drd\theta$ is the proper volume. When weak interactions are neglected, as will be the case in this study (see Sec.~\ref{sec_wkc}), the conservation of charged particles and neutrons can be enforced separately:
\begin{gather}
    \delta N_n = \int_{\mathcal{V_{\mathrm{core}}}}\delta n_n \,d\mathcal{V} = 0\,,\label{eq:cons_n}\\
    \delta N_c = \int_{\mathcal{V_{\mathrm{core}}}}\delta n_c \,d\mathcal{V} = 0 \,.\label{eq:cons_c}
\end{gather}
\begin{figure}[htbp]
    \centering
    \includegraphics[width=8.6cm]{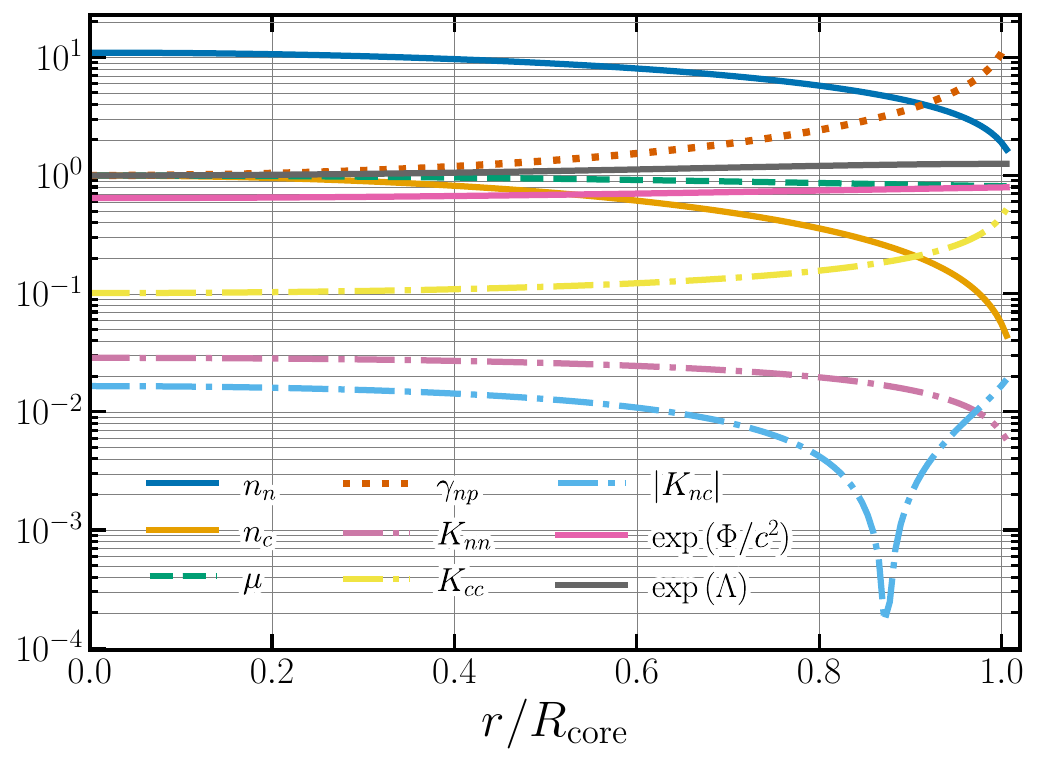}
    \caption{Radial functions derived from the HHJ EoS for an NS with a total radius of $R=12.2 \,\text{km}$, a core radius of $R_{\mathrm{core}}=11.2\,\text{km}$, and a total mass of $M=1.4\,\text{M}_\odot$. The functions have been normalized according to the values listed in Table~\ref{table_code_units}. Metric coefficients, $\mathrm{e}^{\Phi(r)/c^2}$ and $\mathrm{e}^{\Lambda(r)}$, defined in Eq.~\eqref{ds2} are also plotted.} 
    \label{fig_radial_func}
\end{figure}

\subsection{The two-fluid model}\label{Sec:two_fluid} 
This multifluid system evolves under the influence of electromagnetic, pressure, gravitational, and collisional forces. The governing equations have been formulated and studied extensively by various authors throughout the last decades (see e.~g. Refs.~\cite{goldreich1992magnetic,su95,hoyos2008magnetic,Hoyos2010,gusakov2017evolution,ofegeim2018}), with a more recent derivation, accounting for general relativity, provided by \citet{DommesGusakov2021}. The processes of interest involve small velocities that change over timescales much longer than the Alfv\'en crossing time; consequently, the equations of motion can be expressed in the slow-motion approximation, neglecting inertial terms. Also, throughout this work, we adopt the Cowling approximation, assuming that perturbations to the gravitational potential are negligible. Then, the force balance reads
\begin{equation}
    \vec{f}_n +\vec{f}_c +\vec{f}_B=0, \label{eqProtons}
\end{equation}
where
\begin{equation}
    \vec{f}_{i} \equiv -n_{i}\mathrm{e}^{-\Phi/c^{2}}\hat{\nabla}\delta \mu^{\infty}_{i}\quad (i=n,c),
\end{equation}
are forces acting on neutrons and charged particles (including pressure and gravity), and
\begin{equation}\label{eq_fB}
    \vec{f}_B \equiv \dfrac{\vec{J}}{c}\times \vec{B}
\end{equation}
is the Lorentz force, with the electric current density given by
\begin{equation}
     \vec{J}=e n_c(\vec{v}_p -\vec{v}_e)  = \mathrm{e}^{-\Phi/c^{2}}\dfrac{c}{4\pi}\hat{\nabla}\times \left(\mathrm{e}^{\Phi/c^{2}}\vec{B}\right),
\end{equation}
where $e$ is the charge of the proton, and $\vec{v}_{i}$ 
denotes the velocity field of particle species $i$ (more precisely, 
$\vec{v}_{i}$ is the spatial part of the four-velocity defined as in section V of \citet{DommesGusakov2021}; note that $\vec{v}_{i}$ reduces to the ordinary velocity in the Newtonian limit). 
Here, the operator $\hat{\nabla}$ implicitly includes the metric factor $\mathrm{e}^{-\Lambda}$:
\begin{align}
    \hat{\nabla} \equiv & \left(\mathrm{e}^{-\Lambda}\dfrac{\partial}{\partial r}, \dfrac{1}{r}\dfrac{\partial}{\partial \theta}, \dfrac{1}{r \sin \theta} \dfrac{\partial}{\partial \phi}\right).
\end{align}
The Euler equation for the neutrons reads as
\begin{equation}
      0  =\vec{f}_n +\gamma_{np} n_{n}n_{c}\vec{v}_{ad},\label{eqNeutrons}
\end{equation}
with the ambipolar diffusion velocity defined as 
\begin{equation}\label{eqVad0}
\vec{v}_{ad}  \equiv \vec{v}_c -\vec{v}_n  
= \mathrm{e}^{-\Phi/c^{2}}\dfrac{\hat{\nabla}\delta \mu^{\infty}_{n}}{\gamma_{np}n_c}.
\end{equation}
Here, we have used the fact that for a strong magnetic field the velocity difference $\vec{v}_p-\vec{v}_e$ is much smaller than each of the other relevant velocities in order to introduce a single velocity field $\vec{v}_c\equiv\vec{v}_p\approx\vec{v}_e$ for the charged particle fluid. This approximation will also allow us to neglect Ohmic decay in Sec.~\ref{sec:Magnetic_evolution}; however, it is important to note that the relative velocity is small but not entirely negligible, as it gives rise to the electric currents that sustain the magnetic field in the core. Also, here and hereafter, the interaction between electrons and neutrons is neglected as it is much weaker than that between electrons and protons or protons and neutrons. %
The drag coefficients $\gamma_{ij}$ quantify the effect of the particle collisions and are constrained to be symmetric, $\gamma_{ij}=\gamma_{ji}$, in order to satisfy the total momentum conservation. For the interaction between neutrons and protons, it reads \citep{Yakovlev1991}
\begin{equation}\label{eqGammacn}
    \gamma_{np}\approx 5.0\times 10^{-44}\left(\dfrac{T}{10^{9}\text{K}}\right)^2\left(\dfrac{\rho_{\text{nuc}}}{\rho}\right)^{\frac{1}{3}}
    \left(\dfrac{n_{\text{nuc}}}{n_{n}}\right)
    \text{g\,cm}^{3}\text{s}^{-1},
\end{equation}
where $n_{\text{nuc}}=0.16\,\text{fm}^{-3}$ is the nuclear saturation number density, and $T$ is the local temperature.

Finally, the Euler equation for the electrons reads
\begin{align}
  -n_{c}e\left(\vec{E}+\dfrac{\vec{v}_e}{c}\times \vec{B}\right) 
  -n_{c}\mathrm{e}^{-\Phi/c^{2}} \hat{\Grad} \delta \mu_{e}^{\infty} \notag \\
  -\gamma_{ep} n_{c}^{2}(\vec{v}_e- \vec{v}_p) &= 0
  \label{eqElectrons}
\end{align}
where $\vec{E}$ is the electric field.

The velocity fields are further constrained to satisfy the continuity equations, which can be written in the anelastic approximation, namely,
\begin{align}
\mathrm{e}^{-\Phi/c^{2}}\hat{\Grad}\cdot (\mathrm{e}^{\Phi/c^{2}}n_{n}\vec{v}_{n}) &= + \Delta \Gamma,\label{eqCont1} \\
\mathrm{e}^{-\Phi/c^{2}}\hat{\Grad} \cdot \left(\mathrm{e}^{\Phi/c^{2}}n_{c}\vec{v}_{c}\right) &= -\Delta \Gamma,\label{eqCont2}
\end{align}
where we neglected the time derivatives of the number density perturbations, $\delta n_{i}$ ($i=c,n$). 
This eliminates uninteresting dynamics happening on short time scales, allowing for an efficient numerical implementation (see \citet{Moraga2024}). Here, $\Delta \Gamma$ is the net conversion rate per unit volume of charged particles to neutrons by non-equilibrium Urca processes (direct or modified).

\subsection{Magnetic evolution} \label{sec:Magnetic_evolution}
The magnetic evolution is obtained from Faraday's induction equation 
\begin{equation}
     \dfrac{\partial \vec{B}}{\partial t} = -c \hat{\vec{\nabla}}\times \left(\mathrm{e}^{\Phi/c^{2}} \vec{E}\right).\label{eqInduction}
\end{equation}
 
To express $\vec{E}$ as a function of $\vec{B}$ in the form of a generalized Ohm's law, we first derive 
\vec{E} from the quasi-stationary equation for electrons [Eq.~\eqref{eqElectrons}]. By setting 
$\vec{v}_p=\vec{v}_e=\vec{v}_c$ and substituting this into Eq.~\eqref{eqInduction}, we obtain:
\begin{equation}\label{eq_Induction_vc}
    \dfrac{\partial \vec{B}}{\partial t} = \hat{\Grad} \times \left(\mathrm{e}^{\Phi/c^{2}}\vec{v}_{c}\times \vec{B}\right).
\end{equation}

\subsection{Magnetic energy dissipation}

As the magnetic field perturbs the background equilibrium, it induces local deviations from chemical equilibrium and relative motions between neutrons and charged particles. This leads to the dissipation of magnetic energy through non-equilibrium Urca reactions and collisions between charged particles and neutrons. \citet{Moraga2024} derived a detailed expression in the Newtonian limit for the time derivative of the total magnetic energy in the core,
\begin{equation}
    E_B = \int_{\mathcal{V}_{\mathrm{core}}}\mathrm{e}^{\Phi/c^{2}}\dfrac{B^{2}}{8\pi}\,d\mathcal{V} \,,
\end{equation}
that now includes relativistic effects for the above metric. Accounting for all these effects,
\begin{equation}\label{eq_dEBdt_GR}
    \dfrac{dE_B}{dt} = -L_{ad}^\infty-L_{H\nu}^\infty -L_{P}^\infty,
\end{equation}
where
\begin{gather}
    L_{ad}^\infty = \int_{\mathcal{V}_\mathrm{core}} \mathrm{e}^{2\Phi/c^{2}} \gamma_{np} n_n n_c |\vec{v}_{ad} |^{2}\,d\mathcal{V},
\label{L_ad}
\\
    L_{H\nu}^\infty = \int_{\mathcal{V}_\mathrm{core}} \mathrm{e}^{2\Phi/c^{2}} \Delta \Gamma \Delta \mu\,d\mathcal{V},
\label{L_Hnu}
\\
    L_P^\infty = \dfrac{1}{4\pi}\oint_{\partial \mathcal{V}_\mathrm{core}} \mathrm{e}^{2\Phi/c^{2}}  \vec{B}\times \left(\vec{v}_{c}\times \vec{B}\right)\cdot \vec{dS},
\label{L_p}
\end{gather}
are the contributions from ambipolar diffusion, non-equilibrium Urca reactions, and the Poynting flux through the core-crust interface (taken to be $>0$ when going outwards), respectively, and $\Delta\mu\equiv\delta\mu_c-\delta\mu_n$ is the chemical imbalance.

\subsection{Thermal evolution}\label{sec_Tev_0}

During the first $\sim 10-100\,\mathrm{yr}$ after their birth, NSs are non-isothermal, with colder cores due to strong neutrino emission and lower thermal conductivity in the crust. 
Once internal thermal relaxation ends, the redshifted internal temperature, $T
_\infty = T(r)\mathrm{e}^{\Phi(r)/c^2}$, 
becomes uniform in the absence of strong dissipative heating, and its evolution is given by the heat balance equation 
\citep{Thorne1966,Thorne1977,GlenSutherland80,yakovlev2004neutron}
\begin{equation}\label{eqDTdt}
    \dfrac{dT_\infty}{dt}=\dfrac{1}{C}\left(L_\mathrm{H}^\infty-L_{\nu}^\infty-L_\mathrm{cb}^\infty \right),
\end{equation}
where $C$ is the total heat capacity of the NS core, $L_\mathrm{H}^\infty$ is the redshifted total power released by heating mechanisms, $L_{\nu}^\infty$ is the redshifted total neutrino luminosity, and $L_\mathrm{cb}^\infty$ is the redshifted outward heat flux through the core boundary. Here, we assumed that magnetic field dissipation does not generate strong temperature gradients, which would otherwise modify this picture (see Sec.~\ref{sec_conclusions} for a more detailed discussion about the validity of this assumption).

The redshifted luminosities $L_\mathrm{H}^\infty$ and $L_{\nu}^\infty$ are related to the power densities of heating $Q_H(\bm{r})$ and cooling $Q_\nu(\bm{r})$, respectively, measured in the local reference frame, by the integrals over the proper volume of the core: 
\begin{equation}
    L_\mathrm{j}^\infty = \int_{\mathcal{V}_\mathrm{core}} Q_{\mathrm{j}}(\bm{r}) \,\mathrm{e}^{2\Phi(r)/c^2}\,\mathrm{d}\mathcal{V},
\end{equation}
where the index $\mathrm{j}=\nu,H$ denotes either neutrino cooling or heating, respectively. 

In the present case, we are interested in the weak-coupling regime (see Sec.~\ref{sec_wkc}), i.~e., $L^{\infty}_{H} = L^{\infty}_{ad}$, and $L_{H\nu}^{\infty} = 0$. Also, for the HHJ EoS we implemented, the direct Urca mechanism is allowed in the core only for stellar masses $M>1.83\,M_\odot$. Hence, we focus exclusively on the modified Urca process for the neutrino emission,
so that Eq.~\eqref{eqDTdt} reads\begin{equation}\label{eqDTdt_final}
    \dfrac{dT_\infty}{dt}=\dfrac{1}{\tilde{C}}\left(\dfrac{\tilde{L}_\mathrm{ad}^\infty}{T^{3}_{\infty}}-\tilde{L}_{\nu}^\infty T_{\infty}^{7}-\tilde{L}_\mathrm{cb}^{\infty}\,T_{\infty}^{4\delta-1} \right),
\end{equation}
where
\begin{align}
     \tilde{L}_\mathrm{cb}^{\infty} & \equiv L_\mathrm{cb}^{\infty}\,T_{\infty}^{-4\delta} \\
    \tilde{L}^{\infty}_{ad} & \equiv L^{\infty}_{ad}\,T^{2}_{\infty} =\int_{\mathcal{V_{\mathrm{core}}}}\dfrac{n_n}{n_c\tilde{\gamma}_{np}}(\hat{\Grad} \delta \mu_n^{\infty})^2 \,\mathrm{e}^{2\Phi(r)/c^2}\,\mathrm{d}\mathcal{V},\label{eq_Lad_tilde}\\
    \tilde{L}^{\infty}_{\nu} &\equiv \dfrac{L^{\infty}_{\nu}}{T^{8}_{\infty}} = \int_{\mathcal{V}_\mathrm{core}} \tilde{Q}_{\nu}(\bm{r}) \,\mathrm{e}^{-6\Phi(r)/c^2}\,\mathrm{d}\mathcal{V} \\\nonumber
    &= 5.2\times 10^{-32}\,\mathrm{erg}\,\mathrm{s}^{-1}\,\mathrm{K}^{-8}, \\ \label{eq_Lnu_tilde}
    \tilde{C} &\equiv\dfrac{C}{T_{\infty}} = \dfrac{k_B^2}{3\hbar^{3}}\sum_{i=n,p,e}\int_{\mathcal{V_{\mathrm{core}}}} m^{*}_i(n_b)p_{\mathrm{F}i}(n_b)\,\mathrm{e}^{-\Phi(r)/c^{2}}d\mathcal{V}\\\nonumber
      &= 2.6\times 10^{30}\,\mathrm{erg}\,\mathrm{K}^{-2}, \label{eq_C_infty} 
\end{align} 
are temperature-independent coefficients. (Numerical evaluations are presented for the specific NS model used in this paper.) Here, $k_B$ and $\hbar$ denote the Boltzmann and reduced Planck constants, respectively, while $m_{i}^{*}$ and $p_{\mathrm{F}i}$ represent the effective mass and Fermi momentum of each particle species. The latter two quantities depend locally on the baryon number density, given by $n_b = n_c+n_n$. Additionally, we assumed that the redshifted outward heat flux follows a power-law dependence on the redshifted core temperature, $L^{\infty}_{cb}\propto T_{\infty}^{4\delta}$, where the exponent $\delta$ depends on the envelope model (see Appendix~\ref{sec:env} for more details). Furthermore, we defined the rescaled coefficients:
\begin{gather}
    \tilde{\gamma}_{np}\equiv \,\mathrm{e}^{2\Phi(r)/c^2}T^{-2}_{\infty}\,\gamma_{np},\\
    \tilde{Q}_{\nu} \equiv \mathrm{e}^{8\Phi(r)/c^2}\,T_{\infty}^{-8}Q_{\nu}.
\end{gather}

At this point, we highlight a key assumption used in this paper: while the inclusion of General Relativity (GR) is likely negligible for the magnetic evolution -- since the redshift factors are smooth radial functions with typical values $\mathrm{e}^{\Phi/c^{2}} \sim 0.8$, and $\mathrm{e}^{\Lambda} \sim 1.3$ 
(see Fig.~\ref{fig_radial_func}) -- it could play a more significant role in the thermal evolution. In the thermal evolution, the equilibrium neutrino luminosity from Urca reactions plays a crucial role. Since it scales as 
$T^8$ in the modified Urca process, it undergoes a relevant modification by a factor of $\mathrm{e}^{-8\Phi/c^{2}}$, leading to a potentially significant quantitative effect on the thermal history (see e.~g. Ref.~\citep{Ofengeim2017NeutrinoLum}).
Hence, we adopt a hybrid scheme: while we neglect GR effects in the magnetic evolution, we account for them in the thermal evolution. The latter will be implemented by using the redshifted temperature in Eq.~\eqref{eqDTdt_final}, and explicitly incorporating all the redshift factors in Eqs.~\eqref{eq_Lad_tilde}--\eqref{eq_C_infty}, while the rest of the integrand of Eq.~\eqref{eq_Lad_tilde} (the function $\delta \mu_n$) will come from the evolution of the magnetic field that assumes Newtonian gravity. Specifically, in Sec.~\ref{sec_results_magnetothermal}, $\tilde{L}^{\infty}_{ad}$ is computed as
\begin{equation}
   \tilde{L}^{\infty}_{ad} =\int_{\mathcal{V_{\mathrm{core}}}}\dfrac{n_n \mu^{2}}{n_c\tilde{\gamma}_{np}}\left[\Grad \left( \dfrac{\delta \mu_n}{\mu}\right)\right]^2 \,\mathrm{e}^{2\Phi(r)/c^{2}}\mathrm{d}\mathcal{V},
\end{equation}
where $\Grad$ denotes the flat-space gradient operator, while  $\delta \mu_n$ comes from the evolution of the magnetic field (see Sec.~\ref{sec_wkc} for further details).

\subsection{``Weak-coupling'' regime}\label{sec_wkc}

The microphysical processes governing the magneto-thermal evolution in the core are the collisional coupling between charged particles and neutrons, as well as non-equilibrium Urca reactions. 
 
\textit{At high temperatures}, $T\gtrsim 5\times 10^{8}\,\mathrm{K}$, the collisional coupling and Urca reactions are strong, thus the NS core is in the ``strong-coupling'' regime, where the core matter behaves as a single, stably stratified, non-barotropic fluid that locally adjusts its chemical composition, overcoming stable stratification to transport magnetic flux. This regime was extensively studied by \citet{Moraga2024}, who demonstrated 
that the rapid decline of the Urca reaction rates does not allow the magnetic field to evolve significantly, and global chemical equilibrium is not reached in the stellar core before transitioning to the ``weak-coupling'' regime. Thus, the initial hydromagnetic equilibrium field remains essentially frozen during this period.

\textit{At low temperatures}, $T\lesssim 5\times 10^{8}\,\mathrm{K}$, the core is in the ``weak-coupling'' regime, where both the collisional coupling and Urca reactions become significantly less effective. The weakened collisional coupling allows a relative motion between neutrons and charged particles, a process known as ambipolar diffusion. Also, Urca reactions can be completely neglected [$\Delta\Gamma=0$ in Eqs.~\eqref{eqCont1} and \eqref{eqCont2}], as they are very sensitive to temperature. Consequently, neutrons and charged particles behave as two distinct fluid components, coupled through collisions and with the charged component also interacting with the magnetic field. 
This regime is governed by Eqs.~\eqref{eqProtons} and \eqref{eqVad0}, Eqs.~\eqref{eqCont1} and \eqref{eqCont2} with $\Delta \Gamma = 0$, Eq.~\eqref{eq_Induction_vc} and Eq.~\eqref{eqDTdt_final}. Since weak interactions are not considered, the total number of charged particles and neutrons must be conserved independently, as dictated by Eqs.~\eqref{eq:cons_n} and \eqref{eq:cons_c}.

The Newtonian gravity approximation of the equations that describe 
magnetic field evolution in this regime,
and which are used for numerical solutions in 
the present work, is obtained by setting $\mathrm{e}^{\Phi(r)/c^{2}}=\mathrm{e}^{\Lambda(r)} = 1$, and  replacing $\hat{\Grad}  $ with the standard flat-space gradient operator $\Grad$. Under these assumptions, the governing equations take the form: 
\begin{gather}
 \dfrac{\partial \vec{B}}{\partial t} = \Grad \times \left(\vec{v}_{c}\times \vec{B}\right), \label{eq_IndB_NGR}\\
 \vec v_c = \vec v_n + \vec v_{ad},\\ 
 \vec{v}_{ad}   = \dfrac{\mu\Grad \chi_n}{\gamma_{np}n_c},\label{eq_Vad_NGR}\\
  \vec{f}_B+\vec f_ c + \vec f_n =0,\\
  \vec f_i =-n_i \mu \Grad \chi_i,\quad \,(i=n,c),\label{eqForces_NGR}\\
\Grad\cdot (n_{n}\vec{v}_{n})= 0,\label{eqCont1_NGR}\\
   \Grad \cdot \left(n_c \vec{v}_{c}\right)= 0.\label{eqCont2_NGR}
\end{gather}
Here, we define $\chi_i \equiv \delta\mu_i/\mu$. To derive the expressions for the forces 
in Eq.~\eqref{eqForces_NGR}, we used the approximation $\delta \mu^{\infty}_{i} \approx (1+\Phi/c^{2})\delta \mu_{i}$ valid in the Newtonian limit, where $|\Phi|/c^{2}\ll 1$, along with Eq.~\eqref{eqEqBaackground}. Thus, the forces include both pressure and gravitational contributions for each fluid component.

The expression \eqref{eq_dEBdt_GR} for magnetic energy dissipation in the Newtonian gravity approximation is given by
\begin{equation}\label{eqEdotB}
    \dfrac{dE_B}{dt}=-L_{H\nu}-L_{ad}-L_{P} - \dot{E}_c -\dot{E}_n,
\end{equation}
where
\begin{gather}
    L_{H\nu}=\int_{\mathcal{V}_\mathrm{core}}  \Delta \Gamma \Delta \mu\,d\mathcal{V},\label{eqLHnu}\\
      L_{ad}=\int_{\mathcal{V}_\mathrm{core}} \gamma_{np}n_{c}n_{n}|\vec{v}_{\text{ad}}|^{2}\,d\mathcal{V},\label{eqLa}\\
      L_P = \dfrac{1}{4\pi}\oint_{\partial \mathcal{V}_\mathrm{core}} \vec{B}\times \left(\vec{v}_{c}\times \vec{B}\right)\cdot \vec{dS},\\\label{eqLp}
      \dot{E}_c =\int_{\mathcal{V}_\mathrm{core}} n_c\frac{\delta\mu_c}{c^2}\Grad\Phi\cdot \vec{v}_c \,d\mathcal{V}, \, \\
      \dot{E}_n =\int_{\mathcal{V}_\mathrm{core}} n_n\frac{\delta\mu_n}{c^2}\Grad\Phi\cdot \vec{v}_n \,d\mathcal{V} \,.
\end{gather}

As in the relativistic case, strictly dissipative terms -- those that always reduce the magnetic energy in the NS core -- appear in Eq.~\eqref{eqEdotB}, corresponding to energy dissipation due to non-equilibrium Urca reactions ($L_{H\nu}$) and ambipolar diffusion ($L_{ad}$). In contrast, the Poynting flux through the core-crust interface ($L_{P}$), as well as the terms $\dot{E}_c$ and $\dot{E}_{n}$, can have either sign.
The terms $\dot{E}_c$ and $\dot{E}_{n}$ 
are not physically meaningful and arise purely as an artifact of the Newtonian approximation. \citet{Moraga2024} showed that, in the strong-coupling regime, $\dot{E}_c$ and $\dot{E}_n$ are negligible. However, as we will see in Sec.~\ref{sec_constantT}, these terms become necessary in the weak-coupling regime to ensure the numerical consistency with Eq.~\eqref{eqEdotB}.

\subsection{Timescale}

The typical time over which the magnetic field evolves in this regime can be estimated from Eq.~\eqref{eq_IndB_NGR} as 
\begin{equation}\label{eq_tB}
t_{B}\sim \frac{\ell_B}{|\vec v_c|}=\frac{\ell_B}{|\vec v_n+\vec v_{ad}|}\,,
\end{equation}
where $\ell_B$ is the characteristic length-scale of the magnetic field (typically a fraction of the core radius), and
\begin{equation}
    v_{ad} \sim  \frac{f_n}{\gamma_{np}n_c n_n} 
    \sim \frac{f_B}{\gamma_{np}n_c n_n} \sim \frac{B^2}{4\pi\ell_B\gamma_{np}n_c n_n}
\end{equation}
[note that we set $f_n \sim f_B$ in this estimate, which is generally not the case, see Eq.\ \eqref{eq:va estimate} below].
It is often assumed that the neutron velocity (or the net baryon velocity, which is very similar) can be ignored \citep{goldreich1992magnetic,Pons2009,beloborodov2016,passamonti2017magnetic,igoshev2023,Skiathas2024}, so the typical time-scale of magnetic field evolution reduces to 
\begin{equation}
t_{ad} \sim \frac{\ell_B}{v_{ad}} \sim \frac{4\pi \gamma_{np}n_c n_n \ell_B^{2}}{B^2}\,.
\end{equation}
This estimate, however, overestimates the true evolutionary time-scale in the two-fluid model. As first shown by Ref.~\cite{gusakov2017evolution} and \cite{ofegeim2018}, and later corroborated with detailed numerical simulations by Refs.~\cite{castillo2020twofluid,castillo2025AA}, the bulk neutron velocity is larger than the ambipolar velocity by a factor $ v_n/v_{ad} \sim 10-100$ depending on the magnetic field geometry and EoS in the core. 

A better estimate might be done by following \citet{Moraga2024}; namely, for an arbitrary magnetic field, the fluid forces plus gravity can be larger than the Lorentz force, i.~e. $f_c \sim f_n \sim (\ell_c/\ell_B)f_{B}$ (as explicitly demonstrated in \citet{Moraga2024}), where we introduced a length scale
\begin{equation}\label{eq:lc}
    \ell_{c}\equiv -[d\ln(n_{c}/n_{b})/dr]^{-1}
\end{equation}
with the baryon number density $n_b = n_n + n_c$. 
Equation~\eqref{eq:lc} generally satisfies $\ell_c \gtrsim \ell_B$ as the radial profiles $n_c(r)$ and $n_n(r)$, although not identical, exhibit similar smooth decreases over a typical length-scale $\sim \ell_c \sim R_{\text{core}}$. 
Taking this into account, a better estimate of the ambipolar diffusion velocity is
\begin{equation}
    v_{ad} \sim  \frac{f_n}{\gamma_{np}n_c n_n} 
    \sim \frac{\ell_c}{\ell_B}\frac{f_B}{\gamma_{np}n_c n_n} \label{eq:va estimate}\,,
\end{equation}
and, as shown by Ref.~\cite{castillo2025AA}, the neutron velocity can be estimated as
\begin{equation}
    v_n\sim \frac{\ell_{c}}{\ell_B} \frac{f_n}{ \gamma_{np} n_c n_n} \sim \frac{\ell_{c}}{\ell_B}v_{ad}\,. \label{eq:vn-va}
\end{equation}
Thus, by substituting Eqs.~\eqref{eq:va estimate} and \eqref{eq:vn-va} into Eq.~\eqref{eq_tB}, we obtain a more accurate estimate

\begin{align}\label{eq_tBf}
    t_B \sim &\left(\dfrac{\ell_B}{\ell_c}\right)^{2}t_{ad} 
    \sim \frac{4\pi \gamma_{np}n_c n_n \ell_B^{4}}{\ell_c^{2}B^2} \notag\\
    \sim &\, 5.6\times 10^{3}\,\left(\dfrac{T}{10^{9}\,\mathrm{K}}\right)^{2}\,
    \left(\dfrac{B}{10^{15}\,\mathrm{G}}\right)^{-2} \notag\\
    &\quad\times 
     \left(\dfrac{\ell_B}{2\,\mathrm{km}}\right)^{4}\left(\dfrac{\ell_c}{10\,\mathrm{km}}\right)^{-2}\,\mathrm{yr} .
\end{align}

\begin{table}
\centering
\begin{tabular}{ |p{1.9cm}|p{2.1cm}|p{4cm}|}
 \hline 
 Variable & Normalization  & Value\\
\hline
 $B$ & $B_{\mathrm{init}}$ &\,\,\,\,\,--------- \vspace{0.15cm}\\
 $\ell_c,\,\ell_B,\,r$ & $R_{\mathrm{core}}$ & $11.2\,\text{km}$ \\
 $\alpha$ & $B_{\mathrm{init}}R_{\mathrm{core}}^{2}$ & \,$1.2\times 10^{27}B_{\mathrm{init},15}\,\text{G cm}^{2}$ \\
 $\beta$ & $B_{\mathrm{init}}R_{\mathrm{core}}$ & \,$1.1\times 10^{21}B_{\mathrm{init},15}\,\text{G cm}$ \\
 $\gamma_{np}$     & $\gamma_{np0}$ &  $1.13\times 10^{-44}\,T_{9}^{2}\,\mathrm{g}\,\mathrm{ cm}^{3}\,\mathrm{s}^{-1}$\\
 $n_c , n_n$     & $n_{c0}$ & $4.2\times 10^{37}\,\text{cm}^{-3}$ \\
$\delta\mu_j$ & $\frac{B_{\mathrm{init}}^{2}}{4\pi n_{c0}}$ & $1.9\times 10^{-9}B^{2}_{\mathrm{init},15}\,\text{erg}$ \\
$t$ &$\frac{4\pi \gamma_{np0}n_{c0}^{2}R_{\mathrm{core}}^{2}}{B_{\mathrm{init}}^{2}}$ & $1.0\times 10^{7}\,B^{-2}_{\mathrm{init},15}T^{2}_{9}\,\text{yr}$  \\
$\zeta$ &  $n_{c0} \gamma_{np0}$ & $4.7 \times 10^{-7}\,T_{9}^{2}\,\mathrm{g}\,\mathrm{s}^{-1}$ \\
$E_{B}$ & $ \frac{B_{\mathrm{init}}^{2}R_{\mathrm{core}}^{3}}{6}$ & $2.3 \times 10^{47}B_{\mathrm{init},15}^{2}\,\text{erg}$  \\
$L_{ad}\,,L_\mathrm{cb},\,L_{P}$ & $\frac{B_{\mathrm{init}}^{4}R_{\mathrm{core}}}{24\pi \gamma_{np0}n_{c0}^{2}}$ & $7.3\times 10^{32}B_{\mathrm{init},15}^{4}T^{-2}_{9}\,\text{erg}\text{ s}^{-1}$ \\

 \hline
\end{tabular}
\vspace{0.2cm}
 \caption{Summary of the code units, showing the normalization of the different variables, their notations and physical values obtained from the HHJ EoS, for a NS that has a total radius $R=12.2\,\text{km}$, a core radius $R_{\mathrm{core}}=11.2\,\text{km}$, and a total mass $M=1.4\,\textup{M}_\odot$. Here, the free parameters are the initial rms magnetic field strength, $B_{\mathrm{init}}= \langle \vec{B}(t=0)\rangle_{\mathrm{rms}}\equiv[(\int_{\mathcal{V}_{\mathrm{core}}}\vec{B}^{2}\,d\mathcal{V})/\mathcal{V}_{\mathrm{core}}]^{0.5}$, and the temperature, $T$, which are needed to recover the physical units from the simulation output. The notation $B_{\mathrm{init},15}$ denotes $B_{\mathrm{init}}/10^{15}\, \text{G}$, and the subscript 0 indicates evaluation at the stellar center, i.e., $n_{c0} \equiv n_c(r=0)$ and $\gamma_{np0} \equiv \gamma_{np}(r=0)$.
 }
 \label{table_code_units}
\end{table}

\begin{figure*}
    \centering
    \includegraphics[width=9.5cm]{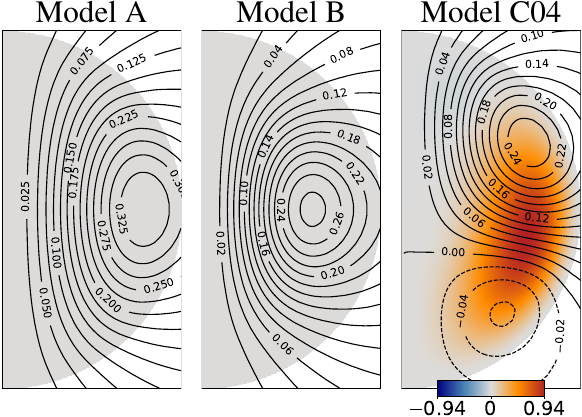}
    
    \caption{Initial magnetic field configurations given by the potentials listed in Tables~\ref{tab_initial_B} and~\ref{tab_B_model}.
    The lines represent the poloidal magnetic field, labeled by the magnitude of $\alpha_{\mathrm{init}}$, while the colors indicate the strength of the function $\beta_{\mathrm{init}}$, which is related to the toroidal field.
    These variables are plotted in code units; see Table~\ref{table_code_units}.}
    \label{fig1}
\end{figure*}

\begin{table*}
    \centering
    \begin{tabular}{||c|c||}
    \hline
   
   &\,Poloidal flux function \\
     \hline
     Abbreviation & $\alpha(r,\theta)$ \\
     \hline
       $ \alpha_{1}$ & $-1.34\,r^{2}(1-\frac{6}{5} r^{2}+\frac{3}{7}r^{4})P^{1}_{1}(\cos \theta)\sin \theta$ \\
    $ \alpha_{2}$ & $-1.24\,r^{3}(1-\frac{10}{7} r^{2}+\frac{5}{9}r^{4})P^{1}_{2}(\cos \theta)\sin \theta$ \\
    $ \alpha_{3}$ & $\sqrt{0.3}\,\alpha_1 + \sqrt{0.7}\,\alpha_2$ \\
    $ \alpha_{4}$ & $(2.05\,r^2-5.20\,r^4+5.80\,r^6-3.10\,r^8+0.65\,r^{10})P^{1}_{1}(\cos \theta)\sin \theta$ \\
     \hline	
   &  \,Poloidal current function   \\
   \hline
   Abbreviation & $\beta(r,\theta)$ \\
   \hline
    $\beta_1$ & $112.54\,r^{5}(1-r)^{2}\sin^{2}\theta \sin\left(\theta-\frac{\pi}{5}\right)$  \\
    \hline
    \end{tabular}
    \caption{Scalar functions $\alpha(r,\theta)$ and $\beta(r,\theta)$ for the initial magnetic field configurations used in our simulations. Each configuration satisfies the normalization conditions $\langle \vec{B}^{\,\text{pol}}_{\mathrm{init}} \rangle =1$ and $\langle \vec{B}^{\,\text{tor}}_{\mathrm{init}} \rangle =1$, respectively, where $\langle \ldots \rangle$ denotes an root-mean-square (rms) average over the stellar core. Here, $P^{1}_{\ell}(\cos \theta)$ is the associated Legendre polynomial of order $\ell$ with azimuthal index $m=1$. The poloidal flux functions, $\alpha$, match the core magnetic field with the external multipolar expansion corresponding to a current-free magnetic configuration. In particular, $\alpha_1$ and $\alpha_2$ correspond to one of the simplest types of magnetic field configurations that can be analytically constructed, subject only to the additional condition that the azimuthal current density vanishes at the stellar surface, $J_{\phi}(r=R,\theta)=0$ (see, e.~g., Refs.~\cite{akgun2013} and \cite{Armaza_2015}). The magnetic field configuration corresponding to $\alpha_4$ is more constrained, also satisfying the \textit{non-penetration} condition for the ambipolar and neutron velocity at the crust-core interface, Eq.~(\ref{eq:non-penetration}) (see Ref.~\cite{ofegeim2018} for more details).} 
    \label{tab_initial_B}    
\end{table*}
\begin{table*}
    \centering
    \begin{tabular}{||c|c|c|c||}
    \hline
         Model & Poloidal flux function & Poloidal current function & $E_{B}^{\mathrm{tor}}/E^{\mathrm{total}}_{B}$\\
         \hline
         A& $\alpha_1$ & 0 & 0 \\
         B& $\alpha_4$ & 0 &  0\\
         C0& $\alpha_3$ & 0 & 0\\
         C04& $\sqrt{0.6}\, \alpha_3$ & $\sqrt{0.4}\, \beta_1$ & 0.4\\
        C06 &  $\sqrt{0.4}\,\alpha_3$ &$\sqrt{0.6}\,\beta_1$ & 0.6 \\
        C08 &  $\sqrt{0.2}\,\alpha_3$ &$\sqrt{0.8}\,\beta_1$ & 0.8  \\
        \hline
    \end{tabular}
    \caption{ Initial magnetic field models used in our simulations. The poloidal flux and current functions are provided in Table~\ref{tab_initial_B}. Here, $E^{\mathrm{tor}}_{B}/E_{B}^{\mathrm{total}}$ is the initial toroidal energy fraction. Some of these models are shown in Fig.~\ref{fig1}.}
    \label{tab_B_model}
\end{table*}

\subsection{Axially symmetric magnetic field}\label{sec:AxiallSym}

We consider an axially symmetric magnetic field, which can be written as
\begin{equation}\label{eqPoloidaToroidalB}
    \vec{B} = \Grad \alpha \times \Grad \phi+\beta \Grad\phi,
\end{equation}
where the scalar potentials $\alpha(r,\theta,t)$ and $\beta(r,\theta,t)$ generate the poloidal (meridional) and toroidal (azimuthal) magnetic field components, respectively.
Here, $r$ is the radial coordinate, and $\theta$ and $\phi$ are the polar and azimuthal angles, respectively, so $\Grad\phi=\hat{\phi}/(r\sin \theta)$, where $\hat\phi$ is the unit vector in the $\phi$ direction. The functions $\alpha(r,\theta,t)$ and $\beta(r,\theta,t)$ are known as the poloidal flux and poloidal current functions, respectively, because $2\pi \alpha(r,\theta,t)$ is the magnetic flux and $c\beta(r, \theta, t)/2$ is the electric current enclosed by an azimuthal circle at given $r$ and $\theta$. The curves $\alpha=\text{const}$ are the poloidal magnetic field lines. 

We note that GR effects are not considered here, so $\mathrm{e}^{\Phi(r)/c^{2}} = \mathrm{e}^{\Lambda(r)} = 1$ and $\hat{\Grad} \rightarrow \Grad$, consistent with the assumption adopted in Sec.~\ref{sec_Tev_0}.

The Lorentz force has poloidal and toroidal components, which read
\begin{equation}
\vec{f}_{B}^{\,\mathrm{pol}}=-\frac{\Delta^*\alpha\Grad\alpha+\beta\Grad\beta}{4\pi r^2\sin^2\theta},\label{eqfBPol}
\end{equation}
\begin{equation}
\vec{f}_{B}^{\,\mathrm{tor}}=\frac{\Grad\beta\times\Grad\alpha}{4\pi r^2\sin^2\theta} .\label{eqfBTor}
\end{equation}
Here, 
\begin{align}
    \Delta^{*} & \equiv r^{2}\sin^{2}\theta\, \Grad \cdot\left(\dfrac{1}{r^{2}\sin^{2}\theta  }\Grad\right) \\\nonumber & = \dfrac{\partial^{2} }{\partial r^{2}}+\dfrac{\sin \theta}{r^{2}}\dfrac{\partial}{\partial \theta}\left(\dfrac{1}{\sin \theta}\dfrac{\partial}{\partial \theta}\right)\label{GS_operator}
\end{align}
is the ``Grad-Shafranov (GS) operator''. In axial symmetry, there is no pressure gradient or gravitational force available to balance the Lorentz force in the azimuthal direction. Therefore, a necessary condition for hydromagnetic equilibrium is that the toroidal magnetic force must vanish at all times, $\vec{f_{B}}^{\,\mathrm{tor}}=0$. Thus, from Eq.~\eqref{eqfBTor}, $\Grad \alpha \parallel \Grad \beta$, so one potential is (at least locally) a function of the other, $\beta = \beta (\alpha)$. As a consequence, Eq.~(\ref{eqfBPol}) can be simplified as 
\begin{equation}\label{eqfBPol1}
    \vec{f}_{B}^{\,\text{pol}}=-\frac{\Delta^*\alpha+\beta\beta'}{4\pi r^2\sin^2\theta}\Grad\alpha.
\end{equation}
(Here and below, primes denote derivatives with respect to $\alpha$.)

\section{Numerical method}\label{sec_numerical_method}
\subsection{Artificial friction method}
For a given initial magnetic field configuration and a specified equation of state, the system of Eqs.~\eqref{eqProtons}--\eqref{eqNeutrons} and \eqref{eqCont1}--\eqref{eqCont2} allows for the calculation of $v^{\,\mathrm{\,pol}}_n$ and $v^{\,\mathrm{\,pol}}_{ad}$ using the method proposed by \citet{gusakov2017evolution} and then improved by \citet{ofegeim2018}. However, enforcing the \textit{non-penetration} boundary condition, 
\begin{equation}
    \vec{v}_{ad} \cdot \vec{\hat r}\big|_{r=R_{\text{core}}} = \vec{v}_n \cdot \vec{\hat r} \big|_{r=R_{\text{core}}} = 0, \label{eq:non-penetration}
\end{equation}
is not trivial in this approach. To satisfy this condition, the initial magnetic field configuration must be carefully constructed so that the resulting velocity fields inherently respect the boundary condition, serving as a natural consistency requirement. In some simple cases, such configurations can be determined for a given equation of state. Furthermore, in this approach $\vec{v}_{n}$ depends on several spatial derivatives of the Lorentz force, which makes this scheme very challenging to implement numerically in order to evolve the magnetic field in time. 

To avoid these difficulties, we use the artificial friction method originally implemented for NS cores by Refs.~\cite{hoyos2008magnetic,Hoyos2010} in one-dimensional simulations, then in axial symmetry by Ref.~\cite{castillo2020twofluid} and \cite{Moraga2024}, and recently validated by Ref.~\cite{castillo2025AA}. This method allows to self-consistently determine the chemical potential perturbations and the velocity fields once a magnetic field configuration has been specified. 

In this approach, a modified net force balance is introduced:
\begin{equation}
    \vec{f}_c +\vec{f}_n +\vec{f}_B -\zeta n_n\vec v_n=0,\label{eqVn}
\end{equation}
where we have included an artificial force $-\zeta n_n \vec{v}_{n}$ acting on the neutrons, parameterized by the small artificial friction coefficient $\zeta$. This force helps to mimic the short-term dynamics driven by inertial terms in the Euler equations, effectively filtering out Alfv\'en waves and allowing for the relaxation of non-equilibrium poloidal-toroidal configurations (i.~e., those that do not satisfy $\beta = \beta(\alpha)$; see Sec.~\ref{sec:AxiallSym}) into equilibrium states over an effective Alfv\'en timescale, 
\begin{equation}\label{eqTBz} 
t_{\zeta B} \sim \dfrac{\ell_{B}}{v_{n}} \sim \frac{4 \pi n_{n} \ell_{B}^{2} \zeta}{B^{2}}. 
\end{equation}

This scheme offers several advantages: It not only enables one to self-consistently solve for the velocity fields, but also allows one to enforce the non-penetration boundary condition at each integration time step. Moreover, it has been shown to converge to the analytical solution derived by \citet{ofegeim2018} in the limit $\zeta \rightarrow 0$ ($\zeta \ll n_c \gamma_{np}$; see e.~g Ref.~\cite{castillo2025AA}).

\subsection{Numerical implementation}

The numerical implementation is divided into three steps: First, we solve a boundary value problem to determine the scalar functions $\chi_c$ and $\chi_{n}$, which allows us to compute the ambipolar velocity drift $\vec{v}_{ad}$ and the neutron velocity $\vec{v}_{n}$ for a given magnetic field configuration $\vec{B}(t^{*})$ at a given time $t^{*}$. Secondly, we integrate the induction equation to evolve the magnetic field in time, updating  $\vec{B}(t^{*}) \rightarrow \vec{B}(t^{*}+\Delta t)$. Finally, we repeat the process using the updated magnetic field configuration, $\vec{B}(t^{*}+\Delta t)$, closing the iterative loop.

The first step is done by substituting the expressions for $\vec{v}_{n}$ [Eq.~\eqref{eqVn}] and $\vec{v}_{ad}$ [Eq.~\eqref{eq_Vad_NGR}] into the continuity equations [Eqs.~\eqref{eqCont1_NGR}--\eqref{eqCont2_NGR}], namely,
\begin{gather}	\Div\left(n_n\mu\Grad \chi_n+n_c\mu\Grad \chi_c\right)
 = \Div\vec{f}_{B}\label{eq:continuity_n4}\,,\\
\Grad\cdot\left[n_c\mu\Grad \chi_n +  g(\zeta)n_c\mu\Grad \chi_c    \right]
  = \Div\left[ g(\zeta) \vec{f}_{B}\right]  \,, \label{eq:continuity_c4}
\end{gather}
where 
\begin{equation}
     g(\zeta,r)= \frac{\zeta}{\gamma_{np} n_n} + \frac{n_c}{n_n}.
\end{equation}
This is a set of parabolic equations for $\chi_c(r,\theta)$ and $\chi_n(r,\theta)$ for a given magnetic field configuration (which determines the right-hand side of both equations). It is supplemented with the constraint of neutron and charged-particle number conservation [Eqs.~\eqref{eq:cons_n} and \eqref{eq:cons_c}], and with the non-penetration boundary conditions [Eq.~\eqref{eq:non-penetration}], which read as
\begin{gather}
    \frac{\partial \chi_c}{\partial r}\Big |_{r=R_{\text{core}}} = \frac{\vec{f}_B}{ n_c \mu}\cdot\vec{\hat{r}}\Big |_{r=R_{\text{core}}},\\
    \frac{\partial \chi_n}{\partial r}\Big |_{r=R_{\text{core}}} =0.
\end{gather}
At each time step, this system is solved through a finite-difference method, and the velocity fields are evaluated from Eqs.~\eqref{eq_Vad_NGR} and \eqref{eqVn}. 

The time evolution of the magnetic field is performed using a conservative approach for the toroidal component, i.~e., employing a finite-volume scheme to evolve $\beta$. 
For the poloidal component, i.~e., to evolve $\alpha$, the time derivative is calculated using a finite-difference method, ensuring second-order accuracy in time. This approach guarantees that the solenoidal condition $\Grad\cdot\vec{B}=0$ is satisfied up to machine precision throughout the entire simulation. 

The code used in this work is the same as in Ref.~\cite{Moraga2024}, designed to evolve the magnetic field under axial symmetry within the Newtonian gravity framework. Earlier versions of this code can be traced back to Refs.~\cite{Castillo2017,castillo2020twofluid}. For more details on the numerical methods, we refer the interested reader to these references.

\subsection{Code units}

The code is written in dimensionless units according to Table~ \ref{table_code_units}. Figure~\ref{fig_radial_func} shows all the background relevant radial functions in dimensionless units.
For the time scales, we used as a reference the expressions 
\begin{align}
    t_{\zeta B} &\equiv \dfrac{4\pi n_{n0} \ell_B^{2} \zeta}{ B_{\mathrm{init}}^{2}}, \\
    t_{ad} &\equiv \frac{4\pi \gamma_{np0}n_{c0}n_{n0}R_{\mathrm{core}}^{2}}{B_{\mathrm{init}}^{2}}\left(\dfrac{\ell_B}{R_{\mathrm{core}}}\right)^{2}.  \label{eqtlambdaB_ref}
\end{align}
In dimensionless code units of Table \ref{table_code_units}, these timescales are
given by
\begin{align}\label{eqtzBdiless2}
t_{\zeta B} & 
= t_{ad}\, \dfrac{\zeta}{n_{c0}\gamma_{np0}},
\end{align}
\begin{align}
t_{ad} & = \dfrac{n_{n0}}{n_{c0}}\left(\dfrac{\ell_B}{R_{\mathrm{core}}}\right)^{2} = 0.684\,.\label{eqtadBdiless}
\end{align}
To derive \eqref{eqtadBdiless} we set $\ell_B /R_{\mathrm{core}} \approx 1/4$. The results in Secs.~\ref{sec_constantT} and~\ref{sec_Tev} will be presented in code units and physical units, respectively.

\subsection{Initial magnetic field configurations}

Figure~\ref{fig1} and Table~\ref{tab_initial_B} illustrate and provide details of the different initial magnetic field configurations used in our simulations. We evolved three distinct large-scale poloidal magnetic configurations, two dipolar (models A and B) and one consisting of a dipole and a quadrupole component contributing 30\% and 70\%, respectively, of the poloidal magnetic energy (model C0). The latter was further evolved with varying ratios of initial toroidal to total magnetic energy, denoted as C04, C06, and C08. All simulations were run using the same radial and angular resolution, namely a grid with $N_r = 60$ radial points and $N_{\theta} = 90$ angular points, and a value of $\zeta/(n_c \gamma_{np}) = 10^{-4}$.
\begin{figure}
    \centering
    \includegraphics[width=8cm]{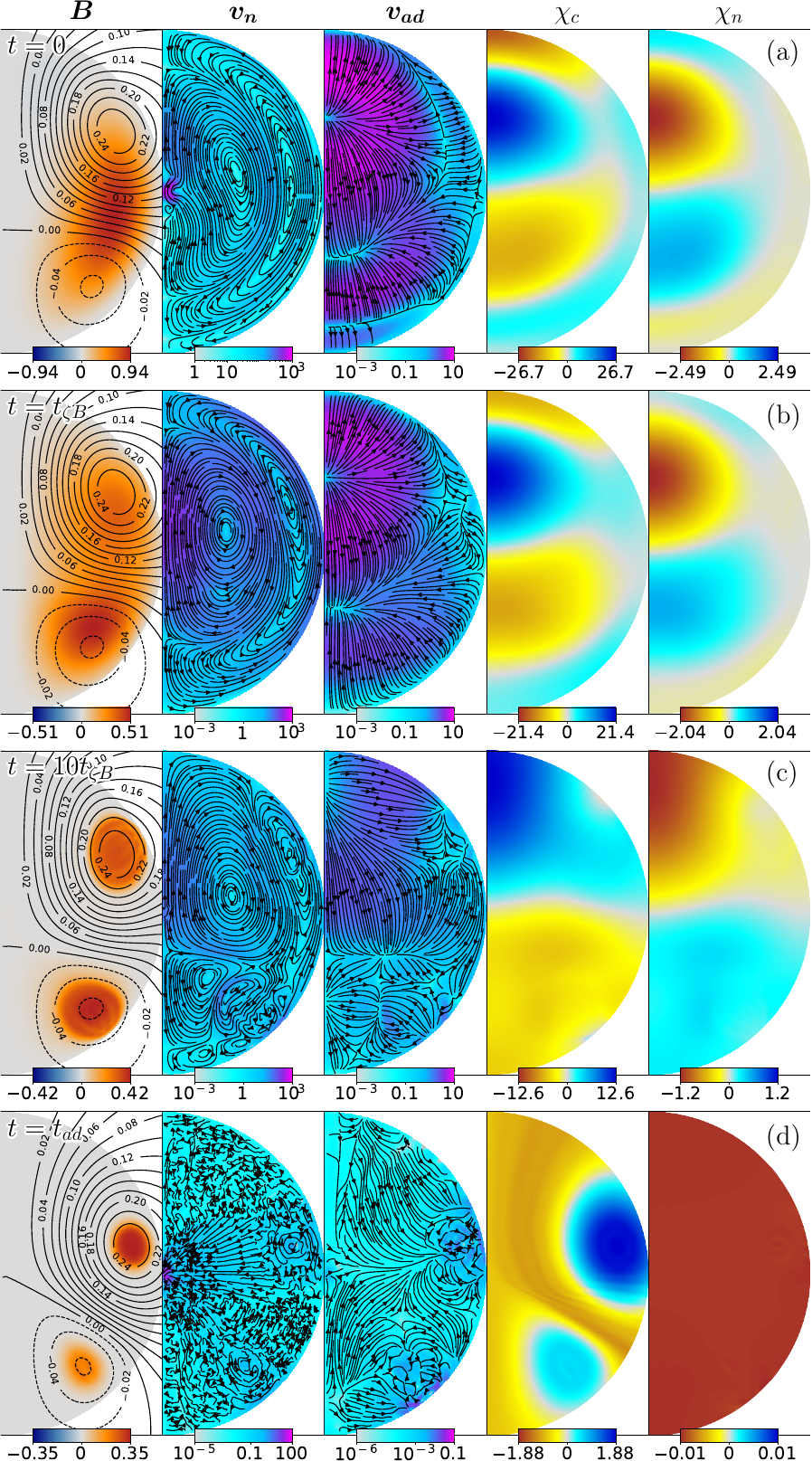}
    \caption{Magnetic evolution at constant temperature for model C04 with $\zeta/(n_{c0}\gamma_{np0})=10^{-4}$, corresponding to $t_{\mathrm{ad}}/t_{\zeta B}=10000$.  
    All panels are meridional cross-sections of the star. Rows (a), (b), (c), (d) correspond to different times, $t=0,t_{\zeta B},10\,t_{\zeta B}$, and $\,t_{\mathrm{ad}}$, respectively. In each row, the five panels display, from left to right: the magnetic field configuration, where lines represent the poloidal magnetic field, labeled by the magnitude of $\alpha$, and colors represent the potential associated to the toroidal field, $\beta$; the poloidal component of the neutron velocity field, $\vec{v}_n$; the poloidal component of the ambipolar velocity field, $\vec{v}_{ad}$; the chemical potential perturbation of the charged particles, $\chi_c$; and the chemical potential perturbation of the neutrons, $\chi_n$. All quantities are given in code units, as defined in Table~\ref{table_code_units}.}
    \label{fig2}
\end{figure}

\begin{figure}
    \centering    \includegraphics[width=8.6cm]{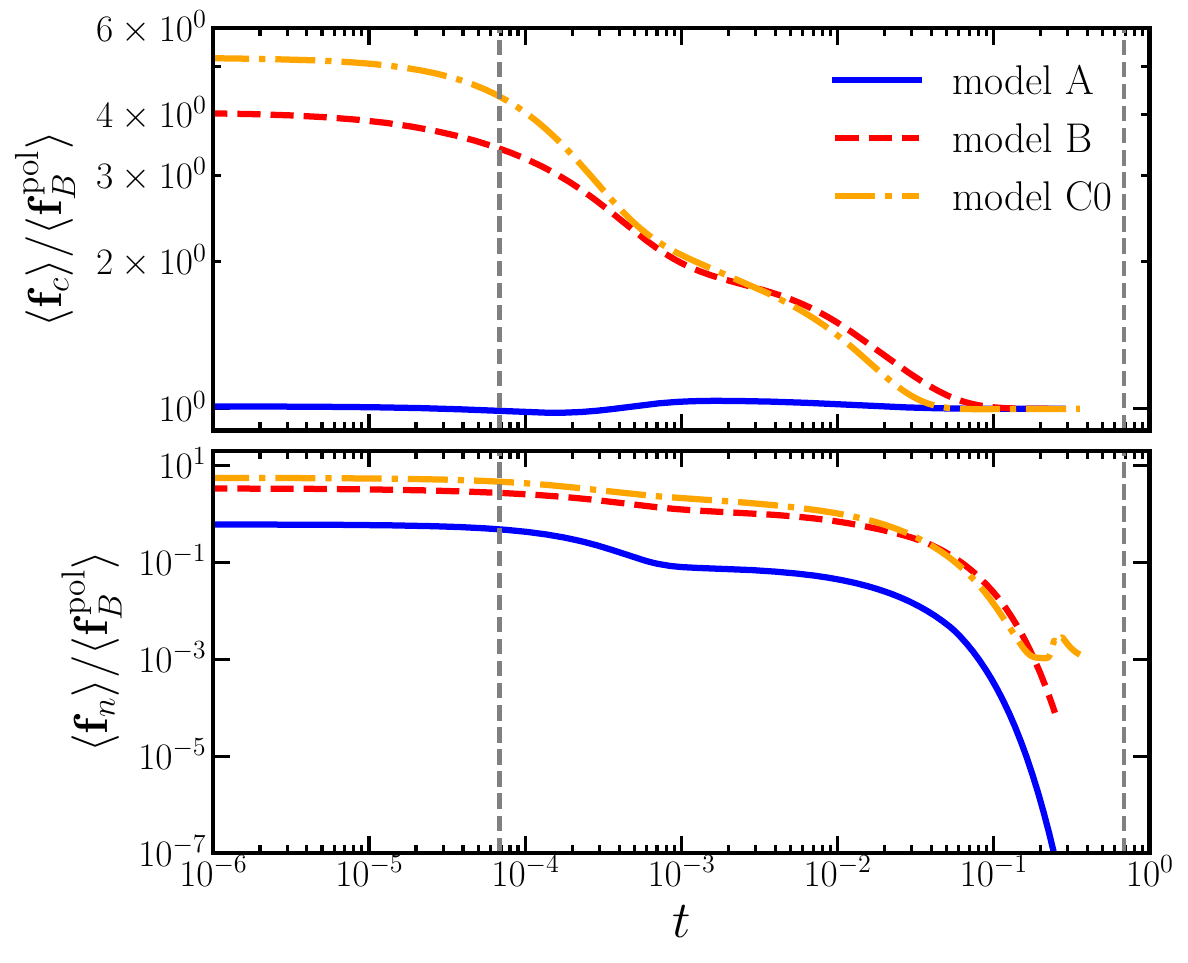}
    \caption{Ratios of the root mean square (rms) values of the different forces, $\langle \vec{f}_n \rangle /\langle \vec{f}^{\,\mathrm{pol}}_B\rangle$ and $\langle \vec{f}_c \rangle /\langle \vec{f}^{\,\mathrm{pol}}_B\rangle$, for models A (solid blue), B (dashed red), and C0 (dashed-dotted orange). The vertical dashed lines indicates, form left to right, the timescale $t_{\zeta B}$ and $t_{ad}$, respectively. The rms averages, denoted by $\langle \rangle$, are computed over the core's volume.} 
    \label{fig1FGS}
\end{figure}

%%%%%%%%%%%%%%%%%%%%%%%%%%%%%%%%%%%%%%%%%%%%
\section{Magnetic field evolution at constant temperature} \label{sec_constantT}

In this section, we discuss our simulations of the magnetic evolution of a NS core influenced by ambipolar diffusion with time-independent friction coefficients $\gamma_{np}$ and $\zeta$, that is, effectively at constant temperature. This problem has been extensively studied within the two-fluid model in axial symmetry by \cite{castillo2020twofluid}. Here, we highlight the main aspects of this evolution, particularly how the magnetic field rearranges itself towards a final equilibrium state.

\begin{figure}
    \centering
    \includegraphics[width=8.6cm]{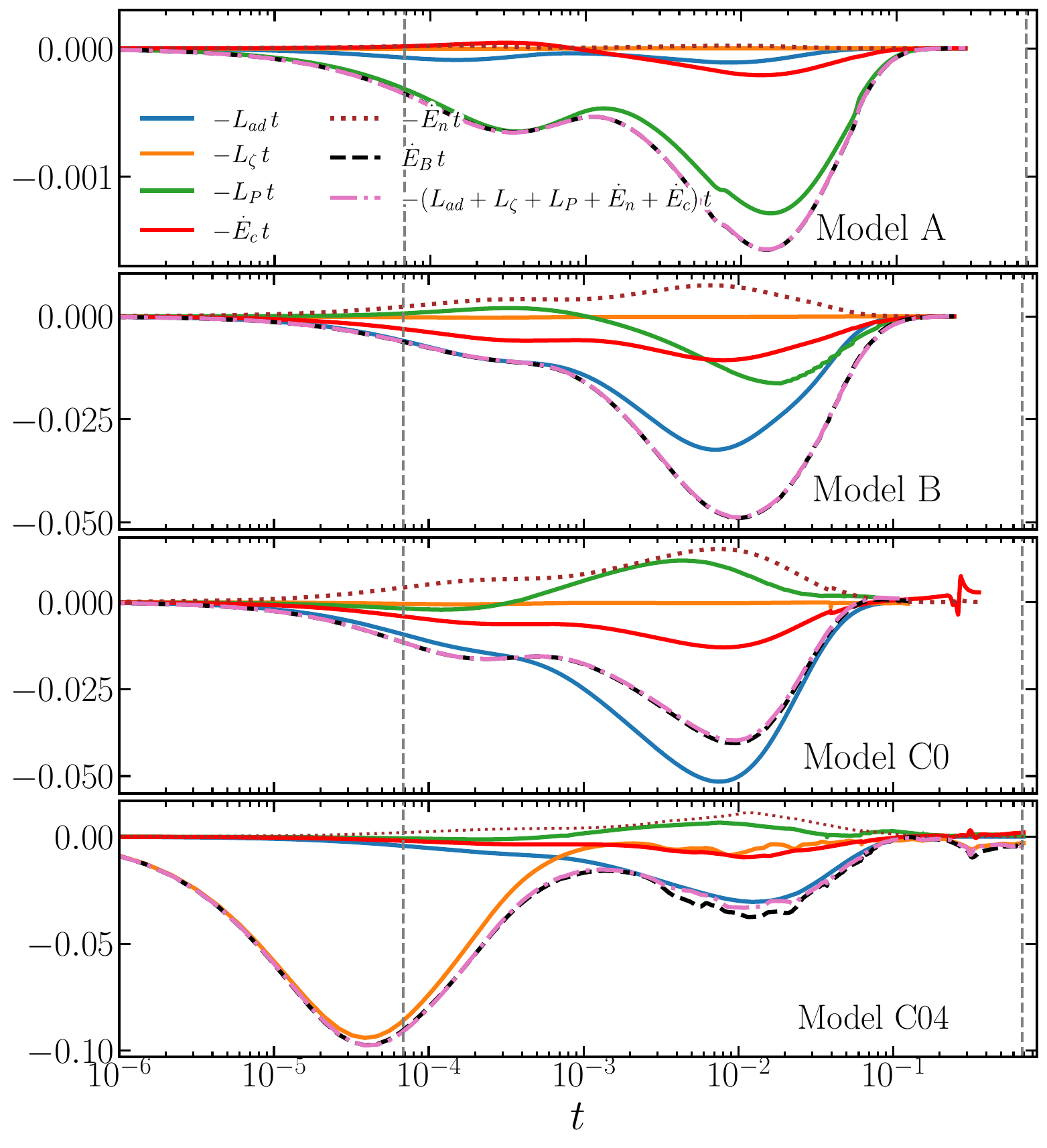}
    \caption{Magnetic energy dissipation for models A, B, C0 and C04 shown from top to bottom, respectively. The plot shows all the dissipation terms given in Eq.~\eqref{eqEdotB}, including the dissipation due to the artificial friction force, which appears as $-L_\zeta$ on the right-hand-side, with $L_{\zeta} \equiv \int_{\mathcal{V}_{\mathrm{core}}}\zeta n_n \vec{v}_n ^{2} d\mathcal{V}$, when this force is included. The curves correspond to: the power released per unit $\ln t$ due to the ambipolar heating, $-L_{ad}t$ (solid -- light blue); the artificial friction dissipation, $-L_{\zeta}t$ (solid -- orange); the net Poynting flux from the crust (modeled as a vacuum) to the core, $-L_{P}t$ (solid -- green); the terms, $\dot{E}_c\,t$ (solid -- red) and $\dot{E}_n\,t$ (dotted -- brown), respectively; the total time derivative of the magnetic energy inside the core, $(dE_{B}/dt) t$ (dashed -- black) and the combination $-(L_{H\nu}+L_{\zeta}+L_{P}+\dot{E}_c +\dot{E}_n)t$ (dashed-dotted -- pink), which is used to check energy conservation, as it should be equal to $(dE_{B}/dt)t$. The vertical dashed lines indicates, form left to right, the timescale $t_{\zeta B}$ and $t_{ad}$, respectively.} 
    \label{figEnerDiss}
\end{figure}

\begin{figure*}
    \centering    \includegraphics[width=\linewidth]{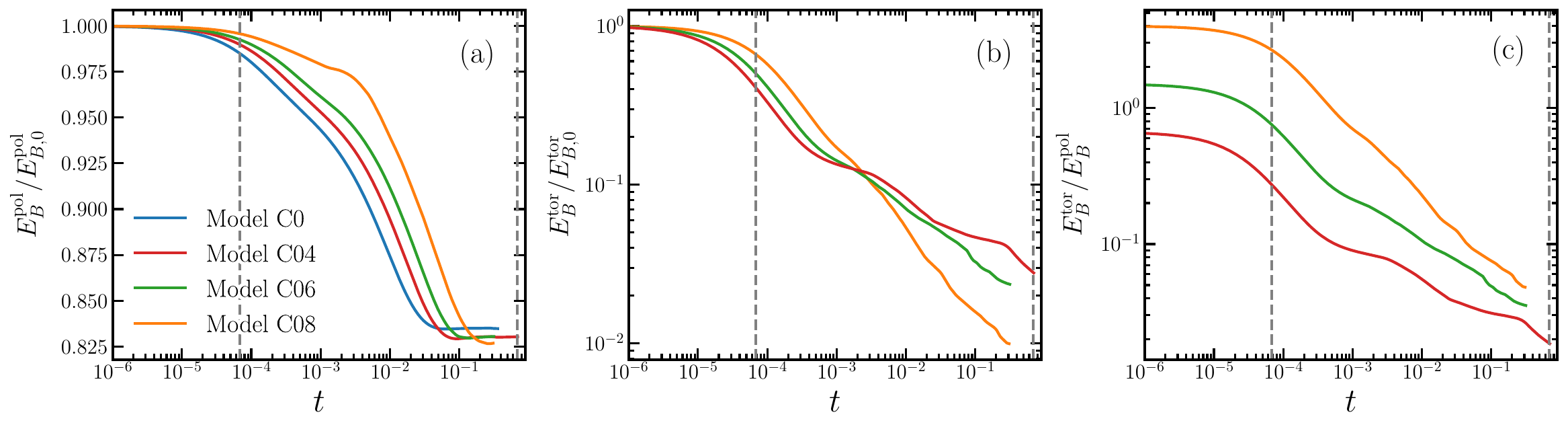}
    \caption{Evolution at constant temperature for the different initial magnetic field configurations corresponding to models C0, C08, C06, and C04 listed in Table~\ref{tab_B_model}, with $\zeta=10^{-4}$, showing:  (a) poloidal magnetic energy, (b) toroidal magnetic energy, both normalized to their initial values; (c) quotient between the toroidal and poloidal magnetic energy. The vertical dashed lines indicates, form left to right, the timescale $t_{\zeta B}$ and $t_{ad}$, respectively.} 
    \label{fig_dif_Epol_E_tor}
\end{figure*}

Figure~\ref{fig2} illustrates the evolution of the most general magnetic field configuration simulated, i.~e., a mixed poloidal and toroidal magnetic field, in this case model C04 (although the evolution is very similar for C06 and C08, which are not shown here).

Initially, the system evolves through a short-term dynamical phase driven mostly by the presence of a non-equilibrium toroidal magnetic field ($\Grad \alpha \nparallel \Grad \beta$, thus $\vec{f}^{\mathrm{tor}}_B \neq 0$). The unbalanced toroidal Lorentz force generates strong toroidal fluid motions, $\vec{v}_n^{\mathrm{tor}}$, which are counteracted by the fictitious friction force $-\zeta n_n \vec{v}_n$. These opposing forces persist until the system gradually settles into a toroidal hydromagnetic equilibrium state, where $\vec{f}^{\mathrm{tor}}_B=0$ throughout the core, on a characteristic timescale of $\sim t_{\zeta B}$.

At the end of this initial phase, the condition $\Grad \alpha \parallel \Grad \beta$ is satisfied since $\vec{f}^{\mathrm{tor}}_B=0$, meaning that one potential becomes (at least locally) a function of the other, $\beta = \beta(\alpha)$. As a result, $\beta$ becomes constant along each poloidal field line (at which $\alpha =  \text{constant}$). Additionally, due to the imposition of a current-free field outside the core, $\beta$ must vanish along all open field lines that extend beyond the core, thereby confining the toroidal magnetic field to regions with closed poloidal field lines, reaching the ``twisted-torus'' configurations expected in axially symmetric hydromagnetic equilibria \citep{Prendergast1956, braithwaite2004fossil}, as seen in the sequence of snapshots of the magnetic field in Fig.~\ref{fig2}. 

In a real NS, the previous process should be identified with the Alfv\'en timescale $t_{\mathrm{Alf}}$. However, here, we introduce the artificial friction force that mimics this process, and the effective Alfv\'en time corresponds to $t_{\zeta B}$. In doing so, we adjust the ratio of timescales $t_{\zeta B}: t_{ad}$ to a feasible value, though not a realistic one, allowing us to efficiently study ambipolar diffusion while using a reasonable amount of computational resources. To achieve this, the fictitious friction parameter introduced in Eq.~(\ref{eqVn}) must satisfy $\zeta \ll \gamma_{np} n_c$, so that a net force imbalance on a fluid element is reduced by bulk motions (with velocity $\vec{v}_{n}^{\mathrm{tor}}$) much more quickly than an imbalance of the partial forces on the charged-particle component is reduced by ambipolar diffusion (with relative velocity $\vec{v}_{ad}$). This is well satisfied for the value we set for our simulations, namely $\zeta/(n_{c0}\gamma_{np0}) =10^{-4}$, for which $t_{\zeta B}: t_{ad} = 1:10000$.
 
After this initial short-term dynamics, the long-term evolution driven by ambipolar diffusion takes over. In the two-fluid model, the background density profiles of neutrons, $n_n(r)$, and charged particles, $n_c(r)$, differ, and the combined motion of both components is strongly constrained by buoyancy forces. In our formalism, this is manifested by the fact that $\vec v_n$ and $\vec v_c$ have to satisfy different continuity equations [Eqs.~\eqref{eqCont1_NGR}--\eqref{eqCont2_NGR}], so the velocity fields cannot be equal. As a result, the relative motion between the two species naturally arises (ambipolar diffusion), whose magnitude is inversely proportional to the collision coefficient: $v_{ad}\propto\gamma_{np}^{-1}$. Nevertheless, the bulk velocities, $\vec v_n$ and $\vec v_c$, remain similar to each other and therefore substantially larger than their difference, $\vec v_{ad}$, as seen in the snapshots for $\vec v_n$ and $\vec{v}_{ad}$ in Fig.~\ref{fig2}, accelerating the motion with respect to the popular approach in which the motion of the neutrons (or the bulk baryon motion) is neglected. 

The evolution proceeds until a final equilibrium state in which all velocities vanish, i.~e., $\vec v_n=\vec v_c=\vec{v}_{ad} = 0$. At this point, the poloidal Lorentz force is balanced by the pressure plus gravity force on the charged particles, 
\begin{equation}
    \vec{f}^{\text{pol}}_B = n_c \mu \Grad \chi_c,\label{eq_GSfB}
\end{equation}
while the neutrons reach diffusive equilibrium, i.~e., 
\begin{equation}
    \Grad \chi_n  = 0.\label{eqchin_GS}
\end{equation}
In this equilibrium state, the magnetic field configuration is a solution of the ``Grad-Shafranov (GS) equation''
\citep{gradrubin54, shafranov66}:
\begin{equation}\label{eq:GS}
\Delta^{*}\alpha+\beta\frac{d\beta}{d\alpha} + 4\pi r^2\sin^2\theta\, n_c\mu\frac{d\chi_c}{d\alpha} = 0,
\end{equation}
which is obtained using the condition $\beta = \beta (\alpha)$, along with Eqs.~\eqref{eqfBPol} and ~\eqref{eq_GSfB}.

Figure~\ref{fig1FGS} shows how the ratios of the rms average of the forces $\vec f_c$ and $\vec f_n$ to the poloidal Lorentz force $\vec{f}^{\mathrm{pol}}_{B}$ evolve in time. While $\langle\vec f_c \rangle/\langle\vec{f}^{\mathrm{pol}}_{B}\rangle$ converges to unity for all the models as they settle into the final equilibrium state, $\langle\vec f_n \rangle/\langle\vec{f}^{\mathrm{pol}}_{B}\rangle$ becomes progressively smaller, consistent with Eqs.~\eqref{eq_GSfB} and \eqref{eqchin_GS}, respectively. On the other hand, row (d) in Fig.~\ref{fig2} illustrates how $\chi_n$ becomes uniform in the later stages of evolution, while the velocity fields have significantly diminished compared to the initial state in row (a), indicating an approach to equilibrium.

In terms of magnetic energy dissipation, Fig.~\ref{figEnerDiss} shows that for the purely poloidal models A, B, and C0, $L_{ad}$ is the main source of magnetic dissipation. In contrast, for model C04 (as well as C06 and C08, not shown here), the initial dissipation is dominated by artificial friction, $L_{\zeta}\equiv \int_\mathcal{V_{\mathrm{core}}}\zeta n_n\vec v_n^{2}$ (see the caption of Fig.~\ref{figEnerDiss}), until the toroidal field relaxes into equilibrium. After this phase ($t \gtrsim 10 \,t_{\zeta B}$), ambipolar diffusion takes over as the primary dissipation mechanism. This behavior is expected -- purely poloidal fields exhibit little evolution at $t\lesssim t_{\zeta B}$, unlike mixed poloidal-toroidal configurations, where the early dynamics corresponds to the relaxation of an initially out-of-equilibrium toroidal component.
It is an interesting fact that model A is extremely close to the GS equilibrium as it requires a very small dissipation in terms of $L_{ad}$, and most of the field rearrangement comes from small changes in the field lines along the crust-core interface, as indicated by the relatively large Poynting flux. 
This magnetic field configuration can be traced back to \cite{akgun2013}, and is one of the simplest axially symmetric configurations constructed by imposing regularity, requiring zero toroidal current density at the crust-core interface, and matching to an exterior dipole field at this boundary.

Although the initial ratio $E^{\mathrm{tor}}_{B}/E^{\mathrm{pol}}_{B}$ varies among these models, it consistently settles to $\lesssim 0.1$ at  $t\sim 0.1\,t_{ad}$ (see Fig.~\ref{fig_dif_Epol_E_tor}[c]), in agreement with previous studies that solved the Grad-Shafranov equation directly and were unable to find configurations with a larger ratio (e.~g., \citealt{Armaza_2015} and references therein). 
While the poloidal magnetic energy stabilizes at an equilibrium value at about the same time (see panel [a]), the toroidal energy continues to decay, albeit at a slower rate beyond that point (see panel [b]).
By $t \sim 0.1\,t_{ad}$, the Grad-Shafranov equilibrium is practically achieved, fluid motions have slowed down substantially, and no significant evolution of the magnetic field (either poloidal or toroidal) is expected.
In this respect the continued decay of $E_{B}^{\mathrm{tor}}$ at late times appears to be due to noisy structures near the torus boundary in $\vec{v}_{n}^{\mathrm{tor}}$ at this stage, which were also reported by \cite{Moraga2024}. However, this numerical artifact remains localized and stable in time, so it should not affect the main conclusions 
of this work.

We end this section by remarking that 
here, as in Refs.~\cite{castillo2020twofluid,castillo2025AA}, we tested several initial magnetic field configurations beyond those shown here, all of which evolve toward a final GS equilibrium state. 

%%%%%%%%%%%%%%%%%%%%%%%%%%%%%%%%%%%%%%%%%%%%%
\section{Magnetothermal evolution}\label{sec_Tev}

\begin{figure}
    \centering    \includegraphics[width=8.6cm]{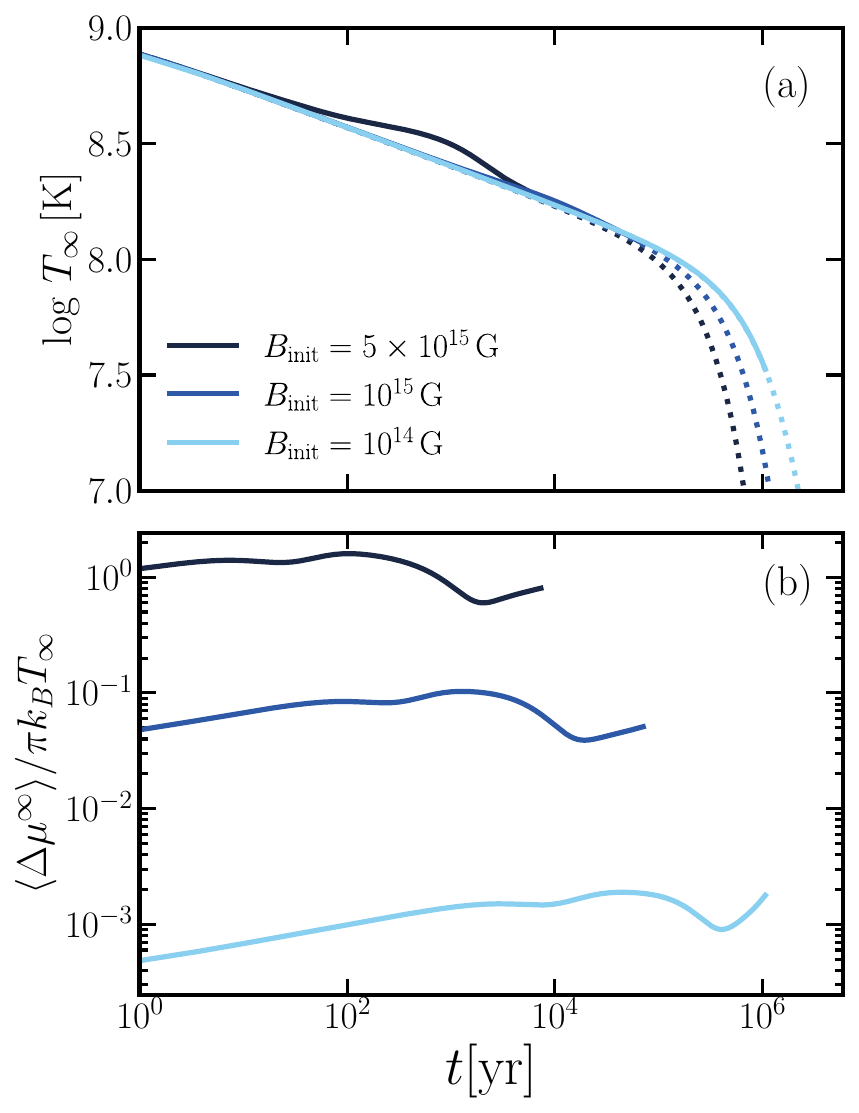}
    \caption{Magneto-thermal time evolution of two relevant variables after applying the time reparametrization procedure described in Sec.~\ref{sec_dtdt'} for model C0 using three different initial rms magnetic field strengths: $5\times10^{15}\,\mathrm{G}$ (dark blue), $10^{15}\,\mathrm{G}$ (blue), and $10^{14}\,\mathrm{G}$ (light blue). Panel (a): Evolution of the redshifted core temperature, $T_\infty $, as a function of time. The dotted lines correspond to the cases with no heating, where the stellar core cools passively. Panel (b): Evolution of the quantity $\langle \Delta \mu ^{\infty}\rangle/(\pi k_{B} T_{\infty})$, where $\langle \,\rangle$ represents the rms-average.} 
    \label{fig_magnetothermal_final}
\end{figure}

\begin{figure}
    \centering    \includegraphics[width=8.1cm]{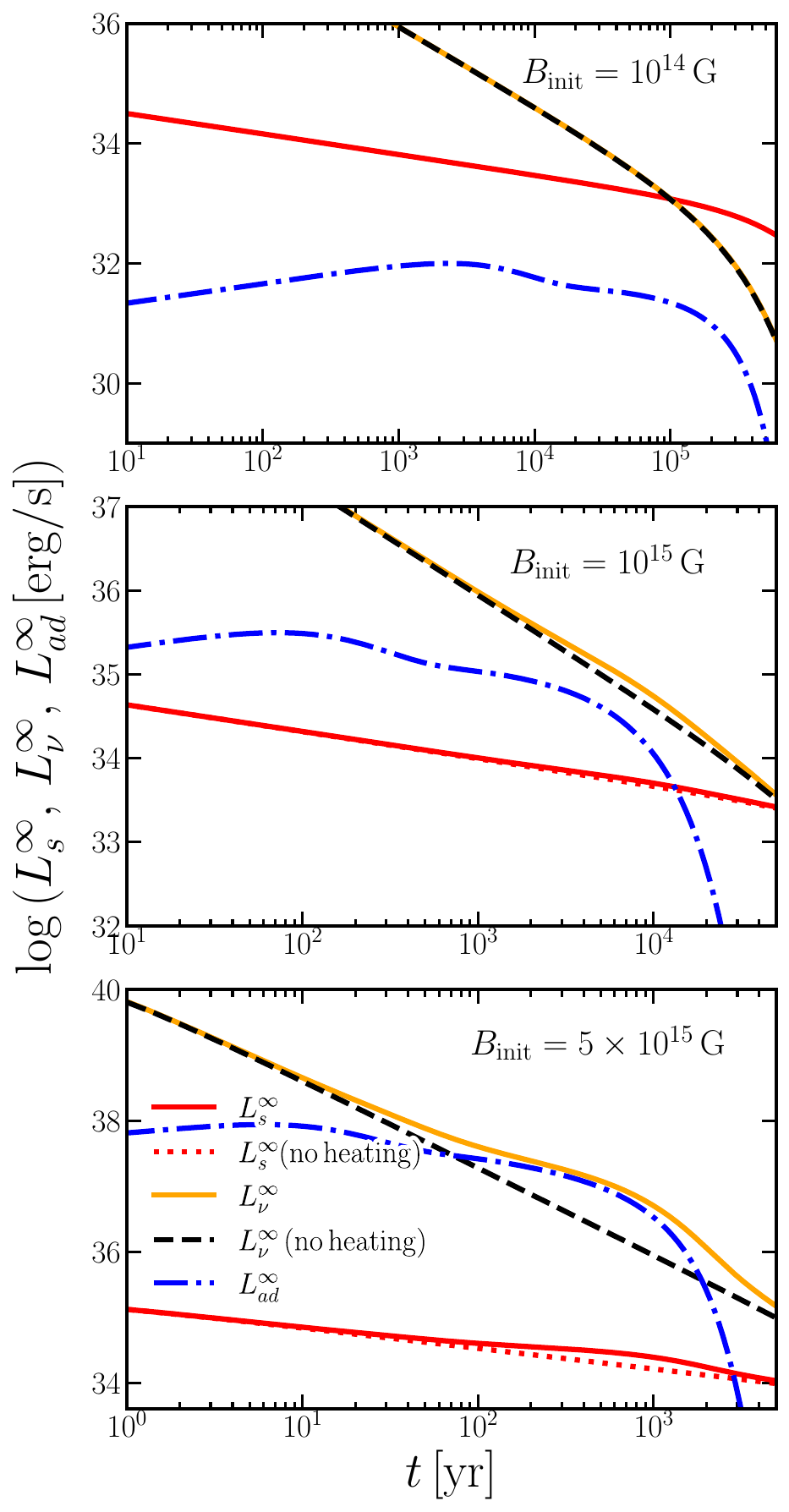}
   \caption{Magneto-thermal evolution of the relevant luminosities for model C0 after applying the time reparametrization procedure explained in Sec.~\ref{sec_dtdt'} for three different rms initial magnetic field strengths, and with the initial temperature $T^{\infty}_{\mathrm{init}} = 10^{9}\,\text{K}$. The curves correspond to the power released by ambipolar diffusion, $L^{\infty}_{ad}$ (dash-dotted-blue), the equilibrium neutrino luminosity, $L^{\infty}_{\nu}$, for both the case with ambipolar heating (solid-orange) and the case with no heating (dashed-black), and the surface photon luminosity $L^{\infty}_{s}$ (solid-red) and the case with no heating (dotted-red).} 
    \label{figHeatBalance}
\end{figure}

The results presented in the previous section are unphysical in the sense that the thermal evolution is not considered, and the collision coefficient $\gamma_{np}$ remains constant, although in reality it depends on time via the core temperature.
In this section, we address the thermal evolution by applying the same procedure as used by \citet{Moraga2024} in the strong-coupling regime. In that work, the results obtained at constant temperature were extended to the case where the temperature evolves through a temperature-dependent reparameterization of the time variable.

\subsection{Time reparametrization}\label{sec_dtdt'}
We start by noting that the system of equations \eqref{eq_IndB_NGR}–\eqref{eq_Vad_NGR}, and \eqref{eqForces_NGR}–\eqref{eqCont2_NGR}, along with (\ref{eqVn}), which describe the evolution of the magnetic field, remains invariant under the change of variables:
\begin{gather}
    dt = 
    \frac{\gamma_{np}(t)}{\gamma_{np}'}dt',\label{rescale-time}\\
    \vec v_{ad} = 
    \frac{\gamma_{np}'}{\gamma_{np}(t)}\vec v_{ad}', \label{rescale-velocity-ad}\\
    \vec v_{n} = \frac{\gamma_{np}'}{\gamma_{np}(t)}\vec v_{n}', \label{rescale-velocity-n}\\
    \zeta(t) =
    \frac{\gamma_{np}(t)}{\gamma_{np}'}\zeta', \label{rescale-zeta}
\end{gather}
where primed variables ($\gamma_{np}',t',\vec {v}'_n,\vec {v}'_{ad},\zeta',T'$) correspond to the previous simulations at an arbitrary constant temperature $T'$, while unprimed variables ($\gamma_{np},t,\vec {v}_n,\vec {v}_{ad},\zeta, T$) take into account the thermal evolution and its effect on the parameters. We note that, according to Eq.~(\ref{eqGammacn}), $\gamma_{np}(t)/\gamma_{np}'=[T_{\infty}(t)/T'_{\infty}]^2$.

Based on this, our approach is as follows:
\begin{enumerate}
    \item We run simulations that evolve the system of equations \eqref{eq_IndB_NGR}–\eqref{eq_Vad_NGR}, \eqref{eqVn}, and \eqref{eqForces_NGR}–\eqref{eqCont2_NGR} using the artificial friction method at constant temperature (thus time-independent $\gamma'_{np}(r)$ and $\zeta'$), calling the time variable $t'$.
    \item Then, using these results, we solve the equation for the temperature [Eq.~\eqref{eqDTdt_final}] as a function of $t^{\prime}$, which reads as
        \begin{equation}\label{e71}
            \dfrac{d \bar{T}}{dt'} = \dfrac{1}{C^{\prime}T^{\prime}_{\infty}}\left[\dfrac{L^{\infty\,\prime}_{ad}}{\bar{T}}- L^{\infty\,\prime}_{\nu}\bar{T}^{9}-L^{\infty\,\prime}_\mathrm{cb}\bar{T}^{4 \delta+1}\right].
    \end{equation} 
  Here, we used $\bar{T} \equiv T_{\infty}/T_{\infty}^{\prime}$, where $T'_{\infty}$ is a reference temperature. We also used $\gamma_{np}/\gamma'_{np}=\bar{T}^{2}$ and used the scaling in Eqs.~(\ref{rescale-time})--(\ref{rescale-zeta}). As a result, the equilibrium neutrino luminosity, ambipolar heating, and energy flux through the boundary evaluated at the constant temperature $T'_{\infty}$ are $L^{\infty\,\prime}_{\nu} =\tilde{L}^{\infty}_{\nu}(T_{\infty}^{\prime})^{8}$, $L^{\infty\,\prime}_{ad} = \tilde{L}^{\infty}_{ad}(T'_{\infty})^{-2}$, $L_{cb}^{\infty\,\prime} = \tilde{L}_{cb}^{\infty}(T_{\infty}^{\prime})^{4\delta}$, respectively. Additionally, we set $C^{\prime} = \tilde{C}\,T_{\infty}^{\prime}$.

    \item Finally, we obtain the physical time variable that includes the effects of the temperature evolution by integrating Eq.~(\ref{rescale-time}). Thus, we can plot any variable of interest as a function of $t$, the real physical time.
\end{enumerate}

\subsection{Simulation results and discussion}\label{sec_results_magnetothermal}
In this subsection, we analyze the magnetothermal evolution in the weak-coupling regime by applying the time reparametrization described in Sec.~\ref{sec_dtdt'} to the simulations run at constant temperature and discussed in Sec.~\ref{sec_constantT}.  
We focus on model C0 to analyze the magnetothermal evolution, as the results for the other models (not shown here) do not differ substantially. We use Eqs.~\eqref{eq_Ls_Tb_14}--\eqref{eq_Ls_Tb_16} to derive the photon luminosity as a function of the core temperature.

Figures~\ref{fig_magnetothermal_final} and \ref{figHeatBalance} illustrate the evolution of all relevant variables describing the magnetothermal evolution after applying the time reparametrization procedure for model C0. The results are shown for three initial rms magnetic field strengths: $B_{\mathrm{init}} = 10^{14}\,\mathrm{G},\,10^{15}\,\mathrm{G}$, and $5\times10^{15}\,\mathrm{G}$. Fig.~\ref{figHeatBalance} shows that only magnetic fields $B_{\mathrm{init}}\gtrsim 5\times10^{15}\,\mathrm{G}$ can generate a significant magnetothermal feedback, leading to a heating-cooling balance ($L^{\infty}_{ad}\approx L^{\infty}_{\nu}$). In Fig.~\ref{fig_magnetothermal_final}(a), the redshifted core temperature stays hotter at approximately $T_{\infty} \sim 3\times10^{8}\,\mathrm{K}$ for around $\sim 10^3\,\mathrm{yr}$, when $L^{\infty}_{ad}$ begins to decrease significantly, marking the end of the heating-cooling balance phase when $L^{\infty}_{ad}\ll L^{\infty}_{\nu}$. 

These results can be interpreted through the following physical considerations and estimates. In the weak-coupling regime, the speed of the magnetic field evolution is primarily controlled by the temperature dependence of the collisional coupling between neutrons and protons, $\gamma_{np} \propto T^{2}$ [see Eq.~\eqref{eqGammacn}]. At high internal temperatures, the coupling is strong, causing the ambipolar velocity $\vec{v}_{ad}$ to be small. As the temperature decreases, $\vec{v}_{ad}$ gradually increases, allowing the system to approach the GS equilibrium described in Sec.~\ref{sec_constantT}. As ambipolar diffusion operates, it releases heat into the core, transforming magnetic into thermal energy through $L^{\infty}_{ad}$. 
This heating mechanism becomes significant for sufficiently strong magnetic fields, where it can counteract the substantial energy loss due to intense neutrino emission, delaying the cooling process. 
The balance between neutrino luminosity and ambipolar heating, $L^{\infty}_{ad}\approx L^{\infty}_{\nu}$, can keep the NS core roughly at a temperature 
\begin{align}
    T^{\mathrm{bal}}_\infty \sim\, & 5.2\times10^{8}\left(\dfrac{B}{5\times10^{15}\,\mathrm{G}}\right)^{\frac{2}{5}} \nonumber \\
    & \times 
    \left(\dfrac{\ell_B}{2\,\mathrm{km}}\right)^{-\frac{2}{5}}
    \left(\dfrac{\ell_c}{10\,\mathrm{km}}\right)^{\frac{1}{5}}\,\mathrm{K}.
\end{align}
The corresponding surface luminosity at this temperature reads (see Eq.~\ref{eq_Ls_Tb_16})\begin{align}
    L_s^\infty \sim\, & 6.7\times 10^{34} \,\left(\dfrac{B}{5\times10^{15}\,\mathrm{G}}\right)^{0.75}\ \nonumber \\
    & \times \left(\dfrac{\ell_B}{2\,\mathrm{km}}\right)^{-0.75}
    \left(\dfrac{\ell_c}{10\,\mathrm{km}}\right)^{0.38}\,\mathrm{erg/s},
\end{align}
until the GS equilibrium is reached in a timescale given by Eq.~\eqref{eq_tBf}, i.~e,
\begin{equation}
    t_{B}\sim 1.5\,\left(\dfrac{B}{5\times10^{15}\,\text{G}}\right)^{-\frac{6}{5}}\left(\dfrac{\ell_B}{2\,\mathrm{km}}\right)^{\frac{16}{5}}\left(\dfrac{\ell_c}{10\,\mathrm{km}}\right)^{-\frac{8}{5}}\,\mathrm{kyr}.
\end{equation}
This can be compared to the time it takes for a star to passively cool down noticeably, say, to half of the equilibrium temperature, $T^{\mathrm{bal}}_{\infty}/2$, 
\begin{align}
    t_{\mathrm{cool}}\sim & 0.22\left(\frac{T_{\infty}^{\mathrm{bal}}/2}{10^9\mathrm{K}}\right)^{-6}\mathrm{yr}\\\nonumber
    \sim & \,7.1 \times10^{2}\,\left(\dfrac{B}{5\times10^{15}\,\text{G}}\right)^{-\frac{12}{5}} \nonumber \\
    &\times \left(\dfrac{\ell_B}{2\,\mathrm{km}}\right)^{\frac{12}{5}}\left(\dfrac{\ell_c}{10\,\mathrm{km}}\right)^{-\frac{6}{5}}\mathrm{yr},
\end{align}
resulting in 
\begin{equation}
    \dfrac{t_{B}}{t_{\mathrm{cool}}} \sim 2.1\,\left(\dfrac{B}{5\times10^{15}\,\text{G}}\right)^{\frac{6}{5}}\left(\dfrac{\ell_B}{2\,\mathrm{km}}\right)^{\frac{4}{5}}\,\left(\dfrac{\ell_c}{10\,\mathrm{km}}\right)^{-\frac{2}{5}}.
\end{equation}
Therefore, a magnetic field of order 
$B\gtrsim 5\times10^{15}\,\mathrm{G}$ can sustain the core temperature at approximately $T_{\infty}\sim T^{\mathrm{bal}}_\infty$ for a duration $t_{B}$, longer than the passive cooling time $t_{\mathrm{cool}}$.

%We note that, once the star reaches the GS equilibrium, the chemical imbalance, $\Delta\mu^\infty \equiv \delta\mu_c^\infty -\delta\mu_n^\infty$, remains constant in time, while the temperature decreases. This raises questions about the potential relevance of non-equilibrium Urca reactions, as the heating associated with these reactions, $L^{\infty}_{H\nu}$, and their effect on the neutrino luminosity, $L^{\infty}_{\nu}$, depend on the variable $\xi \equiv \Delta\mu^{\infty}/(\pi k_B T_{\infty})$ \citep{Reisenegger1995,Fernandez2005}, which should increase once the GS equilibrium has been reached, as the NS keeps cooling. Fig.~\ref{fig_magnetothermal_final}(b) illustrates that for our model and for $B_{\mathrm{init}}=5\times10^{15}\,\mathrm{G}$, $\xi$ is never larger than $\sim 1$ throughout the evolution. This suggests that non-equilibrium Urca reactions are unlikely to play a significant role in our case (see Fig.2 in Ref.~\cite{Fernandez2005}), and justifies why we neglected them. However, this conclusion may not hold for stronger magnetic fields. For this reason, we do
%not model cases with stronger magnetic fields, for which
%the neglect of non-equilibrium Urca reactions would not
%be justified

We note that in our equations we neglect the effect of non-equilibrium reactions, adopting the equilibrium rate of the neutrino luminosity, $L^{\infty}_{\nu}$, and neglecting the heating associated with these reactions, $L^{\infty}_{H\nu}$. Such approximation is valid while the variable $\xi \equiv \Delta\mu^{\infty}/(\pi k_B T_{\infty})$ is $\lesssim 1$ \citep{Reisenegger1995,Fernandez2005} (here $\Delta\mu^\infty \equiv \delta\mu_c^\infty -\delta\mu_n^\infty$ is the chemical potential imbalance). Meanwhile, sufficiently strong magnetic fields may be associated with the chemical imbalance being comparable to the stellar temperature. To check the validity of our approximation we plot $\xi$ as a function of time in Fig.~\ref{fig_magnetothermal_final}(b). One can see that for our model and for $B_{\mathrm{init}}\lesssim 5\times10^{15}\,\mathrm{G}$, $\xi$ is never larger than $\sim 1$ throughout the evolution stage shown in Fig.~\ref{figHeatBalance}. This justifies our assumption
\footnote{Note however, that once the star reaches the GS equilibrium, the chemical imbalance, $\Delta\mu^\infty$, remains constant in time, while the temperature decreases due to stellar cooling. As a result, $\xi$ increases and eventually exceeds unity [the beginning of this stage is marked by the rise of the curves at their right-hand side in Fig.~\ref{fig_magnetothermal_final}(b)]. Our equations are not valid at this stage, in which non-equilibrium reactions should be accounted for.}. %The non-equilibrium reactions would tend to relax $\Delta\mu^\infty$, while the decreasing temperature would tend to increase $\xi$. As a result $\Delta\mu^\infty$ and $T_{\infty}$ will evolve on the cooling timescale keeping $\xi$ at some quasi-equilibrium value.}. 
However, this conclusion may not hold for stronger magnetic fields. For this reason, we do
not model cases with stronger magnetic fields, for which
the neglect of non-equilibrium Urca reactions would not
be justified.

\subsection{Confronting with observational data}

The large quiescent surface X-ray luminosities of some magnetars provide the most natural context for applying the problem studied in this work, as originally suggested by \citet{DuncanThompson1995,DuncanThompson1996}. Therefore, in this section, we confront our results with the available observations.

The results of our simulations are shown in Fig.~\ref{fig_Observ}, where they are compared with the observational data described in Appendix~\ref{sec:observations} and summarized in Table~\ref{tab_Observ}. They indicate that the largest magnetar luminosities cannot be explained solely by the effects of ambipolar diffusion with magnetic fields in the range considered here. Even in the case of 1E 1547--5408, where the curves intersect the observational box, distinguishing between passive cooling and the scenario where ambipolar heating contributes is not feasible due to uncertainties, as the box encompasses both cases.

\begin{figure}
    \centering    \includegraphics[width=\columnwidth]{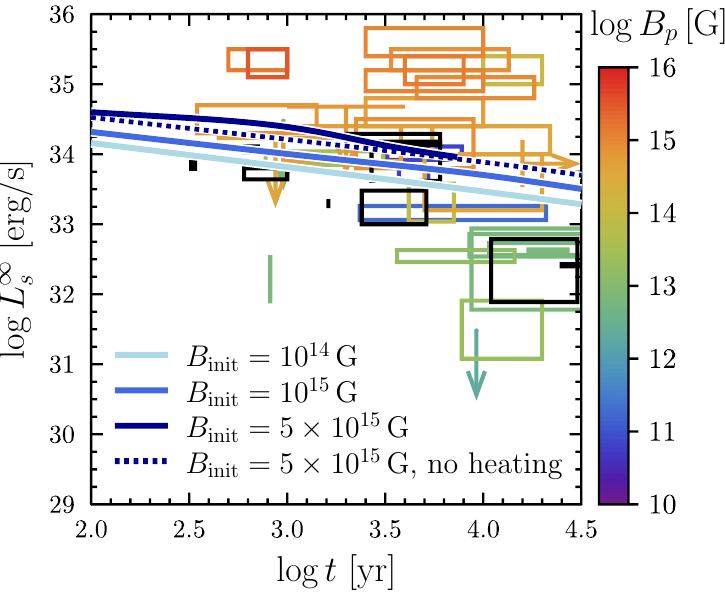}
    \caption{The set of observed cooling NSs (colored boxes) confronted with the same magnetothermal evolution tracks as shown in Fig.~\ref{fig_magnetothermal_final}. Observational data is listed in Table~\ref{tab_Observ}. Box colors show the dipole magnetic field at the pole; black color means that the spin-down estimate for magnetic field is not available.}
    \label{fig_Observ}
\end{figure}

\section{Conclusions}\label{sec_conclusions}
We have modeled the magneto-thermal evolution of a NS core in the weak-coupling regime applicable at relatively low temperatures, $T<T_{\mathrm{trans}}\approx 5\times 10^8\,\mathrm{K}$, including the thermal evolution, which was not covered in the previous models of Refs.~\cite{Castillo2017,castillo2020twofluid}. The latter was included using the same method applied in Ref.~\cite{Moraga2024} for the strong-coupling regime; i.~e., by starting from the magnetic field evolution at constant temperature, using this as an input for the calculation of the thermal evolution, and reparametrizing the time variable to translate these results to the realistic case of evolving temperature. 

For not too strong magnetic fields, $B\lesssim 5\times 10^{15}\mathrm{G}$, this scheme is applicable both in the strong-coupling and in the weak-coupling limit because in each case the magnetic field evolution is controlled by a single temperature-dependent physical process (non-equilibrium Urca reactions and neutron-proton collisions, respectively), which allows the time intervals to be rescaled by a temperature-dependent factor. This would not be possible if Urca reactions and collisions would be considered in the same model, as they have a radically different temperature dependence 
\footnote{%Generalization of this approach to the suprathermal regime (when $\Delta \mu \ll T$) 
%would not be possible for the same reason.
Extending this approach to cases where $\Delta \mu \gtrsim T$ would be similarly infeasible for the same reasons.}. Fortunately, \citet{Moraga2024} showed that there is essentially no magnetic field evolution in the strong-coupling regime, making it reasonable to ignore the non-equilibrium Urca reactions in the evolution of the magnetic field and consider only the weak-coupling regime, as done here. However, for very strong magnetic fields, Urca reactions and ambipolar diffusion can become important simultaneously, so we leave this case for future study.
 
The simulations were performed in axial symmetry, for a NS core composed of two fluids (neutrons and charged particles) coupled by collisions and surrounded by a non-conducting medium (vacuum or perfect resistor), including a fictitious friction force that allows the system to quickly reach a hydromagnetic equilibrium state while ignoring the inertial terms and the time derivatives in the particle conservation laws. The latter, commonly known as the anelastic approximation, is the key difference between our numerical scheme and that of Ref.~\cite{castillo2020twofluid}. This approach results in a more efficient numerical method, enabling us to rescale our results for any initial magnetic field strength, $B_{\mathrm{init}}$.

We used a hybrid approach, modeling the thermal evolution within the framework of GR, while simulating the magnetic evolution using Newtonian gravity, as GR corrections are likely to be relatively unimportant in the latter case.

The main conclusions can be summarized as follows:
\begin{enumerate}
    \item \emph{In the constant-temperature simulations}, we confirmed the previous results reported by Ref.~\cite{castillo2020twofluid}, where the long-term magnetic evolution driven by ambipolar diffusion led the NS core to an equilibrium state. This state is characterized by the neutron fluid reaching diffusive equilibrium, $\chi_n = \text{constant}$, while the Lorentz force balances the pressure and gravity forces acting on the charged particle fluid, $\vec{f}_B = n_c \mu \Grad\chi_c$. Consequently, the magnetic field satisfies the nonlinear Grad-Shafranov equation. This was confirmed for various initial magnetic field configurations and ratios of toroidal to total magnetic energy.

    \item \emph{When the core temperature evolves}, ambipolar heating can effectively counterbalance the immense energy losses due to neutrino emission, but only for
    strong magnetic fields  ($B\gtrsim 5\times10^{15}\,\mathrm{G}$) and for a very short duration ($\lesssim 1\,\mathrm{kyr}$). 
    The resulting surface photon luminosities, $L_{s}^{\infty}$, driven by ambipolar diffusion cannot account for the observed magnetar population. For this mechanism to be viable, it would need to dissipate a greater amount of magnetic energy than in the case of $B_{\mathrm{init}}=5\times10^{15}\,\mathrm{G}$ and over a longer timescale.

\end{enumerate}

Finally, we note some issues that were not included in the present work and should be considered in the future:

\begin{enumerate}
    \item We modeled the magnetothermal evolution of a NS core under the influence of ambipolar diffusion, aiming to explain the surface luminosity of magnetars while neglecting the effects of the crust. The latter was achieved by imposing a vacuum boundary condition on the magnetic field at the crust-core interface. In this approach, the crust is assumed to have negligible conductivity, causing currents to dissipate instantaneously and allowing the magnetic field outside the core to adjust immediately. This approach implicitly assumes that the evolutionary timescale of the magnetic field in the crust is much shorter than that in the core. However, the shortest evolutionary timescale in the crust is set by the Hall effect, which can transfer magnetic energy from large to small scales, where it is more efficiently dissipated through Ohmic dissipation. This process operates on a timescale 
    \begin{align}
        t_{\mathrm{Hall}} \sim &\frac{4\pi n_c e \ell_B^{2}}{c B} \nonumber\\
    \sim &\,1.7 \times 10^{5} 
    \left( \frac{\ell_B}{2\,\mathrm{km}} \right)^{2} \nonumber\\
    &\times\left( \frac{B}{5\times10^{15}\,\mathrm{G}} \right)^{-1} \mathrm{yr},
    \end{align}
    as estimated at the crust-core interface for our stellar model \citep{Cumming_2004}. This timescale is significantly longer than the ambipolar diffusion timescales studied in this work. Consequently, the vacuum boundary condition is not appropriate for the problem considered here; instead, a frozen-field condition would provide a more realistic approximation. Nevertheless, the scenario presented here represents the fastest possible magnetothermal evolution for a magnetic field penetrating both the core and the crust, since a finite crustal conductivity would cause the magnetic field lines to become partially frozen in the crust \citep{blandford2008applications}, thereby slowing the magnetic evolution in the core.

    \item The NS core is expected to become both superfluid and superconducting (for not too strong magnetic fields) relatively early in its evolution (see e.~g., Ref.~\cite{page2013stellar} for a review).
    This suppresses the inter-particle collisions
    \citep{baiko2001thermal}, an effect that could lead to a larger ambipolar velocity, and thus potentially stronger heating during a shorter period. However, there are some caveats. Neutron superfluidity suppresses the number of unpaired neutrons available for scattering as well. The result of such interplay for the dissipation rate is not obvious. Moreover, the formation of a neutron condensate adds a new degree of freedom, allowing neutrons to smooth out $\chi_n$ on dynamical timescales \citep{KantorGusakov2018}, so that they reach diffusive equilibrium, effectively eliminating ambipolar diffusion in pure $npe$-matter—though the presence of muons partially restores it. Proton superconductivity alters the magnetic forces by introducing buoyancy and tension forces acting on vortices in type-II superconductors, which must now be included in the dynamical equations (e.~g., Refs.~\cite{gd16,Gusakovetal2020}). Additionally, complex interactions between particles, neutron vortices, and magnetic flux tubes arise, leading to intricate dynamical effects that can strongly influence both the evolution timescales and the final equilibrium state \citep{glampedakis2011magnetohydrodynamics,gd16,Gusakovetal2020}. Recently, numerical simulations have started incorporating these effects using simplified models (e.~g., Refs.~\cite{Bransgroveetal2018, bransgrove2024giant}). However, some of the microphysical input used in these studies has been subject to criticism (e.~g., Ref.~\cite{gusakov2019force}).

    \item 3D MHD simulations show that magnetic fields are influenced by various types of dynamical instabilities. On one side, it has been theoretically proven that both purely poloidal and purely toroidal fields are inherently unstable in 3D \citep{Tayler1973_AdiabaticStabilityStars, Markey1973_AdiabaticStabilityStars, Flowers1977_EvolutionPulsarMagnetic}. In addition, numerical simulations suggest that barotropic stars do not support stable magnetic equilibria \citep{Braithwaite2009_AxisymmetricMagneticFields, Lander2012_AreThereAny, Mitchell2015_InstabilityMagneticEquilibria, Becerra22a, Becerra2022_StabilityAxiallySymmetric}. 
    Therefore, it remains unclear whether the GS equilibrium is stable against non-axisymmetric perturbations. If an instability arises, the characteristic time when the GS equilibrium is attained would instead correspond to the onset of the instability. This would lead to stronger magnetic energy dissipation and a more complex magnetothermal evolution than the one studied here.

    \item In this work, we adopted the isothermal approximation, which holds in the absence of internal heating sources, as the core's high thermal conductivity allows temperature gradients to relax on a timescale $\sim 100\,\mathrm{yr}$. When heating sources are present, we can estimate the maximum possible heat flux on a global stellar scale within the core as $\sim\kappa R_{\mathrm{core}}^2 \Grad T$, where $\kappa$ is the thermal conductivity in the core (see, e.~g., Ref.~\cite{so22}). The temperature $T$ is constrained 
   by neutrino emission and is unlikely to exceed a few times $10^{9}\,\mathrm{K}$ in any particular region of the star. Thus, we can estimate $\kappa R_{\mathrm{core}}^2 \nabla T \sim 10^{37} \,\mathrm{erg/s}$. This flux efficiently redistributes the dissipated magnetic energy throughout the core as long as the latter remains below this limit. This is the case for magnetic fields $B_{\mathrm{init}}\lesssim 10^{15}\,\mathrm{G}$, as $\dot{E}_B\approx-L_{ad}^{\infty}$ (see Fig.~\ref{figHeatBalance}). 
   Conversely, if the dissipated energy exceeds this threshold,
   most of the energy will be radiated away as neutrinos from the region where the magnetic field is localized, while the rest of the star will be heated at a more moderate rate of $\sim 10^{37} \,\mathrm{erg/s}$. Consequently, temperature gradients could be present in such case. However, our calculations should still remain qualitatively valid and the obtained surface temperature constitutes an upper limit.

   \item Including additional particle species, such as muons ($\mu$), introduces an additional Euler and continuity equation, along with extra collisional couplings ($\gamma_{\mu j}$ with $j=n,p,e$), making the charged-particle component non-barotropic. As a result, the equilibrium configurations become less constrained and do not necessarily correspond to the GS equilibrium. Consequently, the presence of muons could have a significant impact on the magnetic evolution of NSs cores.

   \item We adopted the drag coefficient $\gamma_{np}$ from Ref.~\cite{yakovlev1990electrical}. This was shown to be overestimated by a factor of about $3$ by \citet{DommesGusakov2021}. A weaker interaction implies that the same amount of energy will be dissipated in a shorter time. This difference does not affect our main conclusions.
\end{enumerate}

\begin{acknowledgements}
   This work was supported by ANID doctoral fellowship 21210909 (N.M.), CAS-ANID fund NºCAS220008 (F.C.), FONDECYT Projects 11230837 (F.C.), 1201582 (A.R.), and 1240697 (J.A.V.).  M.G., E.K., and A.P.\ acknowledge support from the Russian Science Foundation (Grant No.\ 22-12-00048-P).
   
\end{acknowledgements}

\appendix  % Start the appendix section

\section{Envelope model}
\label{sec:env}
\begin{figure*}
    \centering    \includegraphics[width=15cm]{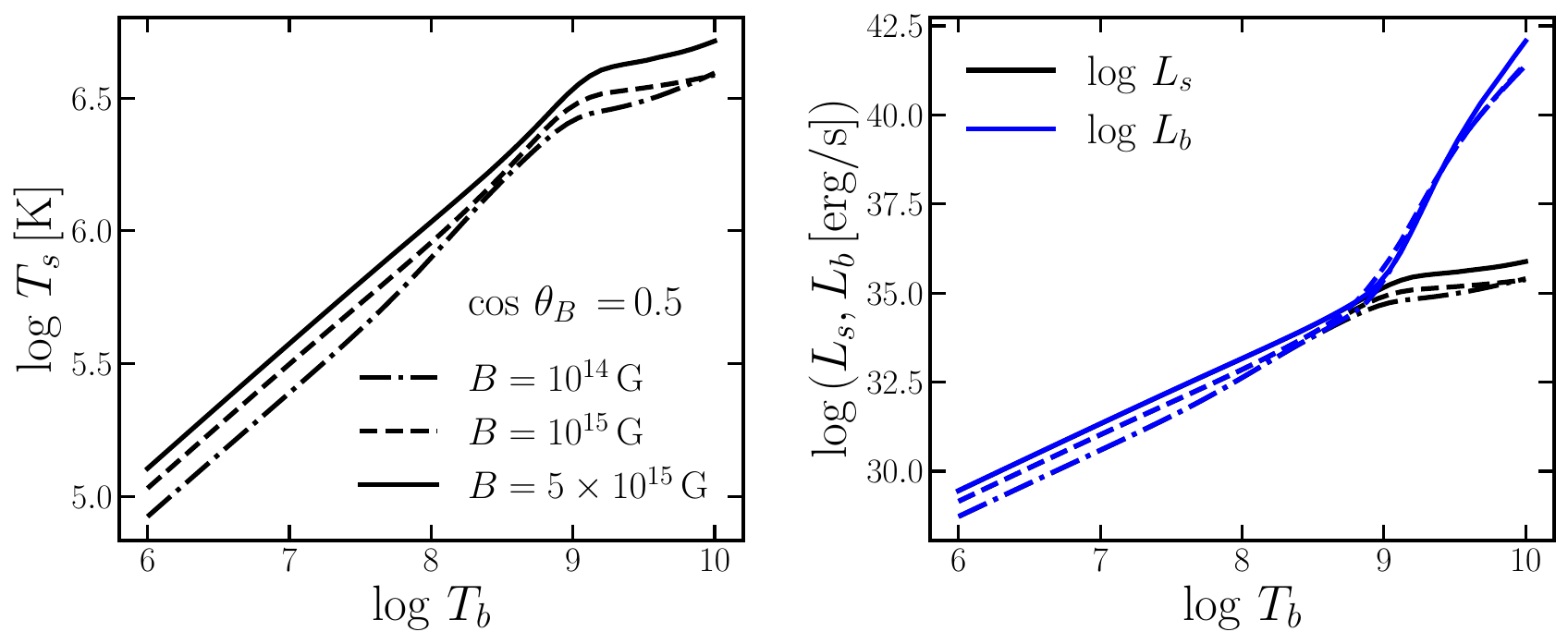}
    \caption{Interpolated heat-blanket calculation results for the HHJ EOS at a fixed angle $\theta_B$ between the magnetic field vector, $\vec{B}$, and the surface normal, given by $\cos \theta_B = 0.5$. The results are shown for different magnetic field strengths: $B = 10^{14}\,\mathrm{G}$ (dash-dotted), $10^{15}\,\mathrm{G}$ (dashed), and $5\times10^{15}\,\mathrm{G}$ (solid). \textit{Left panel}: Effective surface temperature, $T_s$, as a function of the temperature at the base of the heat-blanketing envelope, $T_b$.
    \textit{Right panel}: Photon surface luminosity, $L_s$, and total heat flux at the base of the envelope, as functions of $T_b$.} 
    \label{fig_Ts_Lb_Ls_cth05}
\end{figure*}

Beneath the stellar atmosphere lies a thin layer called the \textit{envelope}, where the steepest
temperature gradient is found (see, e.~g., Ref.~\cite{BEZNOGOV20211}, for review and references). This region has a very short thermal relaxation timescale $\sim1$~yr. Also, an outer part of the envelope is not strongly degenerate; therefore, the EoS is not barotropic there. As a result, it is computationally
expensive to solve the magneto-thermal evolution in this region in detail, coupled to the core
evolution.  The numerical codes that treat the thermal evolution in the core, crust, and envelope
of an NS coupled together (e.~g., Ref.~\cite{PotekhinChabrier18}) can be useful for studying short-scale
processes like superbursts (e.~g., Refs.~\cite{Yakovlev_21,KaminkerPY23}), but in the present paper we are interested in much longer time scales. Therefore, we adopt the common approach \citep{gudmundsson1983}, which relies on stationary envelope models to derive a relationship between the temperature $T_\mathrm{b}$ at the bottom of the envelope and the effective surface temperature $T_\mathrm{s}$. Moreover, we also neglect temperature variability of the inner crust, which has a timescale $\sim100$~yr \citep{GnedinYP01}, and
assume that the redshifted temperature $T_\infty\equiv T \mathrm{e}^{\Phi(r)/c^2}$ in the core equals $T_\mathrm{\infty,b}\equiv
T_\mathrm{b}\mathrm{e}^{\Phi_\mathrm{b}/c^2}$, where $\Phi_\mathrm{b}=\Phi(r_\mathrm{b})$ is the gravitational potential at the bottom of the envelope. We also neglect the neutrino emission from the inner crust, which is justified by its weakness compared to the neutrino
emission from the NS core. This approximation allows us to neglect the difference between the heat flux
through the inner boundary of the envelope $L_\mathrm{b}$ and the heat flux through the core boundary
$L_\mathrm{cb}$, which is needed to solve Eq.~(\ref{eqDTdt}).

By definition, the effective surface temperature $T_\mathrm{s}$  is related to the photon flux density $F_\mathrm{s}$ through the radiative surface of the NS by Stefan's law,
\begin{equation}
   F_\mathrm{s} = \sigma_\mathrm{SB} T_\mathrm{s}^4,
\label{F_s}
\end{equation}  
where $\sigma_\mathrm{SB}$ is the Stefan-Boltzmann constant.
If a strong non-uniform magnetic field is present, $T_\mathrm{s}$ varies over the surface, and the total photon luminosity is given by the integral
\begin{equation}\label{eq_total_Ls}
    L_s = \int_S F_\mathrm{s} \mathrm{d} S
\end{equation}
over the radiative surface $S$.

\begin{figure*}
    \centering    \includegraphics[width=17.5cm]{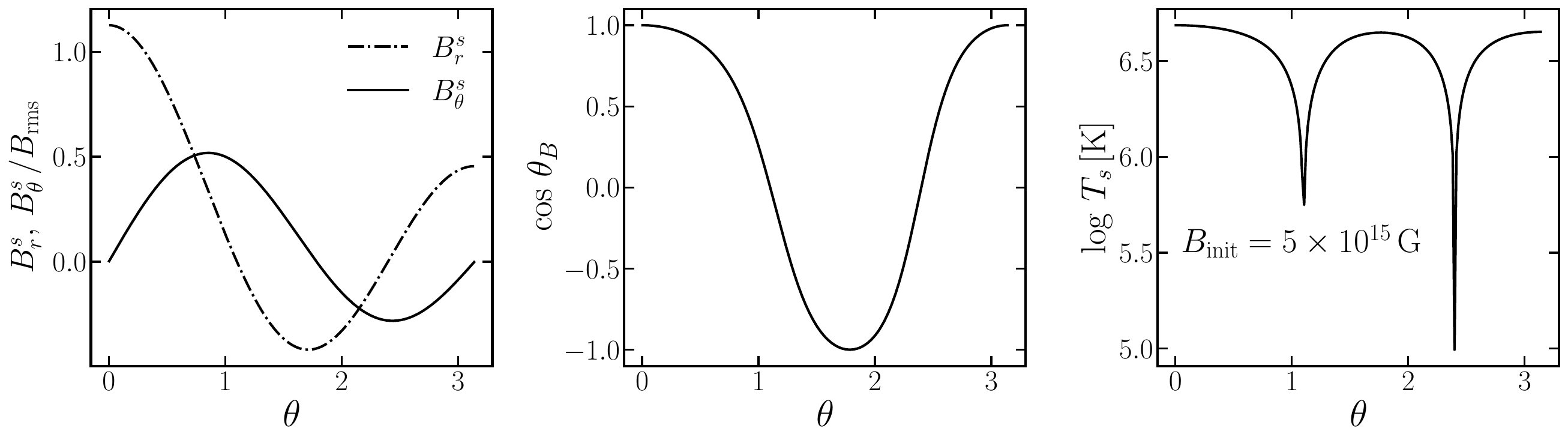}
    \caption{Magnetic field model C0 and its effects on the surface properties of the neutron star. \textit{Left panel}: Magnetic field components in units of $B_{\mathrm{rms}}$ at the surface as a function of the polar angle. \textit{Central panel}: Cosine of the angle between the magnetic field lines and the normal to the surface as a function of the polar angle.\textit{ Right panel}: Surface temperature profile as a function of the polar angle using $B_{\mathrm{init}}=5\times10^{15}\,\mathrm{G}$.} 
    \label{fig_BrBth_cthB_Ts_th_C0}
\end{figure*}

A conventional approach to obtaining the boundary conditions to the heat balance equation (\ref{eqDTdt}) consists in solving the simple equation
\begin{equation}
\frac{\mathrm{d}\ln T}{\mathrm{d}\ln P} =
     \frac{F_r}{T}
     \frac{P}{g\varkappa\rho},
\label{Tsimple}
\end{equation}
where $F_r$ is the thermal flux density toward the NS surface,
$g=(1-2GM/Rc^2)^{-1/2}\,GM/R^2$ is the local gravity at the surface, and
$\varkappa$ is the effective thermal conductivity.
In a strong magnetic field ($B\gtrsim10^{12}\,\mathrm{G}$), the latter is defined as $\varkappa =
\varkappa_\|\cos^2\theta_B + \varkappa_\perp\sin^2\theta_B$, where $\varkappa_\|$ and $\varkappa_\perp$ are the
main components of the conductivity tensor along and across the magnetic field $\bm{B}$, and $\theta_B$ is
the angle between $\bm{B}$ and normal to the surface.
Equation~(\ref{Tsimple}) follows from the TOV equations
by neglecting the variations of the metric functions $\Lambda$ and $\Phi$ inside the envelope and is justified by the smallness of the envelope's thickness and mass, whose lower boundary is usually taken at the mass density $\rho_\mathrm{b}=10^{10}$ g cm$^{-3}$, following Ref.~\cite{gudmundsson1983} (see also Ref.~\cite{BEZNOGOV20211}, and references therein). 
However, as suggested by \citet{PotekhinYakovlev01} and confirmed by \citet{PotekhinCY07},
such a shallow envelope depth may be insufficient
in strong magnetic fields $B \gtrsim 10^{15}$~G, because significant temperature variations can occur at higher densities in this case. Therefore, we shift the envelope boundary deeper and solve
 the heat transport equations
\begin{eqnarray}&&
   \frac{\mathrm{d}\ln T}{\mathrm{d}\ln P} =
     \frac{F_r}{T}
     \frac{P}{g\varkappa\rho} \frac{1}{\mathcal{K}_h \mathcal{K}_g}
      - \frac{\mathrm{d}\Phi}{\mathrm{d}\ln P},
\label{T}
\\&&
   \frac{1}{r^2}\,\frac{\mathrm{d} (r^2 F_r)}{\mathrm{d}\ln P} =
     \frac{PQ_\nu}{\rho g} 
     \frac{1}{\mathcal{K}_h \mathcal{K}_g}
     - 2 F_r \frac{\mathrm{d}\Phi}{\mathrm{d}\ln P},
\label{F_r}
\end{eqnarray}
 coupled to the mechanical structure equations
\begin{eqnarray}&&
   \frac{\mathrm{d}\Phi}{\mathrm{d}\ln P} = - \frac{P}{\rho c^2}\frac{1}{\mathcal{K}_h},
\label{Phi}
\\&&
   \frac{\mathrm{d} r}{\mathrm{d}\ln P} =
     - \frac{P}{\rho g} \frac{\mathcal{K}_r }{\mathcal{K}_h \mathcal{K}_g},
\\&&
   \frac{\mathrm{d} m}{\mathrm{d}\ln P} =
     - \frac{4\pi r^2 P}{g} \frac{ \mathcal{K}_r }{ \mathcal{K}_h \mathcal{K}_g}.
\label{dmdP}
\end{eqnarray}
Here, $g=Gm/(r^2\mathcal{K}_r)$, while
\begin{equation}
   \mathcal{K}_r = \left(1-\frac{2Gm}{rc^2}\right)^{\frac{1}{2}},\,
   \mathcal{K}_h = 1 + \frac{P}{\rho c^2},\,
   \mathcal{K}_g = 1 + 4\pi r^3 \frac{P}{mc^2}
\label{K_g}
\end{equation}
are GR corrections to radius, enthalpy, and surface gravity, respectively (cf.~\cite{Thorne1977}). The boundary conditions at the surface are $T=T_\mathrm{s}$, $F_r=F_\mathrm{s}$ [Eq.~(\ref{F_s})],
$\Phi=(1/2)\ln(1-2GM/Rc^2)$, $r=R$, and $m=M$.

The mass density $\rho$ on the right-hand sides of these
equations is given by the equation of state,  and $\varkappa$ is calculated as a function of $\rho$,
$T$, $B$, and $\theta_B$, as described in Ref.~\cite{potekhin2015neutron}. The equation of state of the envelope is calculated following Ref.~\cite{PotekhinChabrier13} and adopting the ground-state composition provided by the BSk24 model of \citet{Pearson_18}. The value of $P$ at the surface is obtained following Ref.~\cite{gudmundsson1983} from the equation given by the Eddington approximation to the radiative transfer problem, $P =(2/3)g/\kappa_\mathrm{r}$, where $\kappa_\mathrm{r}$ is the radiative opacity. The latter includes contributions due to free-free transitions and scattering. The
calculation of $\kappa_\mathrm{r}$ as a function of $\rho$, $T$, $B$, and $\theta_B$ is mainly based on Ref.~\cite{Potekhin_03}, but includes the updates due to the Compton scattering and electron-positron pairs at high temperatures (see Ref.~\cite{PotekhinChabrier18}).

In the traditional approach \citep{gudmundsson1983}, one neglects $Q_\nu$ and approximately sets
$\mathcal{K}_g =\mathcal{K}_h=1$, $\Phi={}$constant, and $F_r={}$constant. Then the temperature profile (i.e., $T$ as a function of $P$, $B$, and $\theta_B$) is given by Eq.~(\ref{Tsimple}). In this approximation, $L_\mathrm{b} = L_\mathrm{s}$. In the general case, however, $L_\mathrm{b} \neq L_\mathrm{s}$. In particular, $F_\mathrm{b}$ becomes much larger than $F_\mathrm{s}$ at high $T_\mathrm{b} \gtrsim10^9$~K because of the $Q_\nu$ contribution
in Eq.~(\ref{F_r}). We determine both $F_\mathrm{b}$ and $F_\mathrm{s}$ as functions of $T_\mathrm{b}$, 
$B$, and $\theta_B$ by solving the system of equations
(\ref{T})--(\ref{dmdP}), using the classical fourth-order Runge-Kutta method with an adaptive step in $P$ and controlled accuracy. 
\begin{figure}
    \centering    \includegraphics[width=8.6cm]{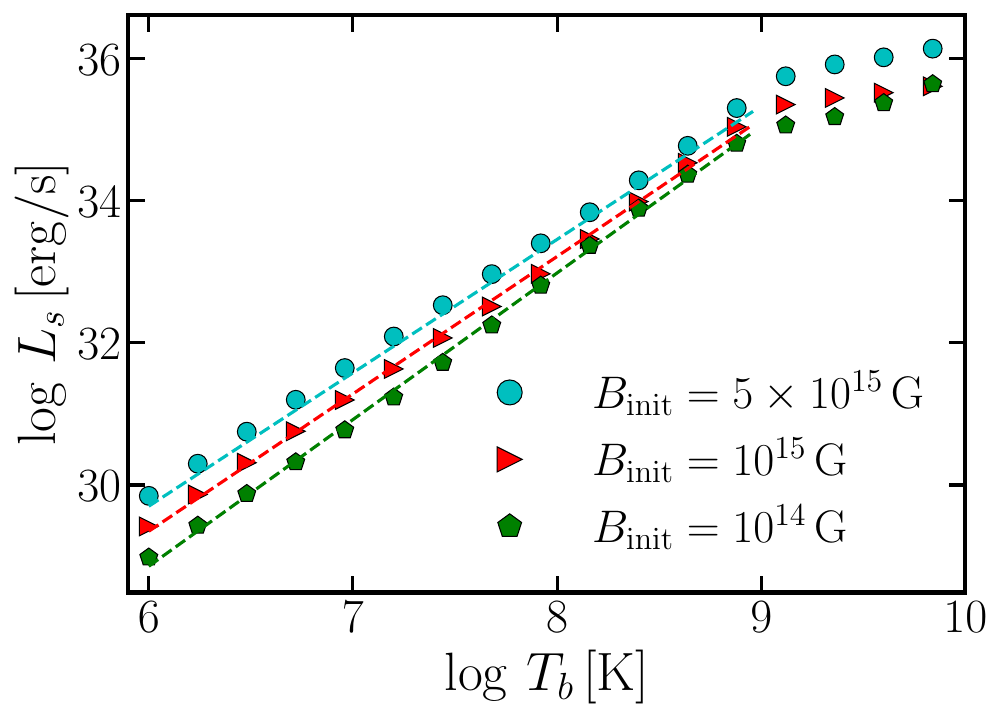}
    \caption{Total surface photon luminosity (Eq.~\ref{eq_total_Ls}) as a function of the blanket (crust-core interface) temperature for model C0. Here, the marker shows the results  obtained using the procedure described in Sec.~\ref{sec:env}, while the dashed lines correspond to a linear fit during the period when $L_{b} = L_s$.} 
    \label{fig_logLs_logTb}
\end{figure}

Previous studies of the thermal structure of NS envelopes have treated convection differently: some 
(see e.~g. Ref.~\cite{potekhin1997,Potekhin_03}) applied the Schwarzschild criterion, valid in the absence of strong magnetic fields, while others (e.~g. Ref.~\cite{potekhin2015neutron}) assumed convection was entirely suppressed by strong magnetic fields. In this work, we adopt a modified Schwarzschild criterion following \citet{GoughTayler66}, which accounts better for the magnetic field’s influence on the onset of convection.

We have computed the tables of $T_\mathrm{s}$ and $F_\mathrm{b}$ as
functions of $T_\mathrm{b}$ and $B$ at $\theta_B=0$ and at
$\theta_B=\pi/2$. To obtain $L_\mathrm{s}$ and $L_\mathrm{b}$, we
integrate $F_\mathrm{s}$ and $F_\mathrm{b}$ over the surface, using
an interpolation in these tables at arbitrary $\theta_B$ with the
fit \citep{Potekhin_03}
\begin{align}
F_\mathrm{s,b}(\theta_B) &=
\left[ F_\mathrm{s,b}^p(0)\cos^2\theta_B 
+ F_\mathrm{s,b}^p\left(\frac{\pi}{2}\right)\sin^2\theta_B \right]^{\frac{1}{p}},
\label{thetaBfit} \\\nonumber
p &= 1 + \left[ F_\mathrm{s}(\pi/2)/F_\mathrm{s}(0) \right]^{\frac{1}{8}}.
\end{align}
In superstrong magnetic fields ($B\gtrsim10^{12}\,\mathrm{G}$), $F_\mathrm{s}(\pi/2) \ll F_\mathrm{s}(0)$. In this case, $p\approx1$,
and Eq.~(\ref{thetaBfit}) reduces to the approximation of \citet{GreensteinHartke83}. 
This limit corresponds to the ground-state, fully catalyzed, or non-accreted (Fe-composition) envelope, which is expected to be realistic for magnetars, as they are young and hot objects.

Figure~\ref{fig_Ts_Lb_Ls_cth05} presents the results of the calculation procedure described above for a fixed angle between the magnetic field lines and the surface normal, $\cos\theta_B =0.5$. To obtain these results, we rescaled the stellar mass and radius to our HHJ EoS so that the envelope is compatible with our stellar model (see Sec.~\ref{sec:background}).

Figure~\ref{fig_BrBth_cthB_Ts_th_C0} shows the surface magnetic field components $B^{s}_r$ and $B^{s}_{\theta}$, along with $\cos \theta_B$ as functions of the polar angle for our initial magnetic field model C0. Additionally, it shows the surface temperature obtained after applying the procedure described above for an rms magnetic field strength of $B_{\mathrm{rms}} = 5 \times 10^{15}\,\mathrm{G}$. 
The influence of a strong magnetic field on the heat conduction within the envelope is clear. The regions where the field lines are perpendicular to the surface normal become thermally insulated, while areas where the field lines are parallel to the normal remain thermally connected to the core.

Figure~\ref{fig_logLs_logTb} displays the logarithm of the total surface photon luminosity, $\log L^{\infty}_s$, as a function of the logarithm of the blanket (crust-core interface) temperature, $\log T_{\infty,b}$, for different rms magnetic field strengths. A linear fit in these variables yields the following relations (with $L^{\infty}_s$ in $\mathrm{erg/s}$ and $T_{\infty,b}$ in K):

\begin{align}
\log\,L^{\infty}_s &= 2.06\,\log\,T_{\infty,b} + 16.48 
&& \text{for } B_{\mathrm{rms}} =10^{14}\,\mathrm{G}, 
\label{eq_Ls_Tb_14}\\
\log\,L^{\infty}_s &= 1.93\,\log\,T_{\infty,b} + 17.75 
&& \text{for } B_{\mathrm{rms}} =10^{15}\,\mathrm{G}, \\
\log\,L^{\infty}_s &= 1.88\,\log\,T_{\infty,b} + 18.44 
&& \text{for } B_{\mathrm{rms}} =5\times10^{15}\,\mathrm{G}.
\label{eq_Ls_Tb_16}
\end{align}
As shown in Fig.~\ref{fig_logLs_logTb}, these linear fits are valid for $T_{b}<10^{9}\,\mathrm{K}$, where $L_{b} \approx L_s$, as discussed above. As explored in Sec.~\ref{sec_results_magnetothermal}, this condition holds in our case. 
In this regard, in all numerical calculations we set $L_{{\rm cb}}^\infty=L_s^\infty$ for simplicity.
This means that the exponents in Eq.~\eqref{e71} are given by $4 \delta  = 2.06,\, 1.93,$  and $1.88$, 
corresponding to $B_\mathrm{rms} = 10^{14},\,10^{15},$ and $5\times10^{15}\, \mathrm{G}$, respectively
[cf. Eqs.\ \eqref{eq_Ls_Tb_14}--\eqref{eq_Ls_Tb_16}].
Finally, we note that despite the significant differences between models A, B, and C0, the $L_s$ vs $T_b$ relations (although not shown here) are nearly identical in all three cases. 
\section{Observational data}\label{sec:observations}

\begin{table*}[]
    \centering
    \begin{tabular}{||l|ccc|l||}
\hline
Name  &  $\log B_p$ [G]  &  $\log t$ [yr]  &  $\log L_s^\infty$ [erg/s]  &   Source \\
\hline
CXOU J2323+5848  &  ---  &  $2.51...2.53$  &  $33.79...33.97$  &  \cite{PotekhinMNRAS2020}  \\ 
XMMU J1720+3726  &  ---  &  $2.78...2.85$  &  $34.03...34.11$  &  \cite{PotekhinMNRAS2020}  \\ 
CXOU J1601+5133  &  ---  &  $2.78...3.00$  &  $33.64...33.80$  &  \cite{PotekhinMNRAS2020}  \\ 
1WGA J1713+3949  &  ---  &  $3.21...3.21$  &  $33.23...33.36$  &  \cite{PotekhinMNRAS2020}  \\ 
CXOU J0852+4617  &  ---  &  $3.38...3.71$  &  $33.00...33.48$  &  \cite{PotekhinMNRAS2020}  \\ 
XMMU J1732+3445  &  ---  &  $3.30...3.78$  &  $34.13...34.29$  &  \cite{PotekhinMNRAS2020}  \\ 
CXOU J1818+1502  &  ---  &  $3.43...3.78$  &  $33.62...34.18$  &  \cite{PotekhinMNRAS2020}  \\ 
2XMM J1046+5943  &  ---  &  $4.04...4.48$  &  $31.89...32.79$  &  \cite{PotekhinMNRAS2020}  \\ 
CXOU J0617+2221  &  ---  &  $4.40...4.54$  &  $32.40...32.43$  &  \cite{PotekhinMNRAS2020}  \\ 
RX J0822--4300  &  $10.76$  &  $3.57...3.72$  &  $33.59...33.79$  &  \cite{PotekhinMNRAS2020}  \\ 
CXOU J1852--0040  &  $10.79$  &  $3.51...3.89$  &  $33.92...34.11$  &  \cite{PotekhinMNRAS2020}  \\ 
1E 1207--5209  &  $11.29$  &  $3.37...4.32$  &  $33.06...33.26$  &  \cite{PotekhinMNRAS2020}  \\ 
PSR J0007+7303  &  $12.34$  &  $3.96...3.97$  &  $<31.48$  &  \cite{PotekhinMNRAS2020}  \\ 
PSR B1823--13  &  $12.75$  &  $4.03...4.63$  &  $32.56...32.73$  &  \cite{PotekhinMNRAS2020}  \\ 
PSR B1706--44  &  $12.79$  &  $3.94...4.54$  &  $31.78...32.94$  &  \cite{PotekhinMNRAS2020}  \\ 
PSR J2021+3651  &  $12.81$  &  $3.93...4.54$  &  $32.54...32.86$  &  \cite{PotekhinMNRAS2020}  \\ 
PSR B0833--45  &  $12.83$  &  $4.23...4.43$  &  $32.61...32.64$  &  \cite{PotekhinMNRAS2020}  \\ 
PSR J0205+6449  &  $12.86$  &  $2.91...2.91$  &  $31.90...32.53$  &  \cite{PotekhinMNRAS2020}  \\ 
PSR B0531+21  &  $12.88$  &  $2.98...2.98$  &  $<34.48$  &  \cite{PotekhinMNRAS2020}  \\ 
PSR J1357--6429  &  $13.19$  &  $3.56...4.16$  &  $32.46...32.63$  &  \cite{PotekhinMNRAS2020}  \\ 
PSR B2334+61  &  $13.30$  &  $3.89...4.30$  &  $31.08...31.91$  &  \cite{PotekhinMNRAS2020}  \\ 
PSR B1509--58  &  $13.48$  &  $2.89...3.49$  &  $33.85...34.04$  &  \cite{PotekhinMNRAS2020}  \\ 
PSR J1119--6127  &  $13.91$  &  $3.62...3.85$  &  $33.04...33.58$  &  \cite{PotekhinMNRAS2020}  \\ 
1E 2259+586  &  $14.08$  &  $4.00...4.30$  &  $35.00...35.40$  &  \cite{vigano2013}  \\ 
SGR 0501+4516  &  $14.57$  &  $3.70...4.30$  &  $33.20...34.00$  &  \cite{vigano2013}  \\ 
Swift J1818--1607  &  $14.57$  &  $2.64...3.24$  &  $<34.23$  &  \cite{HuApJ2020,RajwadeMNRAS2022}  \\ 
XTE J1810--197  &  $14.62$  &  $3.74...4.34$  &  $34.00...34.40$  &  \cite{vigano2013}  \\ 
SGR 1627--41  &  $14.65$  &  $3.40...4.00$  &  $34.40...34.80$  &  \cite{vigano2013}  \\ 
SGR 1745--2900  &  $14.66$  &  $2.98...3.58$  &  $33.90...34.40$  &  \cite{HuNgHoMNRAS2019,CotiZelatiMNRAS2017}  \\ 
1E 1547--5408  &  $14.80$  &  $2.54...3.15$  &  $34.30...34.70$  &  \cite{vigano2013}  \\ 
Swift J1555--5402  &  $14.84$  &  $3.00...3.60$  &  $<34.68$  &  \cite{EnotoApJL2021}  \\ 
1E 1048--5937  &  $14.89$  &  $3.35...3.95$  &  $33.80...34.50$  &  \cite{vigano2013}  \\ 
CXOU J0100--721  &  $14.90$  &  $3.53...4.13$  &  $35.20...35.50$  &  \cite{vigano2013}  \\ 
1RXS J1708--4009  &  $14.97$  &  $3.66...4.26$  &  $34.80...35.10$  &  \cite{vigano2013}  \\ 
CXOU J1714--3810  &  $15.00$  &  $3.40...4.00$  &  $34.90...35.20$  &  \cite{vigano2013}  \\ 
SGR J0526--66  &  $15.04$  &  $3.40...4.00$  &  $35.40...35.80$  &  \cite{vigano2013}  \\ 
1E 1841--045  &  $15.15$  &  $2.70...3.00$  &  $35.20...35.50$  &  \cite{vigano2013}  \\ 
SGR J1900+14  &  $15.15$  &  $3.60...3.90$  &  $35.00...35.40$  &  \cite{vigano2013}  \\ 
SGR J1806--20  &  $15.46$  &  $2.80...3.00$  &  $35.10...35.50$  &  \cite{vigano2013}  \\ 
\hline
\end{tabular}
    \caption{Estimates for magnetic fields $B_p$ (taken at the pole assuming dipole geometry), ages $t$, and surface thermal luminosities $L_s^\infty$ measured by a distant observer for the set of NSs displayed in Fig.~\ref{fig_Observ}. See the text for details.}
    \label{tab_Observ}
\end{table*}

In Table~\ref{tab_Observ}, we list the dataset of isolated cooling NSs that we use to plot Fig.~\ref{fig_Observ}. For weakly-magnetized stars and stars with no magnetic field observed, we mainly take the data from \cite{PotekhinMNRAS2020}. For magnetars, we mainly take the data from Ref.~\cite{vigano2013}, with three exceptions. For Swift~J1818--1607, we take the timing data from \cite{RajwadeMNRAS2022} and adopt the upper limit on the quiescent luminosity from Ref.~\cite{HuApJ2020}. For SGR~1745--2900, we take the timing data from Ref.~\cite{CotiZelatiMNRAS2017} and the data on the quiescent bolometric luminosity from Ref.~\cite{HuNgHoMNRAS2019}. For Swift~J1555--5402, we take both timing data and the upper limit on the luminosity from Ref.~\cite{EnotoApJL2021}.

For the polar magnetic fields, $B_p$, cited values come from pulsar timing estimates, assuming a dipolar geometry. Estimates of ages $t$ are ambiguous for some of the NSs we consider. In case an age estimate not based on spin-down (a kinematic age or an age of the associated supernova remnant) is provided in the literature, we take it as $t$. If only timing data are available, we set lower and upper limits on $t$ as $0.5$ and $2$ times the claimed central value of the characteristic age. We consider only NSs with $t \lesssim 3\times 10^4\,$yr. For the statistical meaning of the intervals and upper limits given in Table~\ref{tab_Observ}, see the cited sources and references therein. 

Notice that, especially for magnetars, the data on thermal luminosities present in the literature are sometimes ambiguous. For instance, Ref. \cite{Park+ApJ2020_0526} yields $L_s^\infty$ for SGR~J0526--66, the hottest source in our set, to be twice lower than in \cite{vigano2013}.

\bibliography{Lib.bib}

\begin{thebibliography}{103}%
\makeatletter
\providecommand \@ifxundefined [1]{%
 \@ifx{#1\undefined}
}%
\providecommand \@ifnum [1]{%
 \ifnum #1\expandafter \@firstoftwo
 \else \expandafter \@secondoftwo
 \fi
}%
\providecommand \@ifx [1]{%
 \ifx #1\expandafter \@firstoftwo
 \else \expandafter \@secondoftwo
 \fi
}%
\providecommand \natexlab [1]{#1}%
\providecommand \enquote  [1]{``#1''}%
\providecommand \bibnamefont  [1]{#1}%
\providecommand \bibfnamefont [1]{#1}%
\providecommand \citenamefont [1]{#1}%
\providecommand \href@noop [0]{\@secondoftwo}%
\providecommand \href [0]{\begingroup \@sanitize@url \@href}%
\providecommand \@href[1]{\@@startlink{#1}\@@href}%
\providecommand \@@href[1]{\endgroup#1\@@endlink}%
\providecommand \@sanitize@url [0]{\catcode `\\12\catcode `\$12\catcode
  `\&12\catcode `\#12\catcode `\^12\catcode `\_12\catcode `\%12\relax}%
\providecommand \@@startlink[1]{}%
\providecommand \@@endlink[0]{}%
\providecommand \url  [0]{\begingroup\@sanitize@url \@url }%
\providecommand \@url [1]{\endgroup\@href {#1}{\urlprefix }}%
\providecommand \urlprefix  [0]{URL }%
\providecommand \Eprint [0]{\href }%
\providecommand \doibase [0]{https://doi.org/}%
\providecommand \selectlanguage [0]{\@gobble}%
\providecommand \bibinfo  [0]{\@secondoftwo}%
\providecommand \bibfield  [0]{\@secondoftwo}%
\providecommand \translation [1]{[#1]}%
\providecommand \BibitemOpen [0]{}%
\providecommand \bibitemStop [0]{}%
\providecommand \bibitemNoStop [0]{.\EOS\space}%
\providecommand \EOS [0]{\spacefactor3000\relax}%
\providecommand \BibitemShut  [1]{\csname bibitem#1\endcsname}%
\let\auto@bib@innerbib\@empty
%</preamble>
\bibitem [{\citenamefont {Mereghetti}\ \emph {et~al.}(2015)\citenamefont
  {Mereghetti}, \citenamefont {Pons},\ and\ \citenamefont
  {Melatos}}]{mereghetti2015}%
  \BibitemOpen
  \bibfield  {author} {\bibinfo {author} {\bibfnamefont {S.}~\bibnamefont
  {Mereghetti}}, \bibinfo {author} {\bibfnamefont {J.~A.}\ \bibnamefont
  {Pons}},\ and\ \bibinfo {author} {\bibfnamefont {A.}~\bibnamefont
  {Melatos}},\ }\href
  {https://doi.org/https://doi.org/10.1007/s11214-015-0146-y} {\bibfield
  {journal} {\bibinfo  {journal} {SSRv}\ }\textbf {\bibinfo {volume} {191}},\
  \bibinfo {pages} {315} (\bibinfo {year} {2015})}\BibitemShut {NoStop}%
\bibitem [{\citenamefont {Kaspi}\ and\ \citenamefont
  {Beloborodov}(2017)}]{kaspi2017magnetars}%
  \BibitemOpen
  \bibfield  {author} {\bibinfo {author} {\bibfnamefont {V.~M.}\ \bibnamefont
  {Kaspi}}\ and\ \bibinfo {author} {\bibfnamefont {A.~M.}\ \bibnamefont
  {Beloborodov}},\ }\href
  {https://doi.org/https://doi.org/10.1146/annurev-astro-081915-023329}
  {\bibfield  {journal} {\bibinfo  {journal} {Annu. Rev. Astron. Astrophys.}\
  }\textbf {\bibinfo {volume} {55}},\ \bibinfo {pages} {261} (\bibinfo {year}
  {2017})}\BibitemShut {NoStop}%
\bibitem [{\citenamefont {{Thompson}}\ and\ \citenamefont
  {{Duncan}}(1995)}]{DuncanThompson1995}%
  \BibitemOpen
  \bibfield  {author} {\bibinfo {author} {\bibfnamefont {C.}~\bibnamefont
  {{Thompson}}}\ and\ \bibinfo {author} {\bibfnamefont {R.~C.}\ \bibnamefont
  {{Duncan}}},\ }\href {https://doi.org/10.1093/mnras/275.2.255} {\bibfield
  {journal} {\bibinfo  {journal} {MNRAS}\ }\textbf {\bibinfo {volume} {275}},\
  \bibinfo {pages} {255} (\bibinfo {year} {1995})}\BibitemShut {NoStop}%
\bibitem [{\citenamefont {{Thompson}}\ and\ \citenamefont
  {{Duncan}}(1996)}]{DuncanThompson1996}%
  \BibitemOpen
  \bibfield  {author} {\bibinfo {author} {\bibfnamefont {C.}~\bibnamefont
  {{Thompson}}}\ and\ \bibinfo {author} {\bibfnamefont {R.~C.}\ \bibnamefont
  {{Duncan}}},\ }\href {https://doi.org/10.1086/178147} {\bibfield  {journal}
  {\bibinfo  {journal} {ApJ}\ }\textbf {\bibinfo {volume} {473}},\ \bibinfo
  {pages} {322} (\bibinfo {year} {1996})}\BibitemShut {NoStop}%
\bibitem [{\citenamefont {Beloborodov}\ and\ \citenamefont
  {Li}(2016)}]{beloborodov2016}%
  \BibitemOpen
  \bibfield  {author} {\bibinfo {author} {\bibfnamefont {A.~M.}\ \bibnamefont
  {Beloborodov}}\ and\ \bibinfo {author} {\bibfnamefont {X.}~\bibnamefont
  {Li}},\ }\href {https://doi.org/10.3847/1538-4357/833/2/261} {\bibfield
  {journal} {\bibinfo  {journal} {ApJ}\ }\textbf {\bibinfo {volume} {833}},\
  \bibinfo {pages} {261} (\bibinfo {year} {2016})}\BibitemShut {NoStop}%
\bibitem [{\citenamefont {Goldreich}\ and\ \citenamefont
  {Reisenegger}(1992)}]{goldreich1992magnetic}%
  \BibitemOpen
  \bibfield  {author} {\bibinfo {author} {\bibfnamefont {P.}~\bibnamefont
  {Goldreich}}\ and\ \bibinfo {author} {\bibfnamefont {A.}~\bibnamefont
  {Reisenegger}},\ }\href {https://doi.org/10.1086/171646} {\bibfield
  {journal} {\bibinfo  {journal} {ApJ}\ }\textbf {\bibinfo {volume} {395}},\
  \bibinfo {pages} {250} (\bibinfo {year} {1992})}\BibitemShut {NoStop}%
\bibitem [{\citenamefont {{Vigan{\`o}}}\ \emph {et~al.}(2013)\citenamefont
  {{Vigan{\`o}}}, \citenamefont {{Rea}}, \citenamefont {{Pons}}, \citenamefont
  {{Perna}}, \citenamefont {{Aguilera}},\ and\ \citenamefont
  {{Miralles}}}]{vigano2013}%
  \BibitemOpen
  \bibfield  {author} {\bibinfo {author} {\bibfnamefont {D.}~\bibnamefont
  {{Vigan{\`o}}}}, \bibinfo {author} {\bibfnamefont {N.}~\bibnamefont {{Rea}}},
  \bibinfo {author} {\bibfnamefont {J.~A.}\ \bibnamefont {{Pons}}}, \bibinfo
  {author} {\bibfnamefont {R.}~\bibnamefont {{Perna}}}, \bibinfo {author}
  {\bibfnamefont {D.~N.}\ \bibnamefont {{Aguilera}}},\ and\ \bibinfo {author}
  {\bibfnamefont {J.~A.}\ \bibnamefont {{Miralles}}},\ }\href
  {https://doi.org/10.1093/mnras/stt1008} {\bibfield  {journal} {\bibinfo
  {journal} {MNRAS}\ }\textbf {\bibinfo {volume} {434}},\ \bibinfo {pages}
  {123} (\bibinfo {year} {2013})}\BibitemShut {NoStop}%
\bibitem [{\citenamefont {Dehman}\ \emph {et~al.}(2023)\citenamefont {Dehman},
  \citenamefont {Vigan{\`o}}, \citenamefont {Pons},\ and\ \citenamefont
  {Rea}}]{dehman20233d}%
  \BibitemOpen
  \bibfield  {author} {\bibinfo {author} {\bibfnamefont {C.}~\bibnamefont
  {Dehman}}, \bibinfo {author} {\bibfnamefont {D.}~\bibnamefont {Vigan{\`o}}},
  \bibinfo {author} {\bibfnamefont {J.~A.}\ \bibnamefont {Pons}},\ and\
  \bibinfo {author} {\bibfnamefont {N.}~\bibnamefont {Rea}},\ }\href
  {https://doi.org/https://doi.org/10.1093/mnras/stac2761} {\bibfield
  {journal} {\bibinfo  {journal} {MNRAS}\ }\textbf {\bibinfo {volume} {518}},\
  \bibinfo {pages} {1222} (\bibinfo {year} {2023})}\BibitemShut {NoStop}%
\bibitem [{\citenamefont {Ascenzi}\ \emph {et~al.}(2024)\citenamefont
  {Ascenzi}, \citenamefont {Vigan{\`o}}, \citenamefont {Dehman}, \citenamefont
  {Pons}, \citenamefont {Rea},\ and\ \citenamefont {Perna}}]{ascenzi2024}%
  \BibitemOpen
  \bibfield  {author} {\bibinfo {author} {\bibfnamefont {S.}~\bibnamefont
  {Ascenzi}}, \bibinfo {author} {\bibfnamefont {D.}~\bibnamefont {Vigan{\`o}}},
  \bibinfo {author} {\bibfnamefont {C.}~\bibnamefont {Dehman}}, \bibinfo
  {author} {\bibfnamefont {J.~A.}\ \bibnamefont {Pons}}, \bibinfo {author}
  {\bibfnamefont {N.}~\bibnamefont {Rea}},\ and\ \bibinfo {author}
  {\bibfnamefont {R.}~\bibnamefont {Perna}},\ }\href
  {https://doi.org/https://doi.org/10.1093/mnras/stae1749} {\bibfield
  {journal} {\bibinfo  {journal} {MNRAS}\ }\textbf {\bibinfo {volume} {533}},\
  \bibinfo {pages} {201} (\bibinfo {year} {2024})}\BibitemShut {NoStop}%
\bibitem [{\citenamefont {Igoshev}\ \emph {et~al.}(2021)\citenamefont
  {Igoshev}, \citenamefont {Hollerbach}, \citenamefont {Wood},\ and\
  \citenamefont {Gourgouliatos}}]{igoshev2021strong}%
  \BibitemOpen
  \bibfield  {author} {\bibinfo {author} {\bibfnamefont {A.~P.}\ \bibnamefont
  {Igoshev}}, \bibinfo {author} {\bibfnamefont {R.}~\bibnamefont {Hollerbach}},
  \bibinfo {author} {\bibfnamefont {T.}~\bibnamefont {Wood}},\ and\ \bibinfo
  {author} {\bibfnamefont {K.~N.}\ \bibnamefont {Gourgouliatos}},\ }\href
  {https://www.nature.com/articles/s41550-020-01220-z} {\bibfield  {journal}
  {\bibinfo  {journal} {Nat. Astron.}\ }\textbf {\bibinfo {volume} {5}},\
  \bibinfo {pages} {145} (\bibinfo {year} {2021})}\BibitemShut {NoStop}%
\bibitem [{\citenamefont {Lander}(2024)}]{lander2024meissner}%
  \BibitemOpen
  \bibfield  {author} {\bibinfo {author} {\bibfnamefont {S.}~\bibnamefont
  {Lander}},\ }\href {https://doi.org/10.1093/mnras/stae2453} {\bibfield
  {journal} {\bibinfo  {journal} {MNRAS}\ }\textbf {\bibinfo {volume} {535}},\
  \bibinfo {pages} {2449} (\bibinfo {year} {2024})}\BibitemShut {NoStop}%
\bibitem [{\citenamefont {Pons}\ \emph {et~al.}(2009)\citenamefont {Pons},
  \citenamefont {Miralles},\ and\ \citenamefont {Geppert}}]{Pons2009}%
  \BibitemOpen
  \bibfield  {author} {\bibinfo {author} {\bibfnamefont {J.~A.}\ \bibnamefont
  {Pons}}, \bibinfo {author} {\bibfnamefont {J.~A.}\ \bibnamefont {Miralles}},\
  and\ \bibinfo {author} {\bibfnamefont {U.}~\bibnamefont {Geppert}},\ }\href
  {https://doi.org/10.1051/0004-6361:200811229} {\bibfield  {journal} {\bibinfo
   {journal} {A\&A}\ }\textbf {\bibinfo {volume} {496}},\ \bibinfo {pages}
  {207} (\bibinfo {year} {2009})}\BibitemShut {NoStop}%
\bibitem [{\citenamefont {De~Grandis}\ \emph {et~al.}(2020)\citenamefont
  {De~Grandis}, \citenamefont {Turolla}, \citenamefont {Wood}, \citenamefont
  {Zane}, \citenamefont {Taverna},\ and\ \citenamefont
  {Gourgouliatos}}]{deGrandis2020}%
  \BibitemOpen
  \bibfield  {author} {\bibinfo {author} {\bibfnamefont {D.}~\bibnamefont
  {De~Grandis}}, \bibinfo {author} {\bibfnamefont {R.}~\bibnamefont {Turolla}},
  \bibinfo {author} {\bibfnamefont {T.~S.}\ \bibnamefont {Wood}}, \bibinfo
  {author} {\bibfnamefont {S.}~\bibnamefont {Zane}}, \bibinfo {author}
  {\bibfnamefont {R.}~\bibnamefont {Taverna}},\ and\ \bibinfo {author}
  {\bibfnamefont {K.~N.}\ \bibnamefont {Gourgouliatos}},\ }\href
  {https://iopscience.iop.org/article/10.3847/1538-4357/abb6f9} {\bibfield
  {journal} {\bibinfo  {journal} {ApJ}\ }\textbf {\bibinfo {volume} {903}},\
  \bibinfo {pages} {40} (\bibinfo {year} {2020})}\BibitemShut {NoStop}%
\bibitem [{\citenamefont {De~Grandis}\ \emph {et~al.}(2021)\citenamefont
  {De~Grandis}, \citenamefont {Taverna}, \citenamefont {Turolla}, \citenamefont
  {Gnarini}, \citenamefont {Popov}, \citenamefont {Zane},\ and\ \citenamefont
  {Wood}}]{deGrandis2021}%
  \BibitemOpen
  \bibfield  {author} {\bibinfo {author} {\bibfnamefont {D.}~\bibnamefont
  {De~Grandis}}, \bibinfo {author} {\bibfnamefont {R.}~\bibnamefont {Taverna}},
  \bibinfo {author} {\bibfnamefont {R.}~\bibnamefont {Turolla}}, \bibinfo
  {author} {\bibfnamefont {A.}~\bibnamefont {Gnarini}}, \bibinfo {author}
  {\bibfnamefont {S.~B.}\ \bibnamefont {Popov}}, \bibinfo {author}
  {\bibfnamefont {S.}~\bibnamefont {Zane}},\ and\ \bibinfo {author}
  {\bibfnamefont {T.~S.}\ \bibnamefont {Wood}},\ }\href
  {https://iopscience.iop.org/article/10.3847/1538-4357/abfdac} {\bibfield
  {journal} {\bibinfo  {journal} {ApJ}\ }\textbf {\bibinfo {volume} {914}},\
  \bibinfo {pages} {118} (\bibinfo {year} {2021})}\BibitemShut {NoStop}%
\bibitem [{\citenamefont {Viganò}\ \emph {et~al.}(2021)\citenamefont
  {Viganò}, \citenamefont {Garcia-Garcia}, \citenamefont {Pons}, \citenamefont
  {Dehman},\ and\ \citenamefont {Graber}}]{vigano2021}%
  \BibitemOpen
  \bibfield  {author} {\bibinfo {author} {\bibfnamefont {D.}~\bibnamefont
  {Viganò}}, \bibinfo {author} {\bibfnamefont {A.}~\bibnamefont
  {Garcia-Garcia}}, \bibinfo {author} {\bibfnamefont {J.~A.}\ \bibnamefont
  {Pons}}, \bibinfo {author} {\bibfnamefont {C.}~\bibnamefont {Dehman}},\ and\
  \bibinfo {author} {\bibfnamefont {V.}~\bibnamefont {Graber}},\ }\href
  {https://doi.org/https://doi.org/10.1016/j.cpc.2021.108001} {\bibfield
  {journal} {\bibinfo  {journal} {Comput. Phys. Commun.}\ }\textbf {\bibinfo
  {volume} {265}},\ \bibinfo {pages} {108001} (\bibinfo {year}
  {2021})}\BibitemShut {NoStop}%
\bibitem [{\citenamefont {Gourgouliatos}\ \emph {et~al.}(2022)\citenamefont
  {Gourgouliatos}, \citenamefont {De~Grandis},\ and\ \citenamefont
  {Igoshev}}]{gourgouliatos+2022magnetic}%
  \BibitemOpen
  \bibfield  {author} {\bibinfo {author} {\bibfnamefont {K.~N.}\ \bibnamefont
  {Gourgouliatos}}, \bibinfo {author} {\bibfnamefont {D.}~\bibnamefont
  {De~Grandis}},\ and\ \bibinfo {author} {\bibfnamefont {A.}~\bibnamefont
  {Igoshev}},\ }\href {https://www.mdpi.com/2073-8994/14/1/130} {\bibfield
  {journal} {\bibinfo  {journal} {Symmetry}\ }\textbf {\bibinfo {volume}
  {14}},\ \bibinfo {pages} {130} (\bibinfo {year} {2022})}\BibitemShut
  {NoStop}%
\bibitem [{\citenamefont {Igoshev}\ \emph {et~al.}(2023)\citenamefont
  {Igoshev}, \citenamefont {Hollerbach},\ and\ \citenamefont
  {Wood}}]{igoshev2023Offsetdipole}%
  \BibitemOpen
  \bibfield  {author} {\bibinfo {author} {\bibfnamefont {A.~P.}\ \bibnamefont
  {Igoshev}}, \bibinfo {author} {\bibfnamefont {R.}~\bibnamefont
  {Hollerbach}},\ and\ \bibinfo {author} {\bibfnamefont {T.}~\bibnamefont
  {Wood}},\ }\href {https://doi.org/10.1093/mnras/stad2404} {\bibfield
  {journal} {\bibinfo  {journal} {MNRAS}\ }\textbf {\bibinfo {volume} {525}},\
  \bibinfo {pages} {3354} (\bibinfo {year} {2023})}\BibitemShut {NoStop}%
\bibitem [{\citenamefont {Igoshev}\ \emph {et~al.}(2025)\citenamefont
  {Igoshev}, \citenamefont {Barrère}, \citenamefont {Raynaud}, \citenamefont
  {Guilet}, \citenamefont {Wood},\ and\ \citenamefont
  {Hollerbach}}]{Igoshev2025}%
  \BibitemOpen
  \bibfield  {author} {\bibinfo {author} {\bibfnamefont {A.}~\bibnamefont
  {Igoshev}}, \bibinfo {author} {\bibfnamefont {P.}~\bibnamefont {Barrère}},
  \bibinfo {author} {\bibfnamefont {R.}~\bibnamefont {Raynaud}}, \bibinfo
  {author} {\bibfnamefont {J.}~\bibnamefont {Guilet}}, \bibinfo {author}
  {\bibfnamefont {T.}~\bibnamefont {Wood}},\ and\ \bibinfo {author}
  {\bibfnamefont {R.}~\bibnamefont {Hollerbach}},\ }\href
  {https://doi.org/10.1038/s41550-025-02477-y} {\bibfield  {journal} {\bibinfo
  {journal} {Nat. Astron.}\ } (\bibinfo {year} {2025})}\BibitemShut {NoStop}%
\bibitem [{\citenamefont {Migdal}(1959)}]{migdal1959superfluidity}%
  \BibitemOpen
  \bibfield  {author} {\bibinfo {author} {\bibfnamefont {A.}~\bibnamefont
  {Migdal}},\ }\href@noop {} {\bibfield  {journal} {\bibinfo  {journal}
  {Nuclear Physics}\ }\textbf {\bibinfo {volume} {13}},\ \bibinfo {pages} {655}
  (\bibinfo {year} {1959})}\BibitemShut {NoStop}%
\bibitem [{\citenamefont {{Page}}\ \emph {et~al.}(2015)\citenamefont {{Page}},
  \citenamefont {{Lattimer}}, \citenamefont {{Prakash}},\ and\ \citenamefont
  {{Steiner}}}]{page2013stellar}%
  \BibitemOpen
  \bibfield  {author} {\bibinfo {author} {\bibfnamefont {D.}~\bibnamefont
  {{Page}}}, \bibinfo {author} {\bibfnamefont {J.~M.}\ \bibnamefont
  {{Lattimer}}}, \bibinfo {author} {\bibfnamefont {M.}~\bibnamefont
  {{Prakash}}},\ and\ \bibinfo {author} {\bibfnamefont {A.~W.}\ \bibnamefont
  {{Steiner}}},\ }\bibinfo {title} {Stellar superfluids},\ in\ \href@noop {}
  {\emph {\bibinfo {booktitle} {Novel Superfluids, vol. 2,}}},\ Vol.\ \bibinfo
  {volume} {157},\ \bibinfo {editor} {edited by\ \bibinfo {editor}
  {\bibfnamefont {K.~H.}\ \bibnamefont {{Bennemann}}}\ and\ \bibinfo {editor}
  {\bibfnamefont {J.~B.}\ \bibnamefont {{Ketterson}}}}\ (\bibinfo  {publisher}
  {International Series of Monographs on Physics, vol. 157, 505, Oxford
  University Press, Oxford},\ \bibinfo {year} {2015})\ pp.\ \bibinfo {pages}
  {505--579}\BibitemShut {NoStop}%
\bibitem [{\citenamefont {Glampedakis}\ \emph {et~al.}(2011)\citenamefont
  {Glampedakis}, \citenamefont {Andersson},\ and\ \citenamefont
  {Samuelsson}}]{glampedakis2011magnetohydrodynamics}%
  \BibitemOpen
  \bibfield  {author} {\bibinfo {author} {\bibfnamefont {K.}~\bibnamefont
  {Glampedakis}}, \bibinfo {author} {\bibfnamefont {N.}~\bibnamefont
  {Andersson}},\ and\ \bibinfo {author} {\bibfnamefont {L.}~\bibnamefont
  {Samuelsson}},\ }\href {https://doi.org/,
  https://doi.org/10.1111/j.1365-2966.2010.17484.x} {\bibfield  {journal}
  {\bibinfo  {journal} {MNRAS}\ }\textbf {\bibinfo {volume} {410}},\ \bibinfo
  {pages} {805} (\bibinfo {year} {2011})}\BibitemShut {NoStop}%
\bibitem [{\citenamefont {{Gusakov}}\ and\ \citenamefont
  {{Dommes}}(2016)}]{gd16}%
  \BibitemOpen
  \bibfield  {author} {\bibinfo {author} {\bibfnamefont {M.~E.}\ \bibnamefont
  {{Gusakov}}}\ and\ \bibinfo {author} {\bibfnamefont {V.~A.}\ \bibnamefont
  {{Dommes}}},\ }\href {https://doi.org/10.1103/PhysRevD.94.083006} {\bibfield
  {journal} {\bibinfo  {journal} {\prd}\ }\textbf {\bibinfo {volume} {94}},\
  \bibinfo {eid} {083006} (\bibinfo {year} {2016})}\BibitemShut {NoStop}%
\bibitem [{\citenamefont {Gusakov}\ \emph {et~al.}(2020)\citenamefont
  {Gusakov}, \citenamefont {Kantor},\ and\ \citenamefont
  {Ofengeim}}]{Gusakovetal2020}%
  \BibitemOpen
  \bibfield  {author} {\bibinfo {author} {\bibfnamefont {M.}~\bibnamefont
  {Gusakov}}, \bibinfo {author} {\bibfnamefont {E.}~\bibnamefont {Kantor}},\
  and\ \bibinfo {author} {\bibfnamefont {D.}~\bibnamefont {Ofengeim}},\ }\href
  {https://doi.org/10.1093/mnras/staa3160} {\bibfield  {journal} {\bibinfo
  {journal} {MNRAS}\ }\textbf {\bibinfo {volume} {499}},\ \bibinfo {pages}
  {4561} (\bibinfo {year} {2020})}\BibitemShut {NoStop}%
\bibitem [{\citenamefont {{Bransgrove}}\ \emph {et~al.}(2025)\citenamefont
  {{Bransgrove}}, \citenamefont {{Levin}},\ and\ \citenamefont
  {{Beloborodov}}}]{bransgrove2024giant}%
  \BibitemOpen
  \bibfield  {author} {\bibinfo {author} {\bibfnamefont {A.}~\bibnamefont
  {{Bransgrove}}}, \bibinfo {author} {\bibfnamefont {Y.}~\bibnamefont
  {{Levin}}},\ and\ \bibinfo {author} {\bibfnamefont {A.~M.}\ \bibnamefont
  {{Beloborodov}}},\ }\href {https://doi.org/10.3847/1538-4357/ad90a3}
  {\bibfield  {journal} {\bibinfo  {journal} {Astrophys. J.}\ }\textbf
  {\bibinfo {volume} {979}},\ \bibinfo {eid} {144} (\bibinfo {year}
  {2025})}\BibitemShut {NoStop}%
\bibitem [{\citenamefont {Braithwaite}\ and\ \citenamefont
  {Spruit}(2004)}]{braithwaite2004fossil}%
  \BibitemOpen
  \bibfield  {author} {\bibinfo {author} {\bibfnamefont {J.}~\bibnamefont
  {Braithwaite}}\ and\ \bibinfo {author} {\bibfnamefont {H.~C.}\ \bibnamefont
  {Spruit}},\ }\href {https://doi.org/https://doi.org/10.1038/nature02934}
  {\bibfield  {journal} {\bibinfo  {journal} {Nature}\ }\textbf {\bibinfo
  {volume} {431}},\ \bibinfo {pages} {819} (\bibinfo {year}
  {2004})}\BibitemShut {NoStop}%
\bibitem [{\citenamefont {{Becerra}}\ \emph
  {et~al.}(2022{\natexlab{a}})\citenamefont {{Becerra}}, \citenamefont
  {{Reisenegger}}, \citenamefont {{Valdivia}},\ and\ \citenamefont
  {{Gusakov}}}]{Becerra22a}%
  \BibitemOpen
  \bibfield  {author} {\bibinfo {author} {\bibfnamefont {L.}~\bibnamefont
  {{Becerra}}}, \bibinfo {author} {\bibfnamefont {A.}~\bibnamefont
  {{Reisenegger}}}, \bibinfo {author} {\bibfnamefont {J.~A.}\ \bibnamefont
  {{Valdivia}}},\ and\ \bibinfo {author} {\bibfnamefont {M.~E.}\ \bibnamefont
  {{Gusakov}}},\ }\href {https://doi.org/https://doi.org/10.1093/mnras/stac102}
  {\bibfield  {journal} {\bibinfo  {journal} {MNRAS}\ }\textbf {\bibinfo
  {volume} {511}},\ \bibinfo {pages} {732} (\bibinfo {year}
  {2022}{\natexlab{a}})}\BibitemShut {NoStop}%
\bibitem [{\citenamefont {{Pethick}}(1992)}]{pethick1992}%
  \BibitemOpen
  \bibfield  {author} {\bibinfo {author} {\bibfnamefont {C.~J.}\ \bibnamefont
  {{Pethick}}},\ }in\ \href@noop {} {\emph {\bibinfo {booktitle} {Structure and
  Evolution of Neutron Stars}}}\ (\bibinfo {year} {1992})\ p.\ \bibinfo {pages}
  {115}\BibitemShut {NoStop}%
\bibitem [{\citenamefont {{Reisenegger}}\ and\ \citenamefont
  {{Goldreich}}(1992)}]{Reisenegger92}%
  \BibitemOpen
  \bibfield  {author} {\bibinfo {author} {\bibfnamefont {A.}~\bibnamefont
  {{Reisenegger}}}\ and\ \bibinfo {author} {\bibfnamefont {P.}~\bibnamefont
  {{Goldreich}}},\ }\href {https://doi.org/10.1086/171645} {\bibfield
  {journal} {\bibinfo  {journal} {Astrophys. J.}\ }\textbf {\bibinfo {volume}
  {395}},\ \bibinfo {pages} {240} (\bibinfo {year} {1992})}\BibitemShut
  {NoStop}%
\bibitem [{\citenamefont {Blandford}\ and\ \citenamefont
  {Thorne}(2008)}]{blandford2008applications}%
  \BibitemOpen
  \bibfield  {author} {\bibinfo {author} {\bibfnamefont {R.~D.}\ \bibnamefont
  {Blandford}}\ and\ \bibinfo {author} {\bibfnamefont {K.~S.}\ \bibnamefont
  {Thorne}},\ }\href@noop {} {\bibfield  {journal} {\bibinfo  {journal}
  {lecture notes, California Institute of Technology}\ ,\ \bibinfo {pages}
  {12}} (\bibinfo {year} {2008})}\BibitemShut {NoStop}%
\bibitem [{\citenamefont {{Moraga}}\ \emph {et~al.}(2024)\citenamefont
  {{Moraga}}, \citenamefont {{Castillo}}, \citenamefont {{Reisenegger}},
  \citenamefont {{Valdivia}},\ and\ \citenamefont {{Gusakov}}}]{Moraga2024}%
  \BibitemOpen
  \bibfield  {author} {\bibinfo {author} {\bibfnamefont {N.~A.}\ \bibnamefont
  {{Moraga}}}, \bibinfo {author} {\bibfnamefont {F.}~\bibnamefont
  {{Castillo}}}, \bibinfo {author} {\bibfnamefont {A.}~\bibnamefont
  {{Reisenegger}}}, \bibinfo {author} {\bibfnamefont {J.~A.}\ \bibnamefont
  {{Valdivia}}},\ and\ \bibinfo {author} {\bibfnamefont {M.~E.}\ \bibnamefont
  {{Gusakov}}},\ }\href {https://doi.org/10.1093/mnras/stad3787} {\bibfield
  {journal} {\bibinfo  {journal} {MNRAS}\ }\textbf {\bibinfo {volume} {527}},\
  \bibinfo {pages} {9431} (\bibinfo {year} {2024})}\BibitemShut {NoStop}%
\bibitem [{\citenamefont {Shapiro}\ and\ \citenamefont
  {Teukolsky}(1983)}]{shapiro1983physics}%
  \BibitemOpen
  \bibfield  {author} {\bibinfo {author} {\bibfnamefont {S.~L.}\ \bibnamefont
  {Shapiro}}\ and\ \bibinfo {author} {\bibfnamefont {S.~A.}\ \bibnamefont
  {Teukolsky}},\ }\href@noop {} {\bibfield  {journal} {\bibinfo  {journal} {New
  York: John Wiley \&Sons}\ } (\bibinfo {year} {1983})}\BibitemShut {NoStop}%
\bibitem [{\citenamefont {{Reisenegger}}\ \emph {et~al.}(2005)\citenamefont
  {{Reisenegger}}, \citenamefont {{Prieto}}, \citenamefont {{Benguria}},
  \citenamefont {{Lai}},\ and\ \citenamefont {{Araya}}}]{Reisenegger2005}%
  \BibitemOpen
  \bibfield  {author} {\bibinfo {author} {\bibfnamefont {A.}~\bibnamefont
  {{Reisenegger}}}, \bibinfo {author} {\bibfnamefont {J.~P.}\ \bibnamefont
  {{Prieto}}}, \bibinfo {author} {\bibfnamefont {R.}~\bibnamefont
  {{Benguria}}}, \bibinfo {author} {\bibfnamefont {D.}~\bibnamefont {{Lai}}},\
  and\ \bibinfo {author} {\bibfnamefont {P.~A.}\ \bibnamefont {{Araya}}},\ }in\
  \href {https://doi.org/10.1063/1.2077190} {\emph {\bibinfo {booktitle}
  {Magnetic Fields in the Universe: From Laboratory and Stars to Primordial
  Structures.}}},\ \bibinfo {series} {American Institute of Physics Conference
  Series}, Vol.\ \bibinfo {volume} {784},\ \bibinfo {editor} {edited by\
  \bibinfo {editor} {\bibfnamefont {E.~M.}\ \bibnamefont {{de Gouveia dal
  Pino}}}, \bibinfo {editor} {\bibfnamefont {G.}~\bibnamefont {{Lugones}}},\
  and\ \bibinfo {editor} {\bibfnamefont {A.}~\bibnamefont {{Lazarian}}}}\
  (\bibinfo {year} {2005})\ pp.\ \bibinfo {pages} {263--273},\ \Eprint
  {https://arxiv.org/abs/astro-ph/0503047} {arXiv:astro-ph/0503047 [astro-ph]}
  \BibitemShut {NoStop}%
\bibitem [{\citenamefont {Reisenegger}(2009)}]{reisenegger2009stable}%
  \BibitemOpen
  \bibfield  {author} {\bibinfo {author} {\bibfnamefont {A.}~\bibnamefont
  {Reisenegger}},\ }\href {https://doi.org/10.1051/0004-6361/200810895}
  {\bibfield  {journal} {\bibinfo  {journal} {A \& A}\ }\textbf {\bibinfo
  {volume} {499}},\ \bibinfo {pages} {557} (\bibinfo {year}
  {2009})}\BibitemShut {NoStop}%
\bibitem [{\citenamefont {Ofengeim}\ and\ \citenamefont
  {Gusakov}(2018)}]{ofegeim2018}%
  \BibitemOpen
  \bibfield  {author} {\bibinfo {author} {\bibfnamefont {D.~D.}\ \bibnamefont
  {Ofengeim}}\ and\ \bibinfo {author} {\bibfnamefont {M.~E.}\ \bibnamefont
  {Gusakov}},\ }\href {https://doi.org/10.1103/PhysRevD.98.043007} {\bibfield
  {journal} {\bibinfo  {journal} {Phys. Rev. D}\ }\textbf {\bibinfo {volume}
  {98}},\ \bibinfo {pages} {043007} (\bibinfo {year} {2018})}\BibitemShut
  {NoStop}%
\bibitem [{\citenamefont {Hoyos}\ \emph {et~al.}(2008)\citenamefont {Hoyos},
  \citenamefont {Reisenegger},\ and\ \citenamefont
  {Valdivia}}]{hoyos2008magnetic}%
  \BibitemOpen
  \bibfield  {author} {\bibinfo {author} {\bibfnamefont {J.}~\bibnamefont
  {Hoyos}}, \bibinfo {author} {\bibfnamefont {A.}~\bibnamefont {Reisenegger}},\
  and\ \bibinfo {author} {\bibfnamefont {J.~A.}\ \bibnamefont {Valdivia}},\
  }\href {https://doi.org/10.1051/0004-6361:200809466} {\bibfield  {journal}
  {\bibinfo  {journal} {A\&A}\ }\textbf {\bibinfo {volume} {487}},\ \bibinfo
  {pages} {789} (\bibinfo {year} {2008})}\BibitemShut {NoStop}%
\bibitem [{\citenamefont {{Passamonti}}\ \emph {et~al.}(2017)\citenamefont
  {{Passamonti}}, \citenamefont {{Akg{\"u}n}}, \citenamefont {{Pons}},\ and\
  \citenamefont {{Miralles}}}]{Passamonti2017ambipolar}%
  \BibitemOpen
  \bibfield  {author} {\bibinfo {author} {\bibfnamefont {A.}~\bibnamefont
  {{Passamonti}}}, \bibinfo {author} {\bibfnamefont {T.}~\bibnamefont
  {{Akg{\"u}n}}}, \bibinfo {author} {\bibfnamefont {J.~A.}\ \bibnamefont
  {{Pons}}},\ and\ \bibinfo {author} {\bibfnamefont {J.~A.}\ \bibnamefont
  {{Miralles}}},\ }\href {https://doi.org/10.1093/mnras/stw2936} {\bibfield
  {journal} {\bibinfo  {journal} {MNRAS}\ }\textbf {\bibinfo {volume} {465}},\
  \bibinfo {pages} {3416} (\bibinfo {year} {2017})}\BibitemShut {NoStop}%
\bibitem [{\citenamefont {Igoshev}\ and\ \citenamefont
  {Hollerbach}(2023)}]{igoshev2023}%
  \BibitemOpen
  \bibfield  {author} {\bibinfo {author} {\bibfnamefont {A.~P.}\ \bibnamefont
  {Igoshev}}\ and\ \bibinfo {author} {\bibfnamefont {R.}~\bibnamefont
  {Hollerbach}},\ }\href
  {https://doi.org/https://doi.org/10.1093/mnras/stac3126} {\bibfield
  {journal} {\bibinfo  {journal} {MNRAS}\ }\textbf {\bibinfo {volume} {518}},\
  \bibinfo {pages} {821} (\bibinfo {year} {2023})}\BibitemShut {NoStop}%
\bibitem [{\citenamefont {{Skiathas}}\ and\ \citenamefont
  {{Gourgouliatos}}(2024)}]{Skiathas2024}%
  \BibitemOpen
  \bibfield  {author} {\bibinfo {author} {\bibfnamefont {D.}~\bibnamefont
  {{Skiathas}}}\ and\ \bibinfo {author} {\bibfnamefont {K.~N.}\ \bibnamefont
  {{Gourgouliatos}}},\ }\href {https://doi.org/10.1093/mnras/stae190}
  {\bibfield  {journal} {\bibinfo  {journal} {MNRAS}\ }\textbf {\bibinfo
  {volume} {528}},\ \bibinfo {pages} {5178} (\bibinfo {year}
  {2024})}\BibitemShut {NoStop}%
\bibitem [{\citenamefont {Gusakov}\ \emph {et~al.}(2017)\citenamefont
  {Gusakov}, \citenamefont {Kantor},\ and\ \citenamefont
  {Ofengeim}}]{gusakov2017evolution}%
  \BibitemOpen
  \bibfield  {author} {\bibinfo {author} {\bibfnamefont {M.}~\bibnamefont
  {Gusakov}}, \bibinfo {author} {\bibfnamefont {E.}~\bibnamefont {Kantor}},\
  and\ \bibinfo {author} {\bibfnamefont {D.}~\bibnamefont {Ofengeim}},\ }\href
  {https://doi.org/https://doi.org/10.1103/PhysRevD.96.103012} {\bibfield
  {journal} {\bibinfo  {journal} {Phys. Rev. D}\ }\textbf {\bibinfo {volume}
  {96}},\ \bibinfo {pages} {103012} (\bibinfo {year} {2017})}\BibitemShut
  {NoStop}%
\bibitem [{\citenamefont {Castillo}\ \emph {et~al.}(2020)\citenamefont
  {Castillo}, \citenamefont {Reisenegger},\ and\ \citenamefont
  {Valdivia}}]{castillo2020twofluid}%
  \BibitemOpen
  \bibfield  {author} {\bibinfo {author} {\bibfnamefont {F.}~\bibnamefont
  {Castillo}}, \bibinfo {author} {\bibfnamefont {A.}~\bibnamefont
  {Reisenegger}},\ and\ \bibinfo {author} {\bibfnamefont {J.~A.}\ \bibnamefont
  {Valdivia}},\ }\href {https://doi.org/10.1093/mnras/staa2543} {\bibfield
  {journal} {\bibinfo  {journal} {MNRAS}\ }\textbf {\bibinfo {volume} {498}},\
  \bibinfo {pages} {3000} (\bibinfo {year} {2020})}\BibitemShut {NoStop}%
\bibitem [{\citenamefont {Hoyos}\ \emph {et~al.}(2010)\citenamefont {Hoyos},
  \citenamefont {Reisenegger},\ and\ \citenamefont {Valdivia}}]{Hoyos2010}%
  \BibitemOpen
  \bibfield  {author} {\bibinfo {author} {\bibfnamefont {J.~H.}\ \bibnamefont
  {Hoyos}}, \bibinfo {author} {\bibfnamefont {A.}~\bibnamefont {Reisenegger}},\
  and\ \bibinfo {author} {\bibfnamefont {J.~A.}\ \bibnamefont {Valdivia}},\
  }\href {https://doi.org/10.1111/j.1365-2966.2010.17237.x} {\bibfield
  {journal} {\bibinfo  {journal} {MNRAS}\ }\textbf {\bibinfo {volume} {408}},\
  \bibinfo {pages} {1730} (\bibinfo {year} {2010})}\BibitemShut {NoStop}%
\bibitem [{\citenamefont {Castillo}\ \emph {et~al.}(2017)\citenamefont
  {Castillo}, \citenamefont {Reisenegger},\ and\ \citenamefont
  {Valdivia}}]{Castillo2017}%
  \BibitemOpen
  \bibfield  {author} {\bibinfo {author} {\bibfnamefont {F.}~\bibnamefont
  {Castillo}}, \bibinfo {author} {\bibfnamefont {A.}~\bibnamefont
  {Reisenegger}},\ and\ \bibinfo {author} {\bibfnamefont {J.~A.}\ \bibnamefont
  {Valdivia}},\ }\href {https://doi.org/10.1093/mnras/stx1604} {\bibfield
  {journal} {\bibinfo  {journal} {MNRAS}\ }\textbf {\bibinfo {volume} {471}},\
  \bibinfo {pages} {507} (\bibinfo {year} {2017})}\BibitemShut {NoStop}%
\bibitem [{\citenamefont {{Tsuruta}}\ \emph {et~al.}(2023)\citenamefont
  {{Tsuruta}}, \citenamefont {{Kelly}}, \citenamefont {{Nomoto}}, \citenamefont
  {{Mori}}, \citenamefont {{Teter}},\ and\ \citenamefont
  {{Liebmann}}}]{Tsuruta2023}%
  \BibitemOpen
  \bibfield  {author} {\bibinfo {author} {\bibfnamefont {S.}~\bibnamefont
  {{Tsuruta}}}, \bibinfo {author} {\bibfnamefont {M.~J.}\ \bibnamefont
  {{Kelly}}}, \bibinfo {author} {\bibfnamefont {K.}~\bibnamefont {{Nomoto}}},
  \bibinfo {author} {\bibfnamefont {K.}~\bibnamefont {{Mori}}}, \bibinfo
  {author} {\bibfnamefont {M.}~\bibnamefont {{Teter}}},\ and\ \bibinfo {author}
  {\bibfnamefont {A.~C.}\ \bibnamefont {{Liebmann}}},\ }\href
  {https://doi.org/10.3847/1538-4357/acbd38} {\bibfield  {journal} {\bibinfo
  {journal} {Astrophys. J.}\ }\textbf {\bibinfo {volume} {945}},\ \bibinfo
  {eid} {151} (\bibinfo {year} {2023})}\BibitemShut {NoStop}%
\bibitem [{\citenamefont {{Castillo}}\ \emph {et~al.}(2025)\citenamefont
  {{Castillo}}, \citenamefont {{Moraga}}, \citenamefont {{Gusakov}},
  \citenamefont {{Valdivia}},\ and\ \citenamefont
  {{Reisenegger}}}]{castillo2025AA}%
  \BibitemOpen
  \bibfield  {author} {\bibinfo {author} {\bibfnamefont {F.}~\bibnamefont
  {{Castillo}}}, \bibinfo {author} {\bibfnamefont {N.~A.}\ \bibnamefont
  {{Moraga}}}, \bibinfo {author} {\bibfnamefont {M.~E.}\ \bibnamefont
  {{Gusakov}}}, \bibinfo {author} {\bibfnamefont {J.~A.}\ \bibnamefont
  {{Valdivia}}},\ and\ \bibinfo {author} {\bibfnamefont {A.}~\bibnamefont
  {{Reisenegger}}},\ }\href {https://doi.org/10.48550/arXiv.2503.11530}
  {\bibfield  {journal} {\bibinfo  {journal} {arXiv e-prints}\ ,\ \bibinfo
  {eid} {arXiv:2503.11530}} (\bibinfo {year} {2025})},\ \Eprint
  {https://arxiv.org/abs/2503.11530} {arXiv:2503.11530 [astro-ph.HE]}
  \BibitemShut {NoStop}%
\bibitem [{\citenamefont {{Chodura}}\ and\ \citenamefont
  {{Schlueter}}(1981)}]{Chodura1981}%
  \BibitemOpen
  \bibfield  {author} {\bibinfo {author} {\bibfnamefont {R.}~\bibnamefont
  {{Chodura}}}\ and\ \bibinfo {author} {\bibfnamefont {A.}~\bibnamefont
  {{Schlueter}}},\ }\href {https://doi.org/10.1016/0021-9991(81)90080-2}
  {\bibfield  {journal} {\bibinfo  {journal} {J. Comput. Phys.}\ }\textbf
  {\bibinfo {volume} {41}},\ \bibinfo {pages} {68} (\bibinfo {year}
  {1981})}\BibitemShut {NoStop}%
\bibitem [{\citenamefont {{Yang}}\ \emph {et~al.}(1986)\citenamefont {{Yang}},
  \citenamefont {{Sturrock}},\ and\ \citenamefont {{Antiochos}}}]{Yang1986}%
  \BibitemOpen
  \bibfield  {author} {\bibinfo {author} {\bibfnamefont {W.~H.}\ \bibnamefont
  {{Yang}}}, \bibinfo {author} {\bibfnamefont {P.~A.}\ \bibnamefont
  {{Sturrock}}},\ and\ \bibinfo {author} {\bibfnamefont {S.~K.}\ \bibnamefont
  {{Antiochos}}},\ }\href {https://doi.org/10.1086/164610} {\bibfield
  {journal} {\bibinfo  {journal} {Astrophys. J.}\ }\textbf {\bibinfo {volume}
  {309}},\ \bibinfo {pages} {383} (\bibinfo {year} {1986})}\BibitemShut
  {NoStop}%
\bibitem [{\citenamefont {{Roumeliotis}}\ \emph {et~al.}(1994)\citenamefont
  {{Roumeliotis}}, \citenamefont {{Sturrock}},\ and\ \citenamefont
  {{Antiochos}}}]{Roumeliotis1994}%
  \BibitemOpen
  \bibfield  {author} {\bibinfo {author} {\bibfnamefont {G.}~\bibnamefont
  {{Roumeliotis}}}, \bibinfo {author} {\bibfnamefont {P.~A.}\ \bibnamefont
  {{Sturrock}}},\ and\ \bibinfo {author} {\bibfnamefont {S.~K.}\ \bibnamefont
  {{Antiochos}}},\ }\href {https://doi.org/10.1086/173862} {\bibfield
  {journal} {\bibinfo  {journal} {Astrophys. J.}\ }\textbf {\bibinfo {volume}
  {423}},\ \bibinfo {pages} {847} (\bibinfo {year} {1994})}\BibitemShut
  {NoStop}%
\bibitem [{\citenamefont {{Vigan{\`o}}}\ \emph {et~al.}(2011)\citenamefont
  {{Vigan{\`o}}}, \citenamefont {{Pons}},\ and\ \citenamefont
  {{Miralles}}}]{Vigano2011}%
  \BibitemOpen
  \bibfield  {author} {\bibinfo {author} {\bibfnamefont {D.}~\bibnamefont
  {{Vigan{\`o}}}}, \bibinfo {author} {\bibfnamefont {J.~A.}\ \bibnamefont
  {{Pons}}},\ and\ \bibinfo {author} {\bibfnamefont {J.~A.}\ \bibnamefont
  {{Miralles}}},\ }\href {https://doi.org/10.1051/0004-6361/201117105}
  {\bibfield  {journal} {\bibinfo  {journal} {Astron. Astrophys.}\ }\textbf
  {\bibinfo {volume} {533}},\ \bibinfo {eid} {A125} (\bibinfo {year}
  {2011})}\BibitemShut {NoStop}%
\bibitem [{\citenamefont {Tolman}(1939)}]{tolman1939static}%
  \BibitemOpen
  \bibfield  {author} {\bibinfo {author} {\bibfnamefont {R.~C.}\ \bibnamefont
  {Tolman}},\ }\href {https://doi.org/https://doi.org/10.1103/PhysRev.55.364}
  {\bibfield  {journal} {\bibinfo  {journal} {Physical Review}\ }\textbf
  {\bibinfo {volume} {55}},\ \bibinfo {pages} {364} (\bibinfo {year}
  {1939})}\BibitemShut {NoStop}%
\bibitem [{\citenamefont {Oppenheimer}\ and\ \citenamefont
  {Volkoff}(1939)}]{TOV}%
  \BibitemOpen
  \bibfield  {author} {\bibinfo {author} {\bibfnamefont {J.~R.}\ \bibnamefont
  {Oppenheimer}}\ and\ \bibinfo {author} {\bibfnamefont {G.~M.}\ \bibnamefont
  {Volkoff}},\ }\href {https://doi.org/10.1103/PhysRev.55.374} {\bibfield
  {journal} {\bibinfo  {journal} {Phys. Rev.}\ }\textbf {\bibinfo {volume}
  {55}},\ \bibinfo {pages} {374} (\bibinfo {year} {1939})}\BibitemShut
  {NoStop}%
\bibitem [{\citenamefont {Heiselberg}\ and\ \citenamefont
  {Hjorth-Jensen}(1999)}]{Heiselberg_1999}%
  \BibitemOpen
  \bibfield  {author} {\bibinfo {author} {\bibfnamefont {H.}~\bibnamefont
  {Heiselberg}}\ and\ \bibinfo {author} {\bibfnamefont {M.}~\bibnamefont
  {Hjorth-Jensen}},\ }\href {https://doi.org/10.1086/312321} {\bibfield
  {journal} {\bibinfo  {journal} {ApJ}\ }\textbf {\bibinfo {volume} {525}},\
  \bibinfo {pages} {L45} (\bibinfo {year} {1999})}\BibitemShut {NoStop}%
\bibitem [{\citenamefont {{Shalybkov}}\ and\ \citenamefont
  {{Urpin}}(1995)}]{su95}%
  \BibitemOpen
  \bibfield  {author} {\bibinfo {author} {\bibfnamefont {D.~A.}\ \bibnamefont
  {{Shalybkov}}}\ and\ \bibinfo {author} {\bibfnamefont {V.~A.}\ \bibnamefont
  {{Urpin}}},\ }\href {https://doi.org/10.1093/mnras/273.3.643} {\bibfield
  {journal} {\bibinfo  {journal} {MNRAS}\ }\textbf {\bibinfo {volume} {273}},\
  \bibinfo {pages} {643} (\bibinfo {year} {1995})}\BibitemShut {NoStop}%
\bibitem [{\citenamefont {Dommes}\ \emph {et~al.}(2020)\citenamefont {Dommes},
  \citenamefont {Gusakov},\ and\ \citenamefont {Shternin}}]{DommesGusakov2021}%
  \BibitemOpen
  \bibfield  {author} {\bibinfo {author} {\bibfnamefont {V.~A.}\ \bibnamefont
  {Dommes}}, \bibinfo {author} {\bibfnamefont {M.~E.}\ \bibnamefont
  {Gusakov}},\ and\ \bibinfo {author} {\bibfnamefont {P.~S.}\ \bibnamefont
  {Shternin}},\ }\href {https://doi.org/10.1103/PhysRevD.101.103020} {\bibfield
   {journal} {\bibinfo  {journal} {Phys. Rev. D}\ }\textbf {\bibinfo {volume}
  {101}},\ \bibinfo {pages} {103020} (\bibinfo {year} {2020})}\BibitemShut
  {NoStop}%
\bibitem [{\citenamefont {Yakovlev}\ and\ \citenamefont
  {Shalybkov}(1991)}]{Yakovlev1991}%
  \BibitemOpen
  \bibfield  {author} {\bibinfo {author} {\bibfnamefont {D.~G.}\ \bibnamefont
  {Yakovlev}}\ and\ \bibinfo {author} {\bibfnamefont {D.~A.}\ \bibnamefont
  {Shalybkov}},\ }\href {https://api.semanticscholar.org/CorpusID:125401679}
  {\bibfield  {journal} {\bibinfo  {journal} {Astrophys. Space Sci.}\ }\textbf
  {\bibinfo {volume} {176}},\ \bibinfo {pages} {191} (\bibinfo {year}
  {1991})}\BibitemShut {NoStop}%
\bibitem [{\citenamefont {{Thorne}}(1966)}]{Thorne1966}%
  \BibitemOpen
  \bibfield  {author} {\bibinfo {author} {\bibfnamefont {K.~S.}\ \bibnamefont
  {{Thorne}}},\ }\href {https://doi.org/10.1086/148595} {\bibfield  {journal}
  {\bibinfo  {journal} {Astrophys. J.}\ }\textbf {\bibinfo {volume} {144}},\
  \bibinfo {pages} {201} (\bibinfo {year} {1966})}\BibitemShut {NoStop}%
\bibitem [{\citenamefont {{Thorne}}(1977)}]{Thorne1977}%
  \BibitemOpen
  \bibfield  {author} {\bibinfo {author} {\bibfnamefont {K.~S.}\ \bibnamefont
  {{Thorne}}},\ }\href {https://doi.org/10.1086/155108} {\bibfield  {journal}
  {\bibinfo  {journal} {ApJ}\ }\textbf {\bibinfo {volume} {212}},\ \bibinfo
  {pages} {825} (\bibinfo {year} {1977})}\BibitemShut {NoStop}%
\bibitem [{\citenamefont {{Glen}}\ and\ \citenamefont
  {{Sutherland}}(1980)}]{GlenSutherland80}%
  \BibitemOpen
  \bibfield  {author} {\bibinfo {author} {\bibfnamefont {G.}~\bibnamefont
  {{Glen}}}\ and\ \bibinfo {author} {\bibfnamefont {P.}~\bibnamefont
  {{Sutherland}}},\ }\href {https://doi.org/10.1086/158154} {\bibfield
  {journal} {\bibinfo  {journal} {Astrophys. J.}\ }\textbf {\bibinfo {volume}
  {239}},\ \bibinfo {pages} {671} (\bibinfo {year} {1980})}\BibitemShut
  {NoStop}%
\bibitem [{\citenamefont {Yakovlev}\ and\ \citenamefont
  {Pethick}(2004)}]{yakovlev2004neutron}%
  \BibitemOpen
  \bibfield  {author} {\bibinfo {author} {\bibfnamefont {D.~G.}\ \bibnamefont
  {Yakovlev}}\ and\ \bibinfo {author} {\bibfnamefont {C.}~\bibnamefont
  {Pethick}},\ }\bibfield  {journal} {\bibinfo  {journal} {ARA \& A}\ }\textbf
  {\bibinfo {volume} {42}},\ \href
  {https://doi.org/https://doi.org/10.1146/annurev.astro.42.053102.134013}
  {https://doi.org/10.1146/annurev.astro.42.053102.134013} (\bibinfo {year}
  {2004})\BibitemShut {NoStop}%
\bibitem [{\citenamefont {Ofengeim}\ \emph {et~al.}(2017)\citenamefont
  {Ofengeim}, \citenamefont {Fortin}, \citenamefont {Haensel}, \citenamefont
  {Yakovlev},\ and\ \citenamefont {Zdunik}}]{Ofengeim2017NeutrinoLum}%
  \BibitemOpen
  \bibfield  {author} {\bibinfo {author} {\bibfnamefont {D.~D.}\ \bibnamefont
  {Ofengeim}}, \bibinfo {author} {\bibfnamefont {M.}~\bibnamefont {Fortin}},
  \bibinfo {author} {\bibfnamefont {P.}~\bibnamefont {Haensel}}, \bibinfo
  {author} {\bibfnamefont {D.~G.}\ \bibnamefont {Yakovlev}},\ and\ \bibinfo
  {author} {\bibfnamefont {J.~L.}\ \bibnamefont {Zdunik}},\ }\href
  {https://doi.org/10.1103/PhysRevD.96.043002} {\bibfield  {journal} {\bibinfo
  {journal} {Phys. Rev. D}\ }\textbf {\bibinfo {volume} {96}},\ \bibinfo
  {pages} {043002} (\bibinfo {year} {2017})}\BibitemShut {NoStop}%
\bibitem [{\citenamefont {Passamonti}\ \emph {et~al.}(2017)\citenamefont
  {Passamonti}, \citenamefont {Akg{\"u}n}, \citenamefont {Pons},\ and\
  \citenamefont {Miralles}}]{passamonti2017magnetic}%
  \BibitemOpen
  \bibfield  {author} {\bibinfo {author} {\bibfnamefont {A.}~\bibnamefont
  {Passamonti}}, \bibinfo {author} {\bibfnamefont {T.}~\bibnamefont
  {Akg{\"u}n}}, \bibinfo {author} {\bibfnamefont {J.~A.}\ \bibnamefont
  {Pons}},\ and\ \bibinfo {author} {\bibfnamefont {J.~A.}\ \bibnamefont
  {Miralles}},\ }\href {https://doi.org/https://doi.org/10.1093/mnras/stx1192}
  {\bibfield  {journal} {\bibinfo  {journal} {MNRAS}\ }\textbf {\bibinfo
  {volume} {469}},\ \bibinfo {pages} {4979} (\bibinfo {year}
  {2017})}\BibitemShut {NoStop}%
\bibitem [{\citenamefont {Akgün}\ \emph {et~al.}(2013)\citenamefont {Akgün},
  \citenamefont {Reisenegger}, \citenamefont {Mastrano},\ and\ \citenamefont
  {Marchant}}]{akgun2013}%
  \BibitemOpen
  \bibfield  {author} {\bibinfo {author} {\bibfnamefont {T.}~\bibnamefont
  {Akgün}}, \bibinfo {author} {\bibfnamefont {A.}~\bibnamefont {Reisenegger}},
  \bibinfo {author} {\bibfnamefont {A.}~\bibnamefont {Mastrano}},\ and\
  \bibinfo {author} {\bibfnamefont {P.}~\bibnamefont {Marchant}},\ }\href
  {https://doi.org/10.1093/mnras/stt913} {\bibfield  {journal} {\bibinfo
  {journal} {MNRAS}\ }\textbf {\bibinfo {volume} {433}},\ \bibinfo {pages}
  {2445} (\bibinfo {year} {2013})}\BibitemShut {NoStop}%
\bibitem [{\citenamefont {Armaza}\ \emph {et~al.}(2015)\citenamefont {Armaza},
  \citenamefont {Reisenegger},\ and\ \citenamefont {Valdivia}}]{Armaza_2015}%
  \BibitemOpen
  \bibfield  {author} {\bibinfo {author} {\bibfnamefont {C.}~\bibnamefont
  {Armaza}}, \bibinfo {author} {\bibfnamefont {A.}~\bibnamefont
  {Reisenegger}},\ and\ \bibinfo {author} {\bibfnamefont {J.~A.}\ \bibnamefont
  {Valdivia}},\ }\href {https://doi.org/10.1088/0004-637x/802/2/121} {\bibfield
   {journal} {\bibinfo  {journal} {ApJ}\ }\textbf {\bibinfo {volume} {802}},\
  \bibinfo {pages} {121} (\bibinfo {year} {2015})}\BibitemShut {NoStop}%
\bibitem [{\citenamefont {{Prendergast}}(1956)}]{Prendergast1956}%
  \BibitemOpen
  \bibfield  {author} {\bibinfo {author} {\bibfnamefont {K.~H.}\ \bibnamefont
  {{Prendergast}}},\ }\href {https://doi.org/10.1086/146186} {\bibfield
  {journal} {\bibinfo  {journal} {ApJ}\ }\textbf {\bibinfo {volume} {123}},\
  \bibinfo {pages} {498} (\bibinfo {year} {1956})}\BibitemShut {NoStop}%
\bibitem [{\citenamefont {Grad}\ and\ \citenamefont
  {Rubin}(1958)}]{gradrubin54}%
  \BibitemOpen
  \bibfield  {author} {\bibinfo {author} {\bibfnamefont {H.}~\bibnamefont
  {Grad}}\ and\ \bibinfo {author} {\bibfnamefont {H.}~\bibnamefont {Rubin}},\
  }\href {https://doi.org/https://doi.org/10.1016/0891-3919(58)90139-6}
  {\bibfield  {journal} {\bibinfo  {journal} {J. Nucl. Phys.}\ }\textbf
  {\bibinfo {volume} {7}},\ \bibinfo {pages} {284 } (\bibinfo {year}
  {1958})}\BibitemShut {NoStop}%
\bibitem [{\citenamefont {{Shafranov}}(1966)}]{shafranov66}%
  \BibitemOpen
  \bibfield  {author} {\bibinfo {author} {\bibfnamefont {V.~D.}\ \bibnamefont
  {{Shafranov}}},\ }\href@noop {} {\bibfield  {journal} {\bibinfo  {journal}
  {Reviews of Plasma Physics}\ }\textbf {\bibinfo {volume} {2}},\ \bibinfo
  {pages} {103} (\bibinfo {year} {1966})}\BibitemShut {NoStop}%
\bibitem [{\citenamefont {{Reisenegger}}(1995)}]{Reisenegger1995}%
  \BibitemOpen
  \bibfield  {author} {\bibinfo {author} {\bibfnamefont {A.}~\bibnamefont
  {{Reisenegger}}},\ }\href {https://doi.org/10.1086/175480} {\bibfield
  {journal} {\bibinfo  {journal} {ApJ}\ }\textbf {\bibinfo {volume} {442}},\
  \bibinfo {pages} {749} (\bibinfo {year} {1995})}\BibitemShut {NoStop}%
\bibitem [{\citenamefont {{Fern{\'a}ndez}}\ and\ \citenamefont
  {{Reisenegger}}(2005)}]{Fernandez2005}%
  \BibitemOpen
  \bibfield  {author} {\bibinfo {author} {\bibfnamefont {R.}~\bibnamefont
  {{Fern{\'a}ndez}}}\ and\ \bibinfo {author} {\bibfnamefont {A.}~\bibnamefont
  {{Reisenegger}}},\ }\href {https://doi.org/10.1086/429551} {\bibfield
  {journal} {\bibinfo  {journal} {ApJ}\ }\textbf {\bibinfo {volume} {625}},\
  \bibinfo {pages} {291} (\bibinfo {year} {2005})}\BibitemShut {NoStop}%
\bibitem [{\citenamefont {Cumming}\ \emph {et~al.}(2004)\citenamefont
  {Cumming}, \citenamefont {Arras},\ and\ \citenamefont
  {Zweibel}}]{Cumming_2004}%
  \BibitemOpen
  \bibfield  {author} {\bibinfo {author} {\bibfnamefont {A.}~\bibnamefont
  {Cumming}}, \bibinfo {author} {\bibfnamefont {P.}~\bibnamefont {Arras}},\
  and\ \bibinfo {author} {\bibfnamefont {E.}~\bibnamefont {Zweibel}},\ }\href
  {https://doi.org/10.1086/421324} {\bibfield  {journal} {\bibinfo  {journal}
  {ApJ}\ }\textbf {\bibinfo {volume} {609}},\ \bibinfo {pages} {999} (\bibinfo
  {year} {2004})}\BibitemShut {NoStop}%
\bibitem [{\citenamefont {Baiko}\ \emph {et~al.}(2001)\citenamefont {Baiko},
  \citenamefont {Haensel},\ and\ \citenamefont {Yakovlev}}]{baiko2001thermal}%
  \BibitemOpen
  \bibfield  {author} {\bibinfo {author} {\bibfnamefont {D.}~\bibnamefont
  {Baiko}}, \bibinfo {author} {\bibfnamefont {P.}~\bibnamefont {Haensel}},\
  and\ \bibinfo {author} {\bibfnamefont {D.}~\bibnamefont {Yakovlev}},\ }\href
  {https://doi.org/https://doi.org/10.1051/0004-6361:20010621} {\bibfield
  {journal} {\bibinfo  {journal} {A\&A}\ }\textbf {\bibinfo {volume} {374}},\
  \bibinfo {pages} {151} (\bibinfo {year} {2001})}\BibitemShut {NoStop}%
\bibitem [{\citenamefont {{Kantor}}\ and\ \citenamefont
  {{Gusakov}}(2018)}]{KantorGusakov2018}%
  \BibitemOpen
  \bibfield  {author} {\bibinfo {author} {\bibfnamefont {E.~M.}\ \bibnamefont
  {{Kantor}}}\ and\ \bibinfo {author} {\bibfnamefont {M.~E.}\ \bibnamefont
  {{Gusakov}}},\ }\href {https://doi.org/10.1093/mnras/stx2682} {\bibfield
  {journal} {\bibinfo  {journal} {MNRAS}\ }\textbf {\bibinfo {volume} {473}},\
  \bibinfo {pages} {4272} (\bibinfo {year} {2018})}\BibitemShut {NoStop}%
\bibitem [{\citenamefont {Bransgrove}\ \emph {et~al.}(2018)\citenamefont
  {Bransgrove}, \citenamefont {Levin},\ and\ \citenamefont
  {Beloborodov}}]{Bransgroveetal2018}%
  \BibitemOpen
  \bibfield  {author} {\bibinfo {author} {\bibfnamefont {A.}~\bibnamefont
  {Bransgrove}}, \bibinfo {author} {\bibfnamefont {Y.}~\bibnamefont {Levin}},\
  and\ \bibinfo {author} {\bibfnamefont {A.}~\bibnamefont {Beloborodov}},\
  }\href {https://doi.org/10.1093/mnras/stx2508} {\bibfield  {journal}
  {\bibinfo  {journal} {MNRAS}\ }\textbf {\bibinfo {volume} {473}},\ \bibinfo
  {pages} {2771} (\bibinfo {year} {2018})}\BibitemShut {NoStop}%
\bibitem [{\citenamefont {Gusakov}(2019)}]{gusakov2019force}%
  \BibitemOpen
  \bibfield  {author} {\bibinfo {author} {\bibfnamefont {M.~E.}\ \bibnamefont
  {Gusakov}},\ }\href
  {https://academic.oup.com/mnras/article/485/4/4936/5372448} {\bibfield
  {journal} {\bibinfo  {journal} {MNRAS}\ }\textbf {\bibinfo {volume} {485}},\
  \bibinfo {pages} {4936} (\bibinfo {year} {2019})}\BibitemShut {NoStop}%
\bibitem [{\citenamefont
  {{Tayler}}(1973)}]{Tayler1973_AdiabaticStabilityStars}%
  \BibitemOpen
  \bibfield  {author} {\bibinfo {author} {\bibfnamefont {R.~J.}\ \bibnamefont
  {{Tayler}}},\ }\href {https://doi.org/10.1093/mnras/161.4.365} {\bibfield
  {journal} {\bibinfo  {journal} {MNRAS}\ }\textbf {\bibinfo {volume} {161}},\
  \bibinfo {pages} {365} (\bibinfo {year} {1973})}\BibitemShut {NoStop}%
\bibitem [{\citenamefont {{Markey}}\ and\ \citenamefont
  {{Tayler}}(1973)}]{Markey1973_AdiabaticStabilityStars}%
  \BibitemOpen
  \bibfield  {author} {\bibinfo {author} {\bibfnamefont {P.}~\bibnamefont
  {{Markey}}}\ and\ \bibinfo {author} {\bibfnamefont {R.~J.}\ \bibnamefont
  {{Tayler}}},\ }\href {https://doi.org/10.1093/mnras/163.1.77} {\bibfield
  {journal} {\bibinfo  {journal} {MNRAS}\ }\textbf {\bibinfo {volume} {163}},\
  \bibinfo {pages} {77} (\bibinfo {year} {1973})}\BibitemShut {NoStop}%
\bibitem [{\citenamefont {{Flowers}}\ and\ \citenamefont
  {{Ruderman}}(1977)}]{Flowers1977_EvolutionPulsarMagnetic}%
  \BibitemOpen
  \bibfield  {author} {\bibinfo {author} {\bibfnamefont {E.}~\bibnamefont
  {{Flowers}}}\ and\ \bibinfo {author} {\bibfnamefont {M.~A.}\ \bibnamefont
  {{Ruderman}}},\ }\href {https://doi.org/10.1086/155359} {\bibfield  {journal}
  {\bibinfo  {journal} {Astrophys. J.}\ }\textbf {\bibinfo {volume} {215}},\
  \bibinfo {pages} {302} (\bibinfo {year} {1977})}\BibitemShut {NoStop}%
\bibitem [{\citenamefont
  {{Braithwaite}}(2009)}]{Braithwaite2009_AxisymmetricMagneticFields}%
  \BibitemOpen
  \bibfield  {author} {\bibinfo {author} {\bibfnamefont {J.}~\bibnamefont
  {{Braithwaite}}},\ }\href {https://doi.org/10.1111/j.1365-2966.2008.14034.x}
  {\bibfield  {journal} {\bibinfo  {journal} {MNRAS}\ }\textbf {\bibinfo
  {volume} {397}},\ \bibinfo {pages} {763} (\bibinfo {year}
  {2009})}\BibitemShut {NoStop}%
\bibitem [{\citenamefont {{Lander}}\ and\ \citenamefont
  {{Jones}}(2012)}]{Lander2012_AreThereAny}%
  \BibitemOpen
  \bibfield  {author} {\bibinfo {author} {\bibfnamefont {S.~K.}\ \bibnamefont
  {{Lander}}}\ and\ \bibinfo {author} {\bibfnamefont {D.~I.}\ \bibnamefont
  {{Jones}}},\ }\href {https://doi.org/10.1111/j.1365-2966.2012.21213.x}
  {\bibfield  {journal} {\bibinfo  {journal} {MNRAS}\ }\textbf {\bibinfo
  {volume} {424}},\ \bibinfo {pages} {482} (\bibinfo {year}
  {2012})}\BibitemShut {NoStop}%
\bibitem [{\citenamefont {{Mitchell}}\ \emph {et~al.}(2015)\citenamefont
  {{Mitchell}}, \citenamefont {{Braithwaite}}, \citenamefont {{Reisenegger}},
  \citenamefont {{Spruit}}, \citenamefont {{Valdivia}},\ and\ \citenamefont
  {{Langer}}}]{Mitchell2015_InstabilityMagneticEquilibria}%
  \BibitemOpen
  \bibfield  {author} {\bibinfo {author} {\bibfnamefont {J.~P.}\ \bibnamefont
  {{Mitchell}}}, \bibinfo {author} {\bibfnamefont {J.}~\bibnamefont
  {{Braithwaite}}}, \bibinfo {author} {\bibfnamefont {A.}~\bibnamefont
  {{Reisenegger}}}, \bibinfo {author} {\bibfnamefont {H.}~\bibnamefont
  {{Spruit}}}, \bibinfo {author} {\bibfnamefont {J.~A.}\ \bibnamefont
  {{Valdivia}}},\ and\ \bibinfo {author} {\bibfnamefont {N.}~\bibnamefont
  {{Langer}}},\ }\href {https://doi.org/10.1093/mnras/stu2514} {\bibfield
  {journal} {\bibinfo  {journal} {MNRAS}\ }\textbf {\bibinfo {volume} {447}},\
  \bibinfo {pages} {1213} (\bibinfo {year} {2015})}\BibitemShut {NoStop}%
\bibitem [{\citenamefont {{Becerra}}\ \emph
  {et~al.}(2022{\natexlab{b}})\citenamefont {{Becerra}}, \citenamefont
  {{Reisenegger}}, \citenamefont {{Valdivia}},\ and\ \citenamefont
  {{Gusakov}}}]{Becerra2022_StabilityAxiallySymmetric}%
  \BibitemOpen
  \bibfield  {author} {\bibinfo {author} {\bibfnamefont {L.}~\bibnamefont
  {{Becerra}}}, \bibinfo {author} {\bibfnamefont {A.}~\bibnamefont
  {{Reisenegger}}}, \bibinfo {author} {\bibfnamefont {J.~A.}\ \bibnamefont
  {{Valdivia}}},\ and\ \bibinfo {author} {\bibfnamefont {M.~E.}\ \bibnamefont
  {{Gusakov}}},\ }\href {https://doi.org/10.1093/mnras/stac2704} {\bibfield
  {journal} {\bibinfo  {journal} {MNRAS}\ }\textbf {\bibinfo {volume} {517}},\
  \bibinfo {pages} {560} (\bibinfo {year} {2022}{\natexlab{b}})}\BibitemShut
  {NoStop}%
\bibitem [{\citenamefont {{Shternin}}\ and\ \citenamefont
  {{Ofengeim}}(2022)}]{so22}%
  \BibitemOpen
  \bibfield  {author} {\bibinfo {author} {\bibfnamefont {P.}~\bibnamefont
  {{Shternin}}}\ and\ \bibinfo {author} {\bibfnamefont {D.}~\bibnamefont
  {{Ofengeim}}},\ }\href {https://doi.org/10.1140/epja/s10050-022-00687-w}
  {\bibfield  {journal} {\bibinfo  {journal} {Eur. Phys. J. A}\ }\textbf
  {\bibinfo {volume} {58}},\ \bibinfo {eid} {42} (\bibinfo {year}
  {2022})}\BibitemShut {NoStop}%
\bibitem [{\citenamefont {Yakovlev}\ and\ \citenamefont
  {Shalybkov}(1990)}]{yakovlev1990electrical}%
  \BibitemOpen
  \bibfield  {author} {\bibinfo {author} {\bibfnamefont {D.}~\bibnamefont
  {Yakovlev}}\ and\ \bibinfo {author} {\bibfnamefont {D.}~\bibnamefont
  {Shalybkov}},\ }\href {https://doi.org/10.1007/BF00646698} {\bibfield
  {journal} {\bibinfo  {journal} {SvA Lett}\ }\textbf {\bibinfo {volume}
  {16}},\ \bibinfo {pages} {86} (\bibinfo {year} {1990})}\BibitemShut {NoStop}%
\bibitem [{\citenamefont {Beznogov}\ \emph {et~al.}(2021)\citenamefont
  {Beznogov}, \citenamefont {Potekhin},\ and\ \citenamefont
  {Yakovlev}}]{BEZNOGOV20211}%
  \BibitemOpen
  \bibfield  {author} {\bibinfo {author} {\bibfnamefont {M.}~\bibnamefont
  {Beznogov}}, \bibinfo {author} {\bibfnamefont {A.}~\bibnamefont {Potekhin}},\
  and\ \bibinfo {author} {\bibfnamefont {D.}~\bibnamefont {Yakovlev}},\ }\href
  {https://doi.org/https://doi.org/10.1016/j.physrep.2021.03.004} {\bibfield
  {journal} {\bibinfo  {journal} {Phys. Rep.}\ }\textbf {\bibinfo {volume}
  {919}},\ \bibinfo {pages} {1} (\bibinfo {year} {2021})}\BibitemShut {NoStop}%
\bibitem [{\citenamefont {{Potekhin}}\ and\ \citenamefont
  {{Chabrier}}(2018)}]{PotekhinChabrier18}%
  \BibitemOpen
  \bibfield  {author} {\bibinfo {author} {\bibfnamefont {A.~Y.}\ \bibnamefont
  {{Potekhin}}}\ and\ \bibinfo {author} {\bibfnamefont {G.}~\bibnamefont
  {{Chabrier}}},\ }\href {https://doi.org/10.1051/0004-6361/201731866}
  {\bibfield  {journal} {\bibinfo  {journal} {Astron. Astrophys.}\ }\textbf
  {\bibinfo {volume} {609}},\ \bibinfo {eid} {A74} (\bibinfo {year}
  {2018})}\BibitemShut {NoStop}%
\bibitem [{\citenamefont {{Yakovlev}}\ \emph {et~al.}(2021)\citenamefont
  {{Yakovlev}}, \citenamefont {{Kaminker}}, \citenamefont {{Potekhin}},\ and\
  \citenamefont {{Haensel}}}]{Yakovlev_21}%
  \BibitemOpen
  \bibfield  {author} {\bibinfo {author} {\bibfnamefont {D.~G.}\ \bibnamefont
  {{Yakovlev}}}, \bibinfo {author} {\bibfnamefont {A.~D.}\ \bibnamefont
  {{Kaminker}}}, \bibinfo {author} {\bibfnamefont {A.~Y.}\ \bibnamefont
  {{Potekhin}}},\ and\ \bibinfo {author} {\bibfnamefont {P.}~\bibnamefont
  {{Haensel}}},\ }\href {https://doi.org/10.1093/mnras/staa3547} {\bibfield
  {journal} {\bibinfo  {journal} {MNRAS}\ }\textbf {\bibinfo {volume} {500}},\
  \bibinfo {pages} {4491} (\bibinfo {year} {2021})}\BibitemShut {NoStop}%
\bibitem [{\citenamefont {{Kaminker}}\ \emph {et~al.}(2023)\citenamefont
  {{Kaminker}}, \citenamefont {{Potekhin}},\ and\ \citenamefont
  {{Yakovlev}}}]{KaminkerPY23}%
  \BibitemOpen
  \bibfield  {author} {\bibinfo {author} {\bibfnamefont {A.~D.}\ \bibnamefont
  {{Kaminker}}}, \bibinfo {author} {\bibfnamefont {A.~Y.}\ \bibnamefont
  {{Potekhin}}},\ and\ \bibinfo {author} {\bibfnamefont {D.~G.}\ \bibnamefont
  {{Yakovlev}}},\ }\href {https://doi.org/10.1134/S1063773723120034} {\bibfield
   {journal} {\bibinfo  {journal} {Astron. Lett.}\ }\textbf {\bibinfo {volume}
  {49}},\ \bibinfo {pages} {824} (\bibinfo {year} {2023})}\BibitemShut
  {NoStop}%
\bibitem [{\citenamefont {Gudmundsson}\ \emph {et~al.}(1983)\citenamefont
  {Gudmundsson}, \citenamefont {Pethick},\ and\ \citenamefont
  {Epstein}}]{gudmundsson1983}%
  \BibitemOpen
  \bibfield  {author} {\bibinfo {author} {\bibfnamefont {E.~H.}\ \bibnamefont
  {Gudmundsson}}, \bibinfo {author} {\bibfnamefont {C.}~\bibnamefont
  {Pethick}},\ and\ \bibinfo {author} {\bibfnamefont {R.~I.}\ \bibnamefont
  {Epstein}},\ }\href {https://doi.org/10.1086/161292} {\bibfield  {journal}
  {\bibinfo  {journal} {ApJ}\ }\textbf {\bibinfo {volume} {272}},\ \bibinfo
  {pages} {286} (\bibinfo {year} {1983})}\BibitemShut {NoStop}%
\bibitem [{\citenamefont {{Gnedin}}\ \emph {et~al.}(2001)\citenamefont
  {{Gnedin}}, \citenamefont {{Yakovlev}},\ and\ \citenamefont
  {{Potekhin}}}]{GnedinYP01}%
  \BibitemOpen
  \bibfield  {author} {\bibinfo {author} {\bibfnamefont {O.~Y.}\ \bibnamefont
  {{Gnedin}}}, \bibinfo {author} {\bibfnamefont {D.~G.}\ \bibnamefont
  {{Yakovlev}}},\ and\ \bibinfo {author} {\bibfnamefont {A.~Y.}\ \bibnamefont
  {{Potekhin}}},\ }\href {https://doi.org/10.1046/j.1365-8711.2001.04359.x}
  {\bibfield  {journal} {\bibinfo  {journal} {MNRAS}\ }\textbf {\bibinfo
  {volume} {324}},\ \bibinfo {pages} {725} (\bibinfo {year}
  {2001})}\BibitemShut {NoStop}%
\bibitem [{\citenamefont {{Potekhin}}\ and\ \citenamefont
  {{Yakovlev}}(2001)}]{PotekhinYakovlev01}%
  \BibitemOpen
  \bibfield  {author} {\bibinfo {author} {\bibfnamefont {A.~Y.}\ \bibnamefont
  {{Potekhin}}}\ and\ \bibinfo {author} {\bibfnamefont {D.~G.}\ \bibnamefont
  {{Yakovlev}}},\ }\href {https://doi.org/10.1051/0004-6361:20010698}
  {\bibfield  {journal} {\bibinfo  {journal} {Astron. Astrophys.}\ }\textbf
  {\bibinfo {volume} {374}},\ \bibinfo {pages} {213} (\bibinfo {year}
  {2001})}\BibitemShut {NoStop}%
\bibitem [{\citenamefont {{Potekhin}}\ \emph {et~al.}(2007)\citenamefont
  {{Potekhin}}, \citenamefont {{Chabrier}},\ and\ \citenamefont
  {{Yakovlev}}}]{PotekhinCY07}%
  \BibitemOpen
  \bibfield  {author} {\bibinfo {author} {\bibfnamefont {A.~Y.}\ \bibnamefont
  {{Potekhin}}}, \bibinfo {author} {\bibfnamefont {G.}~\bibnamefont
  {{Chabrier}}},\ and\ \bibinfo {author} {\bibfnamefont {D.~G.}\ \bibnamefont
  {{Yakovlev}}},\ }\href {https://doi.org/10.1007/s10509-007-9362-6} {\bibfield
   {journal} {\bibinfo  {journal} {Astrophys. Space Sci.}\ }\textbf {\bibinfo
  {volume} {308}},\ \bibinfo {pages} {353} (\bibinfo {year}
  {2007})}\BibitemShut {NoStop}%
\bibitem [{\citenamefont {Potekhin}\ \emph {et~al.}(2015)\citenamefont
  {Potekhin}, \citenamefont {Pons},\ and\ \citenamefont
  {Page}}]{potekhin2015neutron}%
  \BibitemOpen
  \bibfield  {author} {\bibinfo {author} {\bibfnamefont {A.~Y.}\ \bibnamefont
  {Potekhin}}, \bibinfo {author} {\bibfnamefont {J.~A.}\ \bibnamefont {Pons}},\
  and\ \bibinfo {author} {\bibfnamefont {D.}~\bibnamefont {Page}},\ }\href
  {https://doi.org/https://doi.org/10.1007/s11214-015-0180-9} {\bibfield
  {journal} {\bibinfo  {journal} {Space Sci. Rev.}\ }\textbf {\bibinfo {volume}
  {191}},\ \bibinfo {pages} {239} (\bibinfo {year} {2015})}\BibitemShut
  {NoStop}%
\bibitem [{\citenamefont {{Potekhin}}\ and\ \citenamefont
  {{Chabrier}}(2013)}]{PotekhinChabrier13}%
  \BibitemOpen
  \bibfield  {author} {\bibinfo {author} {\bibfnamefont {A.~Y.}\ \bibnamefont
  {{Potekhin}}}\ and\ \bibinfo {author} {\bibfnamefont {G.}~\bibnamefont
  {{Chabrier}}},\ }\href {https://doi.org/10.1051/0004-6361/201220082}
  {\bibfield  {journal} {\bibinfo  {journal} {Astron. Astrophys.}\ }\textbf
  {\bibinfo {volume} {550}},\ \bibinfo {eid} {A43} (\bibinfo {year}
  {2013})}\BibitemShut {NoStop}%
\bibitem [{\citenamefont {{Pearson}}\ \emph {et~al.}(2018)\citenamefont
  {{Pearson}}, \citenamefont {{Chamel}}, \citenamefont {{Potekhin}},
  \citenamefont {{Fantina}}, \citenamefont {{Ducoin}}, \citenamefont
  {{Dutta}},\ and\ \citenamefont {{Goriely}}}]{Pearson_18}%
  \BibitemOpen
  \bibfield  {author} {\bibinfo {author} {\bibfnamefont {J.~M.}\ \bibnamefont
  {{Pearson}}}, \bibinfo {author} {\bibfnamefont {N.}~\bibnamefont {{Chamel}}},
  \bibinfo {author} {\bibfnamefont {A.~Y.}\ \bibnamefont {{Potekhin}}},
  \bibinfo {author} {\bibfnamefont {A.~F.}\ \bibnamefont {{Fantina}}}, \bibinfo
  {author} {\bibfnamefont {C.}~\bibnamefont {{Ducoin}}}, \bibinfo {author}
  {\bibfnamefont {A.~K.}\ \bibnamefont {{Dutta}}},\ and\ \bibinfo {author}
  {\bibfnamefont {S.}~\bibnamefont {{Goriely}}},\ }\href
  {https://doi.org/10.1093/mnras/sty2413} {\bibfield  {journal} {\bibinfo
  {journal} {MNRAS}\ }\textbf {\bibinfo {volume} {481}},\ \bibinfo {pages}
  {2994} (\bibinfo {year} {2018})}\BibitemShut {NoStop}%
\bibitem [{\citenamefont {{Potekhin}}\ \emph {et~al.}(2003)\citenamefont
  {{Potekhin}}, \citenamefont {{Yakovlev}}, \citenamefont {{Chabrier}},\ and\
  \citenamefont {{Gnedin}}}]{Potekhin_03}%
  \BibitemOpen
  \bibfield  {author} {\bibinfo {author} {\bibfnamefont {A.~Y.}\ \bibnamefont
  {{Potekhin}}}, \bibinfo {author} {\bibfnamefont {D.~G.}\ \bibnamefont
  {{Yakovlev}}}, \bibinfo {author} {\bibfnamefont {G.}~\bibnamefont
  {{Chabrier}}},\ and\ \bibinfo {author} {\bibfnamefont {O.~Y.}\ \bibnamefont
  {{Gnedin}}},\ }\href {https://doi.org/10.1086/376900} {\bibfield  {journal}
  {\bibinfo  {journal} {Astrophys. J.}\ }\textbf {\bibinfo {volume} {594}},\
  \bibinfo {pages} {404} (\bibinfo {year} {2003})}\BibitemShut {NoStop}%
\bibitem [{\citenamefont {{Potekhin}}\ \emph {et~al.}(1997)\citenamefont
  {{Potekhin}}, \citenamefont {{Chabrier}},\ and\ \citenamefont
  {{Yakovlev}}}]{potekhin1997}%
  \BibitemOpen
  \bibfield  {author} {\bibinfo {author} {\bibfnamefont {A.~Y.}\ \bibnamefont
  {{Potekhin}}}, \bibinfo {author} {\bibfnamefont {G.}~\bibnamefont
  {{Chabrier}}},\ and\ \bibinfo {author} {\bibfnamefont {D.~G.}\ \bibnamefont
  {{Yakovlev}}},\ }\href@noop {} {\bibfield  {journal} {\bibinfo  {journal}
  {Astron. Astrophys.}\ }\textbf {\bibinfo {volume} {323}},\ \bibinfo {pages}
  {415} (\bibinfo {year} {1997})}\BibitemShut {NoStop}%
\bibitem [{\citenamefont {{Gough}}\ and\ \citenamefont
  {{Tayler}}(1966)}]{GoughTayler66}%
  \BibitemOpen
  \bibfield  {author} {\bibinfo {author} {\bibfnamefont {D.~O.}\ \bibnamefont
  {{Gough}}}\ and\ \bibinfo {author} {\bibfnamefont {R.~J.}\ \bibnamefont
  {{Tayler}}},\ }\href {https://doi.org/10.1093/mnras/133.1.85} {\bibfield
  {journal} {\bibinfo  {journal} {MNRAS}\ }\textbf {\bibinfo {volume} {133}},\
  \bibinfo {pages} {85} (\bibinfo {year} {1966})}\BibitemShut {NoStop}%
\bibitem [{\citenamefont {{Greenstein}}\ and\ \citenamefont
  {{Hartke}}(1983)}]{GreensteinHartke83}%
  \BibitemOpen
  \bibfield  {author} {\bibinfo {author} {\bibfnamefont {G.}~\bibnamefont
  {{Greenstein}}}\ and\ \bibinfo {author} {\bibfnamefont {G.~J.}\ \bibnamefont
  {{Hartke}}},\ }\href {https://doi.org/10.1086/161195} {\bibfield  {journal}
  {\bibinfo  {journal} {Astrophys. J.}\ }\textbf {\bibinfo {volume} {271}},\
  \bibinfo {pages} {283} (\bibinfo {year} {1983})}\BibitemShut {NoStop}%
\bibitem [{\citenamefont {{Potekhin}}\ \emph {et~al.}(2020)\citenamefont
  {{Potekhin}}, \citenamefont {{Zyuzin}}, \citenamefont {{Yakovlev}},
  \citenamefont {{Beznogov}}, \citenamefont {{Shibanov}},\ and\ \citenamefont
  {et~al.}}]{PotekhinMNRAS2020}%
  \BibitemOpen
  \bibfield  {author} {\bibinfo {author} {\bibfnamefont {A.~Y.}\ \bibnamefont
  {{Potekhin}}}, \bibinfo {author} {\bibfnamefont {D.~A.}\ \bibnamefont
  {{Zyuzin}}}, \bibinfo {author} {\bibfnamefont {D.~G.}\ \bibnamefont
  {{Yakovlev}}}, \bibinfo {author} {\bibfnamefont {M.~V.}\ \bibnamefont
  {{Beznogov}}}, \bibinfo {author} {\bibfnamefont {Y.~A.}\ \bibnamefont
  {{Shibanov}}},\ and\ \bibinfo {author} {\bibnamefont {et~al.}},\ }\href
  {https://doi.org/10.1093/mnras/staa1871} {\bibfield  {journal} {\bibinfo
  {journal} {MNRAS}\ }\textbf {\bibinfo {volume} {496}},\ \bibinfo {pages}
  {5052} (\bibinfo {year} {2020})}\BibitemShut {NoStop}%
\bibitem [{\citenamefont {{Hu}}\ \emph {et~al.}(2020)\citenamefont {{Hu}},
  \citenamefont {{Begi{\c{c}}arslan}}, \citenamefont {{G{\"u}ver}},
  \citenamefont {{Enoto}}, \citenamefont {{Younes}},\ and\ \citenamefont
  {et~al.}}]{HuApJ2020}%
  \BibitemOpen
  \bibfield  {author} {\bibinfo {author} {\bibfnamefont {C.-P.}\ \bibnamefont
  {{Hu}}}, \bibinfo {author} {\bibfnamefont {B.}~\bibnamefont
  {{Begi{\c{c}}arslan}}}, \bibinfo {author} {\bibfnamefont {T.}~\bibnamefont
  {{G{\"u}ver}}}, \bibinfo {author} {\bibfnamefont {T.}~\bibnamefont
  {{Enoto}}}, \bibinfo {author} {\bibfnamefont {G.}~\bibnamefont {{Younes}}},\
  and\ \bibinfo {author} {\bibnamefont {et~al.}},\ }\href
  {https://doi.org/10.3847/1538-4357/abb3c9} {\bibfield  {journal} {\bibinfo
  {journal} {Astrophys. J.}\ }\textbf {\bibinfo {volume} {902}},\ \bibinfo
  {eid} {1} (\bibinfo {year} {2020})}\BibitemShut {NoStop}%
\bibitem [{\citenamefont {{Rajwade}}\ \emph {et~al.}(2022)\citenamefont
  {{Rajwade}}, \citenamefont {{Stappers}}, \citenamefont {{Lyne}},
  \citenamefont {{Shaw}}, \citenamefont {{Mickaliger}},\ and\ \citenamefont
  {et~al.}}]{RajwadeMNRAS2022}%
  \BibitemOpen
  \bibfield  {author} {\bibinfo {author} {\bibfnamefont {K.~M.}\ \bibnamefont
  {{Rajwade}}}, \bibinfo {author} {\bibfnamefont {B.~W.}\ \bibnamefont
  {{Stappers}}}, \bibinfo {author} {\bibfnamefont {A.~G.}\ \bibnamefont
  {{Lyne}}}, \bibinfo {author} {\bibfnamefont {B.}~\bibnamefont {{Shaw}}},
  \bibinfo {author} {\bibfnamefont {M.~B.}\ \bibnamefont {{Mickaliger}}},\ and\
  \bibinfo {author} {\bibnamefont {et~al.}},\ }\href
  {https://doi.org/10.1093/mnras/stac446} {\bibfield  {journal} {\bibinfo
  {journal} {MNRAS}\ }\textbf {\bibinfo {volume} {512}},\ \bibinfo {pages}
  {1687} (\bibinfo {year} {2022})}\BibitemShut {NoStop}%
\bibitem [{\citenamefont {{Hu}}\ \emph {et~al.}(2019)\citenamefont {{Hu}},
  \citenamefont {{Ng}},\ and\ \citenamefont {{Ho}}}]{HuNgHoMNRAS2019}%
  \BibitemOpen
  \bibfield  {author} {\bibinfo {author} {\bibfnamefont {C.-P.}\ \bibnamefont
  {{Hu}}}, \bibinfo {author} {\bibfnamefont {C.~Y.}\ \bibnamefont {{Ng}}},\
  and\ \bibinfo {author} {\bibfnamefont {W.~C.~G.}\ \bibnamefont {{Ho}}},\
  }\href {https://doi.org/10.1093/mnras/stz513} {\bibfield  {journal} {\bibinfo
   {journal} {MNRAS}\ }\textbf {\bibinfo {volume} {485}},\ \bibinfo {pages}
  {4274} (\bibinfo {year} {2019})}\BibitemShut {NoStop}%
\bibitem [{\citenamefont {{Coti Zelati}}\ \emph {et~al.}(2017)\citenamefont
  {{Coti Zelati}}, \citenamefont {{Rea}}, \citenamefont {{Turolla}},
  \citenamefont {{Pons}}, \citenamefont {{Papitto}},\ and\ \citenamefont
  {et~al.}}]{CotiZelatiMNRAS2017}%
  \BibitemOpen
  \bibfield  {author} {\bibinfo {author} {\bibfnamefont {F.}~\bibnamefont
  {{Coti Zelati}}}, \bibinfo {author} {\bibfnamefont {N.}~\bibnamefont
  {{Rea}}}, \bibinfo {author} {\bibfnamefont {R.}~\bibnamefont {{Turolla}}},
  \bibinfo {author} {\bibfnamefont {J.~A.}\ \bibnamefont {{Pons}}}, \bibinfo
  {author} {\bibfnamefont {A.}~\bibnamefont {{Papitto}}},\ and\ \bibinfo
  {author} {\bibnamefont {et~al.}},\ }\href
  {https://doi.org/https://doi.org/10.1093/mnras/stx1700} {\bibfield  {journal}
  {\bibinfo  {journal} {MNRAS}\ }\textbf {\bibinfo {volume} {471}},\ \bibinfo
  {pages} {1819} (\bibinfo {year} {2017})}\BibitemShut {NoStop}%
\bibitem [{\citenamefont {{Enoto}}\ \emph {et~al.}(2021)\citenamefont
  {{Enoto}}, \citenamefont {{Ng}}, \citenamefont {{Hu}}, \citenamefont
  {{G{\"u}ver}}, \citenamefont {{Jaisawal}},\ and\ \citenamefont
  {et~al.}}]{EnotoApJL2021}%
  \BibitemOpen
  \bibfield  {author} {\bibinfo {author} {\bibfnamefont {T.}~\bibnamefont
  {{Enoto}}}, \bibinfo {author} {\bibfnamefont {M.}~\bibnamefont {{Ng}}},
  \bibinfo {author} {\bibfnamefont {C.-P.}\ \bibnamefont {{Hu}}}, \bibinfo
  {author} {\bibfnamefont {T.}~\bibnamefont {{G{\"u}ver}}}, \bibinfo {author}
  {\bibfnamefont {G.~K.}\ \bibnamefont {{Jaisawal}}},\ and\ \bibinfo {author}
  {\bibnamefont {et~al.}},\ }\href {https://doi.org/10.3847/2041-8213/ac2665}
  {\bibfield  {journal} {\bibinfo  {journal} {Astrophys. J. Lett.}\ }\textbf
  {\bibinfo {volume} {920}},\ \bibinfo {eid} {L4} (\bibinfo {year}
  {2021})}\BibitemShut {NoStop}%
\bibitem [{\citenamefont {{Park}}\ \emph {et~al.}(2020)\citenamefont {{Park}},
  \citenamefont {{Bhalerao}}, \citenamefont {{Kargaltsev}},\ and\ \citenamefont
  {{Slane}}}]{Park+ApJ2020_0526}%
  \BibitemOpen
  \bibfield  {author} {\bibinfo {author} {\bibfnamefont {S.}~\bibnamefont
  {{Park}}}, \bibinfo {author} {\bibfnamefont {J.}~\bibnamefont {{Bhalerao}}},
  \bibinfo {author} {\bibfnamefont {O.}~\bibnamefont {{Kargaltsev}}},\ and\
  \bibinfo {author} {\bibfnamefont {P.~O.}\ \bibnamefont {{Slane}}},\ }\href
  {https://doi.org/10.3847/1538-4357/ab83f8} {\bibfield  {journal} {\bibinfo
  {journal} {\apj}\ }\textbf {\bibinfo {volume} {894}},\ \bibinfo {eid} {17}
  (\bibinfo {year} {2020})}\BibitemShut {NoStop}%
\end{thebibliography}%
\bibliographystyle{apsrev4-2}

\end{document}